\documentclass[reprint,aps,prx,twocolumn,superscriptaddress,nofootinbib,longbibliography]{revtex4-2}

\usepackage{amsmath,amssymb,amsfonts,mathtools}
\usepackage{physics}
\usepackage{graphicx}
\usepackage{bm}
\usepackage{hyperref}
\usepackage{booktabs}
\usepackage{xcolor}
\usepackage{pifont}
\usepackage{tikz}
\usetikzlibrary{arrows.meta,decorations.markings,calc,patterns}

\makeatletter
\def\Revised@name{revised }\def\Accepted@name{accepted }\def\Published@name{published }
\makeatother

\hypersetup{
    colorlinks=true,
    linkcolor=blue,
    citecolor=blue,
    urlcolor=blue
}

\begin{document}

\title{Topological Phase Transitions and Their Thermodynamic Fate in Arbitrary-$S$ Classical Pyrochlore Spin Ice}

\author{Sena Watanabe}
\email[Contact author: ]{watanabe-sena397@g.ecc.u-tokyo.ac.jp}
\affiliation{Department of Applied Physics, The University of Tokyo, Tokyo 113-8656, Japan}

\author{Yukitoshi Motome}
\email[Contact author: ]{motome@ap.t.u-tokyo.ac.jp}
\affiliation{Department of Applied Physics, The University of Tokyo, Tokyo 113-8656, Japan}

\author{Haruki Watanabe}
\email[Contact author: ]{hwatanabe@ust.hk}
\affiliation{Department of Physics, Hong Kong University of Science and Technology, Clear Water Bay, Hong Kong, China}
\affiliation{Institute for Advanced Study, Hong Kong University of Science and Technology, Clear Water Bay, Hong Kong, China}
\affiliation{Center for Theoretical Condensed Matter Physics, Hong Kong University of Science and Technology, Clear Water Bay, Hong Kong, China}
\affiliation{Department of Applied Physics, The University of Tokyo, Tokyo 113-8656, Japan}

\received{5 April 2026}
\revised{15 August 2026}
\accepted{8 September 2026}
\published{14 October 2026}

\begin{abstract}
We develop a self-contained theoretical framework that classifies the topological phases and critical phenomena of classical pyrochlore magnets with arbitrary spin $S$, subject to competing exchange and single-ion anisotropies. We work first in the monopole-free (ice-rule) limit. In the small-$w$ regime, where the single-ion term favors low spin amplitudes, dualities reveal a dichotomy: Integer spins exhibit a continuous 3D $XY$ deconfinement transition, whereas half-integer spins remain in a $U(1)$ Coulomb liquid without any transition. In the large-$w$ regime, where the local spin amplitudes are maximized ($|S^z| = S$), the macroscopic flux is quantized to multiples of $2S$. By mapping the defect structure to topological loop gases, we prove that the compatibility between the physical ice rule and the modular $\mathbb{Z}_{2S}$ flux conservation holds at the vertex level if and only if $S \le 3/2$. For $S=3/2$, this maps the system to the three-state Potts model, whose symmetry-allowed cubic invariant drives a first-order transition. For $S \ge 2$, no mapping to a discrete clock model exists. Using a decomposition of the partition function, we show that the only processes sensitive to the discrete $\mathbb{Z}_{2S}$ structure carry exponentially small statistical weight and correspond to irrelevant operators at the 3D $XY$ fixed point, so the deconfinement transition is of the 3D $XY$ type. Finally, incorporating thermal monopoles, we show that they act as a symmetry-breaking effective magnetic field that severs defect strings. Consequently, the continuous transitions are rounded into crossovers, whereas the first-order $S=3/2$ transition is predicted to survive at finite temperatures, terminating at a critical end point. Classical Monte Carlo simulations for $S$ up to $7/2$ corroborate the predicted crossover behavior at the accessible temperatures; these temperatures lie above the estimated critical end point, so the predicted survival of the $S=3/2$ first-order transition remains to be tested directly.
\end{abstract}

\maketitle

\section{Introduction}

The study of geometrically frustrated magnetism on the pyrochlore lattice has been a fertile ground for discovering exotic states of matter, most notably classical and quantum spin liquids \cite{Balents2010, Savary2017, Knolle2019, Broholm2020}. The quintessential example is classical spin ice, realized in rare-earth oxides such as $\mathrm{Ho_2Ti_2O_7}$ and $\mathrm{Dy_2Ti_2O_7}$ \cite{Harris1997, Bramwell2001, Ramirez1999, Castelnovo2008}. In these materials, strong crystalline electric fields acting on the large total angular momenta ($J=8$ for $\mathrm{Ho^{3+}}$, $J=15/2$ for $\mathrm{Dy^{3+}}$) isolate a low-energy doublet, allowing the magnetic moments to be treated as effective $S=1/2$ Ising pseudospins oriented along the local $\langle 111 \rangle$ tetrahedral axes \cite{Melko2001}. The frustrated exchange interactions force the spins into a two-in, two-out ice rule \cite{Pauling1935}, mapping the low-energy physics to a fluctuating, divergence-free effective magnetic field. This emergent electromagnetism characterizes the $U(1)$ Coulomb liquid phase, distinguished by algebraic spin correlations and pinch-point singularities \cite{Hermele2004, Isakov2004, Henley2005, Henley2010, Fennell2009}.

Recent theoretical and numerical advances have pushed beyond the rigid $S=1/2$ paradigm by exploring $S=1$ and higher-spin systems in which the dominant easy-axis exchange competes with a single-ion anisotropy \cite{Pandey2025_S1, KunduDamle2025, Watanabe2026, Pandey2026_S32}. Throughout this paper, $S^z_\ell$ denotes the spin component along the local $\langle 111 \rangle$ axis of site $\ell$ (the axis joining the centers of the two tetrahedra sharing that site), the transverse components of the exchange are set to zero, and the model is treated entirely classically. For the $S=1$ case, the competition between a dominant easy-axis exchange and a single-ion anisotropy $\mu(S^z)^2$ ($\mu > 0$) that penalizes large $|S^z|$ yields a rich phase diagram \cite{Pandey2025_S1, KunduDamle2025, Watanabe2026}. Unlike the $S=1/2$ model, the $S=1$ spins can access a nonmagnetic $S^z=0$ state. In the monopole-free (ice-rule) limit, tuning the anisotropy drives the system through three distinct macroscopic phases: a trivial paramagnet, a $U(1)$ Coulomb phase, and a $\mathbb{Z}_2$-confined Coulomb phase (a Coulomb liquid whose macroscopic polarization flux is restricted to even integers, so that strings carrying a single unit of flux are confined while pairs remain free), separated by continuous transitions in the 3D $XY$ and 3D Ising universality classes. More recently, Pandey et al. investigated the $S=3/2$ case, uncovering a first-order $\mathbb{Z}_3$ flux-confinement transition separating two distinct Coulomb liquids \cite{Pandey2026_S32}. These developments connect to the broader understanding of fractionalization and magnetic fragmentation in pyrochlore systems \cite{BrooksBartlett2014, Petit2016, Rehn2017}. The same class of models also arises in a broader range of systems, including water ice under high pressure \cite{Watanabe2026ice} and artificial nanomagnet arrays \cite{Berchialla2026}.

These discoveries raise fundamental questions. How do the macroscopic topological structures and critical phenomena generalize to an arbitrary spin $S$? Why does the $S=3/2$ model exhibit a first-order transition while $S=1$ exhibits continuous transitions, and can the first-order character be derived from a microscopic mapping rather than inferred from symmetry considerations \cite{Pandey2026_S32}? What is the ultimate fate of the phase transitions and universality classes for $S \ge 2$? Finally, what is the thermodynamic fate of these topological phase transitions at finite temperatures, where the strict ice rules are inevitably violated by thermally excited magnetic monopoles?

In this paper, we address these questions by developing a unified theoretical framework for the spin-$S$ pyrochlore magnet, treating the monopole-free limit first and thermal monopoles second. In the monopole-free limit, we first show that the phase diagram contains only the phases already identified above: Uniform backgrounds of intermediate spin amplitude are excluded by the convexity of the single-ion energy, and scenarios of ``partial deconfinement,'' in which strings of higher topological charge would condense before the fundamental strings, are excluded by a free-energy comparison of the competing defect strings (Sec.~\ref{sec:completeness}).

\begin{figure}[t]
\centering
\includegraphics[width=0.8\columnwidth]{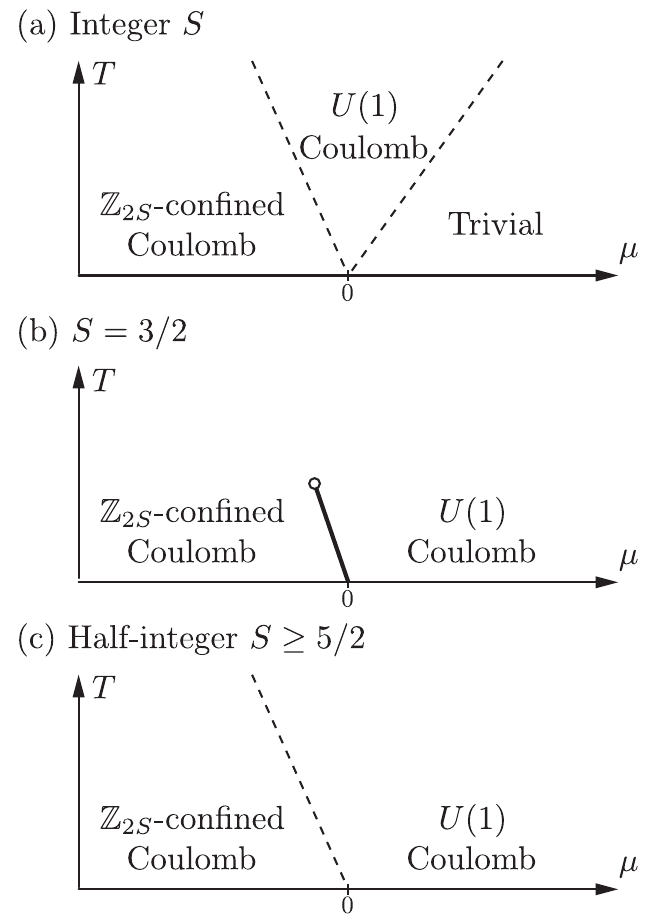}
\caption{Schematic phase diagram in the $(\mu, T)$ plane for (a)~integer $S \ge 1$, (b)~$S = 3/2$, and (c)~half-integer $S \ge 5/2$; the regimes are labeled in each panel, and $\mu$ and $T$ are quoted in units of $J$. Dashed lines denote crossover ridges (no thermodynamic singularity at $T > 0$); the solid line in (b) denotes the first-order coexistence line, terminating at a critical end point (open circle). The ridges and the coexistence line emanate from the origin along the rays $\mu = -T \ln w_c$ fixed by the monopole-free critical fugacities $w_c$ of Tables~\ref{tab:integer_spins} and~\ref{tab:halfinteger_spins}; on each ray $\mu/T$ is constant, and the sharp transitions are realized only in the monopole-free limit $T/J \to 0$ taken at fixed $\mu/T$ (Sec.~\ref{sec:model_and_review}\,C). For integer $S$~(a), the $U(1)$ Coulomb regime is the wedge of finite opening angle bounded by the rays through $w_{c1}$ and $w_{c2}$. An equivalent representation in the $(w, T/J)$ plane, in which the monopole-free transitions appear as separate points on the $T/J \to 0$ axis, is given in Sec.~I of Supplemental Material~\cite{SMarxiv}.}
\label{fig:phase_diagram}
\end{figure}

We then determine the nature of each transition. In the small-$w$ regime, an exact rewriting of the partition function in terms of a dual phase field shows that all integer-$S$ models undergo a 3D $XY$ transition at a common, $S$-independent coupling, whereas all half-integer-$S$ models reduce to the ordinary $S=1/2$ spin ice and therefore have no small-$w$ transition at all (Sec.~\ref{sec:small_w}). In the large-$w$ regime, we map the defect strings onto loop gases (Sec.~\ref{sec:large_w}). For $S = 3/2$, the loop gas coincides with the high-temperature expansion of the three-state Potts model, which explains microscopically the first-order transition observed numerically in Ref.~\cite{Pandey2026_S32} and locates it quantitatively. For $S \ge 2$, we show that no discrete clock model reproduces the loop gas; instead, the loop gas is that of the continuous 3D $XY$ model, up to corrections whose statistical weight is exponentially small and whose corresponding operators are irrelevant at the $XY$ fixed point, so the transition is continuous and of the 3D $XY$ type. The resulting classification in the monopole-free limit is summarized in Tables~\ref{tab:integer_spins} and~\ref{tab:halfinteger_spins} and the schematic phase diagram in Fig.~\ref{fig:phase_diagram}. 

Finally, by incorporating thermal monopoles into the partition function, we show that they enter the dual descriptions as a symmetry-breaking field that allows defect strings to terminate. This rounds all the continuous transitions into crossovers at any $T>0$ and predicts that the $S=3/2$ first-order line is the only boundary that can survive at finite temperature, terminating at a critical end point (Sec.~\ref{sec:finite_T}). We corroborate the crossover behavior with classical Monte Carlo simulations for $S$ up to $7/2$ (Sec.~\ref{sec:MC}); the temperatures accessible to these simulations lie above the estimated critical end point, so the survival of the $S=3/2$ first-order line itself remains an open target for future numerics. Beyond the specific context of spin ice, our analysis illustrates in a quantitative setting how a lattice model whose microscopic conservation law is real valued---and hence $U(1)$---can imitate a discrete gauge structure at low energies, and when this imitation must fail.

For clarity, we state explicitly what each of the recent works on which this paper builds established, and what is new here. Reference~\cite{Pandey2025_S1} mapped out the $S=1$ phase diagram by worm-algorithm simulations performed directly in the ice-rule manifold, locating the two transitions [$w_{c1} = 0.3609(1)$ and $w_{c2} = 1.8904(1)$] and identifying their universality classes through a mapping to an $XY$ partition function combined with finite-size scaling; Ref.~\cite{Watanabe2026} (by the present authors) derived the same $S=1$ results analytically via dual $XY$ and loop-gas representations and quantified the effect of thermal monopoles as an explicit symmetry-breaking field in those dual representations; Ref.~\cite{KunduDamle2025} studied a related two-dimensional model in a strong field; and Ref.~\cite{Pandey2026_S32} discovered the first-order $\mathbb{Z}_3$ transition for $S=3/2$ by the same worm-algorithm approach and rationalized it by a symmetry-based argument.

The present paper contains all $S \ge 1$ within a single framework, and we wish to be equally explicit about where its novelty does and does not lie. The transformations themselves are not new: The dualities, loop-gas expansions, and mean-field evaluations used below are standard tools of lattice statistical mechanics, and we credit their sources where they appear. What is new is the structure these tools reveal once they are set up for arbitrary $S$.

In the small-$w$ regime, the arbitrary-$S$ dual link weights expose a parity dichotomy invisible at $S=1$: The leading $XY$ coupling, and hence the location of the transition, is the same for every integer $S$, whereas for half-integer $S$ the dual weights are antiperiodic---the dual hallmark of half-integer spin---and the model reduces to the ordinary spin ice, with no transition at all. In the large-$w$ regime, the organizing result is the ice-rule compatibility criterion of Sec.~\ref{sec:defect_graphs_compatibility}, an elementary geometric statement that, to our knowledge, has no counterpart in the literature: A discrete $\mathbb{Z}_{2S}$ description can be consistent with the physical ice rule at the vertex level only for $S \le 3/2$, which explains with a single argument why the Ising and Potts mappings succeed at $S=1$ and $S=3/2$ and why every such mapping must fail for $S \ge 2$.

On the $S=3/2$ side of this divide, the criterion upgrades the symmetry-based rationale of Ref.~\cite{Pandey2026_S32} to a microscopic derivation---the loop gas is identified with the high-temperature expansion of the three-state Potts model, so that the first-order character follows a priori from the symmetry-allowed cubic invariant rather than being inferred from the numerics---and a Bethe-Peierls evaluation then locates the transition in closed form, within 3\% of the numerical value. On the $S \ge 2$ side, for which no previous results exist, numerical or analytic, we show that no discrete clock model reproduces the loop gas, bound the residual discrete corrections through a decomposition of the partition function, and thereby obtain both the 3D $XY$ universality class and the closed-form critical line $w_c(S)$ for the entire family.

\nocite{SMarxiv}\nocite{Kikuchi1951, Yedidia2001, Chester2020}%
The completeness argument of Sec.~\ref{sec:completeness} and the $S \ge 2$ Monte Carlo simulations of Sec.~\ref{sec:MC} are likewise reported here for the first time. The same holds for the finite-temperature analysis of Sec.~\ref{sec:finite_T} in the generality given there: Whereas thermal monopoles in spin ice are traditionally treated through the continuum electrostatic analogy \cite{Ryzhkin2005, Castelnovo2008, Henley2010}, we represent the monopole gas exactly in the dual variables, where a charge-$Q$ monopole is the local operator $e^{iQ\theta}$ and the gas therefore acts as an explicit $U(1)$ symmetry-breaking field---a correspondence introduced for $S = 1$ in Ref.~\cite{Watanabe2026} and established here for every $S$, culminating in the zero-field, self-generated critical end point for $S = 3/2$, to our knowledge the first such prediction for a frustrated magnet (the known critical end point of dipolar spin ice, the field-driven monopole liquid-gas transition \cite{Castelnovo2008}, requires a finite $[111]$ magnetic field).

This paper is organized as follows. In Sec.~\ref{sec:model_and_review} we define the model and derive the exact partition function. Section~\ref{sec:completeness} establishes the macroscopic flux quantization and the completeness of the phase diagram. The small-$w$ duality mapping and the integer vs half-integer dichotomy are presented in Sec.~\ref{sec:small_w}, while the large-$w$ loop-gas mappings and the classification of deconfinement transitions are developed in Sec.~\ref{sec:large_w}. The finite-temperature effects of thermal monopoles are analyzed in Sec.~\ref{sec:finite_T}. Section~\ref{sec:MC} presents the finite-temperature Monte Carlo simulations, Sec.~\ref{sec:outlook} presents an experimental outlook, and Sec.~\ref{sec:conclusion} concludes with future directions.

\begin{table*}[t]
\centering
\caption{Phase diagram of integer-spin models in the monopole-free limit ($v = 0$). The macroscopic polarization flux $\bm{P}$ serves as a topological invariant distinguishing the three phases. The transition out of the trivial paramagnet belongs to the 3D $XY$ universality class for all integer $S$. The deconfinement transition out of the $\mathbb{Z}_{2S}$-confined phase is 3D Ising for $S=1$ and 3D $XY$ for $S \ge 2$, the latter supported by the decomposition of Sec.~\ref{sec:S_ge_2} and the irrelevance of the $\mathbb{Z}_{2S}$ fusion cascade. The $S=1$ critical fugacity is $w_{c2} = 2/(3\,\tanh K_c^{\text{Ising}})$, where $K_c^{\text{Ising}}$ is the 3D Ising critical coupling on the diamond lattice~\cite{Watanabe2026}. For $S \ge 2$, $w_{c2} = [2/(3\,\tau_1)]^{1/(2S-1)}$ with $\tau_1 \coloneqq I_1(K_c^{XY})/I_0(K_c^{XY})$ (see Sec.~\ref{sec:S_ge_2}). At finite temperatures ($v > 0$), thermal monopoles round all these transitions into continuous crossovers (see Sec.~\ref{sec:finite_T}). The value $w_{c1} \approx 0.344$ is the leading-order analytic estimate of Sec.~\ref{sec:small_w}; the measured value for $S = 1$ is $0.3609(1)$ \cite{Pandey2025_S1}.}
\label{tab:integer_spins}
\renewcommand{\arraystretch}{1.4}
\begin{tabular}{cccccc}
\toprule
Spin $S$ & Trivial phase & $w_{c1}$ (transition) & $U(1)$ Coulomb phase & $w_{c2}$ (deconfinement) & $\mathbb{Z}_{2S}$-confined phase \\
\midrule
$1$       & $\bm{P}=\bm{0}$ & $\approx 0.344$ (3D $XY$) & $\bm{P} \in \mathbb{Z}^3$ & $\frac{2}{3\,\tanh K_c^{\text{Ising}}} \approx 1.88$ (3D Ising) & $\bm{P} \in (2\mathbb{Z})^3$  \\
$\geq 2$  & $\bm{P}=\bm{0}$ & $\approx 0.344$ (3D $XY$) & $\bm{P} \in \mathbb{Z}^3$ & $\left(\frac{2}{3\,\tau_1}\right)^{\!\frac{1}{2S-1}} \!\!\approx (1.859)^{\frac{1}{2S-1}}$ (3D $XY$)    & $\bm{P} \in (2S\mathbb{Z})^3$ \\
\bottomrule
\end{tabular}
\end{table*}

\begin{table*}
\centering
\caption{Phase diagram of half-integer-spin models in the monopole-free limit ($v = 0$). The absence of the nonmagnetic $S^z=0$ state forbids the trivial paramagnetic phase, reducing the diagram to two phases (except for $S=1/2$, which has only one). For $S=1/2$, the single-ion anisotropy is a constant and the system resides in the $U(1)$ Coulomb phase for all $w$. The $S=3/2$ deconfinement transition is first order, driven by $\mathbb{Z}_3$ defect-string fusion mapped to the three-state Potts model; the quoted $w_c \approx 1.44$ is the closed-form Bethe-Peierls estimate $w_c^2 = \sqrt{2} + 2/3 \approx 2.081$, which overestimates the Monte Carlo value $w^2 \approx 2.02$ of Ref.~\cite{Pandey2026_S32} (i.e., $w_c \approx 1.42$) by $\sim 3.0\%$. For $S \ge 5/2$, the exponential suppression of the fusion cascade restores 3D $XY$ criticality. At finite temperatures ($v > 0$), the continuous transitions become crossovers, while the $S=3/2$ first-order transition is predicted to persist up to a critical end point (see Sec.~\ref{sec:finite_T}).}
\label{tab:halfinteger_spins}
\renewcommand{\arraystretch}{1.4}
\begin{tabular}{cccc}
\toprule
Spin $S$ & $U(1)$ Coulomb phase & $w_{c}$ (deconfinement) & $\mathbb{Z}_{2S}$-confined phase \\
\midrule
$1/2$     & $\bm{P} \in \mathbb{Z}^3$ & None [always $U(1)$ phase] & --- \\
$3/2$     & $\bm{P} \in \mathbb{Z}^3$ & $\approx 1.44$ (first order; 3-Potts) & $\bm{P} \in (3\mathbb{Z})^3$ \\
$\geq 5/2$ & $\bm{P} \in \mathbb{Z}^3$ & $\left(\frac{2}{3\,\tau_1}\right)^{\!\frac{1}{2S-1}} \!\!\approx (1.859)^{\frac{1}{2S-1}}$ (3D $XY$) & $\bm{P} \in (2S\mathbb{Z})^3$ \\
\bottomrule
\end{tabular}
\end{table*}

\section{Model formulation and exact partition function}
\label{sec:model_and_review}

Before embarking on the generalized theory for arbitrary spin $S$, we rigorously define the microscopic Hamiltonian and introduce the exact representation of the partition function that forms the foundation of all subsequent duality mappings.

\subsection{Lattice geometry and effective Hamiltonian}

The pyrochlore lattice is a network of corner-sharing tetrahedra whose sites lie at the midpoints of the links $\ell = \langle \bm{r}_A, \bm{r}_B \rangle$ of an underlying bipartite diamond lattice \cite{Ramirez1999, GingrasMcClarty2014}. The diamond lattice comprises $N$ vertices (equivalently, $N$ tetrahedra of the pyrochlore lattice) and $2N$ links. We place on each site a discrete spin variable $S_\ell^z \in \{-S, -S+1, \dots, S\}$ with arbitrary positive integer or half-integer magnitude $S$. The local quantization axis on every link is oriented from the A-sublattice vertex $\bm{r}_A$ toward the B-sublattice vertex $\bm{r}_B$.

We study the generalized Blume-Capel Hamiltonian on this lattice \cite{Blume1966, Capel1966}, which couples an antiferromagnetic nearest-neighbor exchange $J > 0$ with a single-ion anisotropy $\mu$:
\begin{equation}
    H = \frac{J}{2}\sum_{\bm{r}} \Big(\sum_{\ell \in \bm{r}} S^z_\ell\Big)^2 + \mu \sum_\ell (S_\ell^z)^2,\label{eq:model}
\end{equation}
where the outer sum runs over every diamond-lattice vertex $\bm{r}$ and $\ell \in \bm{r}$ labels its four incident links (coordination number $z=4$). Since both terms are quadratic in $S_\ell^z$, the Hamiltonian is invariant under a simultaneous sign flip $S_\ell^z \to -S_\ell^z$ on \emph{all} links (global time-reversal symmetry). Moreover, because the exchange energy at each vertex is the squared divergence $(\sum_{\ell \in \bm{r}} S_\ell^z)^2 = (-\sum_{\ell \in \bm{r}} S_\ell^z)^2$, its value is independent of the sublattice sign convention adopted for the link orientations. Accounting for the sublattice orientation of each link, the discrete lattice divergence is defined as $\nabla \cdot \bm{S}_{\bm{r}_A} \coloneqq +\sum_{\ell \in \bm{r}_A} S_\ell^z$ and $\nabla \cdot \bm{S}_{\bm{r}_B} \coloneqq -\sum_{\ell \in \bm{r}_B} S_\ell^z$ (the sign flip on the B sublattice arises because the local quantization axis on each link is oriented from A to B), which can be interpreted as a magnetic monopole charge \cite{Castelnovo2008b}.

\subsection{Exact partition function}

At temperature $T$ the canonical partition function is
\begin{equation}
    Z \coloneqq \sum_{\{S_\ell^z\}} e^{-H/T}. \label{eq:partition_basic}
\end{equation}
We parametrize the Boltzmann weights by two dimensionless fugacities: $w \coloneqq e^{-\mu/T}$, governing the single-ion cost, and $v \coloneqq e^{-J/(2T)}$, governing the monopole cost. Physically, $\mu > 0$ ($w < 1$) penalizes large spin amplitudes (easy-plane tendency), whereas $\mu < 0$ ($w > 1$) favors the maximal amplitude $|S^z| = S$ (easy-axis tendency). Introducing an independent monopole charge variable $Q_{\bm{r}} \in \mathbb{Z}$ at each vertex and enforcing $Q_{\bm{r}} = \nabla \cdot \bm{S}_{\bm{r}}$ via a Kronecker delta, we rewrite the partition function as
\begin{equation}
    Z = \sum_{\{Q_{\bm{r}}\}}\sum_{\{S_\ell^z\}} \left( \prod_{\bm{r}} \delta_{\nabla \cdot \bm{S}_{\bm{r}}, Q_{\bm{r}}} \right) v^{\sum_{\bm{r}}Q_{\bm{r}}^2} w^{\sum_\ell(S_\ell^z)^2}, \label{eq:partition_exact}
\end{equation}
which cleanly separates the topological weight $v^{Q^2}$ from the local anisotropy weight $w^{(S^z)^2}$ and holds for arbitrary $S$ and $T$. This representation is the starting point for all analytical results derived below.

\subsection{Monopole-free limit and macroscopic polarization flux}

When $T \ll J$ the monopole fugacity $v$ is exponentially small, so configurations with any $Q_{\bm{r}} \neq 0$ carry a prohibitive Boltzmann penalty. Retaining only the $Q_{\bm{r}}=0$ sector in Eq.~\eqref{eq:partition_exact} imposes the ice rule $\nabla \cdot \bm{S}_{\bm{r}} = 0$ at every vertex. This monopole-free limit ($v = 0$) is the setting of a large fraction of this paper: Sections~\ref{sec:completeness}--\ref{sec:large_w} work exclusively at $v = 0$, and the effects of $v > 0$ are analyzed in Sec.~\ref{sec:finite_T} and probed numerically in Sec.~\ref{sec:MC}. We stress that $v \to 0$ at fixed $w$ is a joint limit on both couplings rather than a zero-temperature limit at fixed couplings: Since $w = e^{-\mu/T}$, holding $w$ fixed as $T/J \to 0$ means holding $\mu/T = -\ln w$ fixed, so that $|\mu|/J$ vanishes together with $T/J$. The variable $w$ therefore describes the physically natural hierarchy $|\mu| \ll J$ with $T$ of order $|\mu|$, in which the residual thermal fluctuations within the ice-rule manifold are controlled by $w$ rather than by $T$ itself. Equivalently, for a material with fixed $\mu$ and $J$, cooling sweeps $w = e^{-\mu/T}$ monotonically away from unity, so that the regimes organized by $w$ are traversed on cooling to $T$ of order $|\mu|$---still far below $J$, where $v$ is exponentially small. This is the limit in which the phase diagrams of Refs.~\cite{Pandey2025_S1, Pandey2026_S32} were obtained. We emphasize that the sharp transitions analyzed in Secs.~\ref{sec:completeness}--\ref{sec:large_w} exist strictly on the $v = 0$ manifold: At any $T > 0$ the monopole fugacity is nonzero, and, as shown in Sec.~\ref{sec:finite_T}, the continuous transitions are then rounded into crossovers, with the sole exception of the first-order line at $S = 3/2$.

The ice-rule manifold further decomposes into topologically distinct sectors labeled by a macroscopic polarization flux $\bm{P}$. Under periodic boundary conditions the component along a cubic axis $z$ reads
\begin{equation}
    P_z = \sum_{\ell \in C_k} \mathrm{sgn}(z_B - z_A) \, S_\ell^z,
\end{equation}
where $C_k$ denotes the set of links intersected by any plane of constant $z$ cutting through the lattice. The sign factor projects each local spin onto the global $z$-direction. Because the divergence-free condition forces all flux entering a slab to exit, $P_z$ is independent of the choice of $C_k$ and therefore constitutes a conserved topological quantum number \cite{Jaubert2013, BrooksBartlett2014}. 

\subsection{Review of the foundational cases ($S=1/2$ and $S=1$)}

For the foundational $S=1/2$ case, individual spins can only take values $\pm 1/2$. The single-ion anisotropy term evaluates to a constant: $\mu (S_\ell^z)^2 = \mu/4$. Because this uniformly shifts the global energy, the parameter $w$ is rendered physically irrelevant. The thermodynamics is dictated solely by the topological ice rule. The macroscopic flux can freely take any integer value, $\bm{P} \in \mathbb{Z}^3$ (since any macroscopic plane intersects an even number of links, and summing an even number of half-integer spins yields an integer). The macroscopic state is the canonical $U(1)$ Coulomb liquid phase \cite{Huse2003, Hermele2004, Henley2010}.

The $S=1$ limit introduces the nonmagnetic state $S_\ell^z = 0$. For $w \ll 1$ ($\mu > 0$), the nonmagnetic state is energetically favored, collapsing the system into a trivial paramagnet with $\bm{P} = \bm{0}$. As $w$ increases, defect strings of $S_\ell^z = \pm 1$ condense through a 3D $XY$ continuous phase transition, yielding a $U(1)$ Coulomb phase.

For $w \gg 1$ ($\mu < 0$), the energetic penalty forces the spins to maximize their amplitude, settling into $S_\ell^z = \pm 1$ on all links. Because any macroscopic plane on the bipartite diamond lattice is intersected by an even number of links, summing an even number of $\pm 1$ states strictly quantizes the macroscopic flux to even integers, $\bm{P} \in (2\mathbb{Z})^3$, stabilizing a $\mathbb{Z}_2$-confined Coulomb phase. In the monopole-free limit, the ice rule forbids isolated defect links, so the elementary excitations are \emph{closed loops} of $S_\ell^z=0$ links. Because $0 = -0$, these loops lack internal orientation. The resulting undirected loop gas maps onto the high-temperature expansion of the 3D Ising model, placing the deconfinement transition in the 3D Ising universality class \cite{Watanabe2026}, in agreement with the independent symmetry-based argument and the finite-size-scaling analysis of Ref.~\cite{Pandey2025_S1}; the loop-gas estimate $w_{c2} \approx 1.884$ \cite{Watanabe2026} agrees with the numerical transition point $w_{c2} = 1.8904(1)$ of Ref.~\cite{Pandey2025_S1} to $0.3\%$.

\section{Macroscopic flux quantization and the completeness of the phase diagram}
\label{sec:completeness}

When generalizing to arbitrary $S \ge 3/2$, the enlarged local state space raises the question of whether more complex phase structures---such as intermediate uniform backgrounds or partial deconfinement cascades---can arise. Throughout this section, we work in the monopole-free limit $v = 0$. We first characterize the macroscopic topology in both limiting regimes of $w$, and then argue that intermediate phases are excluded.

\subsection{Macroscopic flux quantization in the large-$w$ limit}

In the large-$w$ limit, the vacuum resides exclusively at $S_\ell^z = S \sigma_\ell$ with polarities $\sigma_\ell \in \{-1, 1\}$. To characterize the macroscopic topology, we compute the global polarization flux $P_z$ across a macroscopic cross-sectional plane $C_k$:
\begin{equation}
    P_z = S \sum_{\ell \in C_k} \mathrm{sgn}(z_B - z_A) \sigma_\ell = S \sum_{\ell \in C_k} \tilde{\sigma}_\ell,
\end{equation}
where $\tilde{\sigma}_\ell \coloneqq \mathrm{sgn}(z_B - z_A) \sigma_\ell \in \{-1, 1\}$.

A macroscopic plane completely traverses the bipartite diamond lattice. The total number of links $N_\perp$ intersecting this plane is strictly an even integer (e.g., $N_\perp = 4L^2$ for a plane orthogonal to a cubic axis in a periodic lattice of $L^3$ unit cells). Let $N_+$ be the number of links crossing the plane with positive projected polarity ($\tilde{\sigma}_\ell = +1$), and $N_-$ be the number with negative projected polarity ($\tilde{\sigma}_\ell = -1$). We have the geometric identity $N_+ + N_- = N_\perp$. The algebraic sum of the projected polarities evaluates to
\begin{align}
    \sum_{\ell \in C_k} \tilde{\sigma}_\ell &= N_+ (1) + N_- (-1) \nonumber \\
    &= (N_\perp - N_-) - N_- = N_\perp - 2N_-.
\end{align}
Because $N_\perp$ is an even integer, the quantity $N_\perp - 2N_-$ is manifestly an even integer, denoted as $2k$ ($k \in \mathbb{Z}$). Substituting this topological constraint back into the macroscopic flux equation yields:
\begin{equation}
    P_z = S (2k) = 2S k, \quad \text{and hence} \quad \bm{P} \in (2S \mathbb{Z})^3.
\end{equation}

This elementary parity argument shows that the macroscopic flux in the large-$w$ limit is quantized to multiples of $2S$, defining a $\mathbb{Z}_{2S}$-confined Coulomb phase for any $S$. The same fact was noted for $S=1$ in Refs.~\cite{Pandey2025_S1, Watanabe2026} and for $S=3/2$ in Ref.~\cite{Pandey2026_S32}; we have recorded here its form for general $S$.

\subsection{Macroscopic flux quantization in the small-$w$ limit}

In the opposite limit $w \ll 1$ ($\mu > 0$), the single-ion anisotropy favors the smallest accessible spin amplitude on every link. The nature of the resulting ground-state manifold depends qualitatively on whether $S$ is an integer or a half-integer.

For integer $S$, the unique minimum of the single-ion energy $\mu m^2$ is $m = 0$. In this regime every link is occupied by the nonmagnetic state $S_\ell^z = 0$, giving $P_z = 0$ identically. Thermal excitations with $|m| \ge 1$ are dilute (fugacity $w \ll 1$) and form small, closed loops under the ice rule, so the macroscopic flux remains locked at $\bm{P} = \bm{0}$. The system is a trivial paramagnet, with a unique ground state and local (closed-loop) excitations.

For half-integer $S$, the minimum amplitude is $|m| = 1/2$, and the ground-state manifold consists of all ice-rule-satisfying configurations of $S_\ell^z = \pm 1/2$. This is precisely the $S = 1/2$ spin ice, irrespective of the original value of $S$. Substituting $S \to 1/2$ into the large-$w$ result $\bm{P} \in (2S\mathbb{Z})^3$ immediately yields $\bm{P} \in \mathbb{Z}^3$. The small-$w$ phase of any half-integer spin is therefore a fully deconfined $U(1)$ Coulomb liquid, topologically equivalent to the canonical $S = 1/2$ spin ice \cite{Huse2003, Hermele2004, Henley2010}.

To summarize, the two limiting regimes exhibit the following macroscopic flux sectors: For integer $S$, the accessible sectors change from $\bm{P} = \bm{0}$ (small $w$) to $\bm{P} \in (2S\mathbb{Z})^3$ (large $w$); for half-integer $S$, they change from $\bm{P} \in \mathbb{Z}^3$ (small $w$) to $\bm{P} \in (2S\mathbb{Z})^3$ (large $w$). In both cases, the set of thermodynamically accessible flux sectors differs between the two limits, guaranteeing the existence of at least one phase boundary as $w$ is varied: For integer $S$, the nonzero sectors in $(2S\mathbb{Z})^3$ must be activated (deconfinement), while for half-integer $S$, the sectors in $\mathbb{Z}^3 \setminus (2S\mathbb{Z})^3$ must be suppressed (confinement). The remaining question is whether intermediate phases can intervene between these two extremes.

\subsection{Prohibition of intermediate background phases}

Let us evaluate whether a macroscopic uniform background comprised predominantly of intermediate spin amplitudes ($|S_\ell^z| = m$, where $0 < m < S$) can emerge as a thermodynamically stable phase.

In the strict monopole-free limit ($v \to 0$), the thermodynamic competition between different uniform backgrounds is dictated by the local single-ion anisotropy energy function, $\mathcal{E}(m) \propto \mu m^2$. For the small-$w$ regime ($\mu > 0$), this is a strictly convex parabola, minimizing uniquely at the smallest possible amplitude (0 for integer $S$, $\pm 1/2$ for half-integer $S$). For the large-$w$ regime ($\mu < 0$), the concave nature of $\mathcal{E}(m)$ places the minimum at the boundaries $m = \pm S$.

Because any intermediate spin state ($0 < m < S$) carries an extensive energy penalty compared to either the minimal or maximal-amplitude states, both intermediate uniform and mixed backgrounds are thermodynamically forbidden. The system must transit directly between the minimal and maximal-amplitude backgrounds.

\subsection{Hierarchy of string free energies and the absence of partial deconfinement}

Even after excluding intermediate backgrounds, one might envision a cascade of partial deconfinement transitions within the large-$w$ regime, in which defect strings of successively higher topological charge $\phi$ condense at different fugacities. Here, a charge-$\phi$ defect string denotes a connected string of links on which $S^z_\ell$ deviates from its vacuum value by $\phi$ units, $S^z_\ell = \pm(S-\phi)$ with $1 \le \phi \le 2S-1$; the equivalent flux formulation of this charge is given in Sec.~\ref{sec:defect_graphs_compatibility}. We now show that this scenario is also excluded.

In the $\mathbb{Z}_{2S}$-confined phase ($\mu < 0$), the deconfinement transition out of the vacuum ($S_\ell^z = \pm S$) is driven by the thermal proliferation of topological defect strings. A defect string carrying a relative integer topological charge $\phi$ incurs a string tension $\Delta E_\phi$:
\begin{align}
    \Delta E_\phi &= -|\mu| (S-\phi)^2 - \left( -|\mu| S^2 \right) \nonumber \\
    &= |\mu| (2S\phi - \phi^2), \quad (1 \le \phi \le 2S-1).
\end{align}

This tension forms an inverted parabola over $\phi \in [1, 2S-1]$. The absolute minimum tension is strictly achieved at the boundaries: the fundamental unit charge ($\phi = 1$) and its dual ($\phi = 2S-1$).
\begin{equation}
    \Delta E_1 = \Delta E_{2S-1} = |\mu|(2S-1). \label{eq:tension}
\end{equation}

For $S \ge 2$, any intermediate-charge defect ($1 < \phi < 2S-1$) incurs a strictly larger string tension ($\Delta E_{\phi} > \Delta E_1$). (For $S=3/2$, the only available charges are the fundamental string $\phi=1$ and its dual $\phi=2$, which share the identical minimum tension and thus co-condense; no intermediate defect exists.)

The tension comparison above is purely energetic, and the quantity that actually controls which strings proliferate is the free energy per unit length, which also contains the configurational entropy of the string. We therefore formulate the argument in terms of free energies. A charge-$\phi$ string of length $|C|$ contributes a statistical weight $[e^{s_\phi}\, w^{-\phi(2S-\phi)}]^{|C|}$, where $e^{s_\phi}$ collects the entropic factors per unit length: the number of lattice embeddings, governed by the connective constant $\mu_{\mathrm{SAW}} \approx 2.88$ of self-avoiding walks on the diamond lattice, and local vertex-weight and internal-degeneracy factors, which are bounded above and below by $S$-independent constants of order unity. Within Pauling's independent-vertex approximation, each such factor is a ratio of ice-rule state counts at a four-coordinated diamond-lattice vertex times a link degeneracy of at most $2$, and is therefore bounded independently of both $S$ and $\phi$ (Sec.~\ref{sec:entropy_dichotomy} carries out the explicit evaluation for the fundamental strings).

Crucially, all string species live on the same lattice, so the embedding entropy is common to every $\phi$, and the species-dependent part of $s_\phi$ is bounded. A charge-$\phi$ string can proliferate---its free energy per unit length can turn nonpositive---only when $\phi(2S-\phi)\ln w \le s_\phi$. Because the tension exponent $\phi(2S-\phi)$ is strictly minimized at the boundaries $\phi = 1$ and $\phi = 2S-1$, while $s_\phi$ varies only within bounded $O(1)$ limits, the proliferation threshold is reached first---i.e., at the largest value of $w$ as $w$ is lowered from deep inside the confined phase---by the fundamental strings. For an intermediate-charge string to condense first, its entropy per unit length would have to exceed that of the fundamental string by an amount that grows with the tension gap $[\phi(2S-\phi) - (2S-1)]\ln w > 0$, which the bounded local factors cannot supply.

We emphasize the logical status of this argument: The tension hierarchy is exact, while the entropic bounds invoke the same independent-vertex (Pauling) treatment that is used---and validated at the few-percent level against the Monte Carlo data of Refs.~\cite{Pandey2025_S1, Pandey2026_S32}---throughout Sec.~\ref{sec:large_w}. Within this controlled treatment, partial deconfinement is excluded; the simulations of Sec.~\ref{sec:MC}, which show no intermediate \emph{cooperative} anomaly between the two crossover scales, corroborate this conclusion (the broad secondary shoulder present for $S \ge 2$ is a size-independent single-ion Schottky feature, as established in Sec.~\ref{sec:MC}). We also note that the entropic process by which a charge-$2$ string splits into two charge-$1$ strings works in favor of this conclusion, since it converts higher-charge strings into fundamental strings.

Once the fundamental strings proliferate as winding strings around the torus, each such string shifts the macroscopic flux across a cross-sectional plane by $\pm 1$, so $\bm{P} \in \mathbb{Z}^3$---full deconfinement into the $U(1)$ Coulomb liquid in a single step. More generally, condensation of charge-$n$ defect strings would change the accessible flux sectors from $(2S\mathbb{Z})^3$ to $(\gcd(n,2S)\,\mathbb{Z})^3$, since each winding string shifts the flux by $\pm n$. For $n = 1$, $\gcd(1,2S) = 1$ recovers $\mathbb{Z}^3$. Since the fundamental strings reach their proliferation threshold first and already achieve complete deconfinement, intermediate fractionalized phases with partial flux quantization [e.g., $\bm{P} \in (S\mathbb{Z})^3$, which would require $n \ge 2$ strings to condense before $n = 1$] do not occur. Combining this result with the macroscopic flux quantization in both limits (Secs.~\ref{sec:completeness}\,A and~\ref{sec:completeness}\,B) and the prohibition of intermediate backgrounds (Sec.~\ref{sec:completeness}\,C), we conclude that the global phase diagram is limited to exactly three phases for integer $S$ (trivial paramagnet, $U(1)$ Coulomb liquid, $\mathbb{Z}_{2S}$-confined phase) and two phases for half-integer $S$ ($U(1)$ Coulomb liquid, $\mathbb{Z}_{2S}$-confined phase).

\section{Small-$w$ regime: Exact duality mapping and the spin parity dichotomy}
\label{sec:small_w}

When $w \ll 1$ ($\mu > 0$), large spin amplitudes are energetically penalized and the thermodynamics depends qualitatively on the parity of $2S$. We expose this integer vs half-integer dichotomy through an exact duality transformation of the monopole-free partition function \cite{Savit1980, Kogut1979}. We continue to work at $v = 0$ throughout this section. The transformation below is the standard duality between integer-valued, divergence-free link currents and a compact phase field, familiar from the analyses of the $XY$ model by Jos\'e \textit{et al.} \cite{Jose1977} and by Dasgupta and Halperin \cite{DasguptaHalperin1981} (see also the textbook treatment in Ref.~\cite{ChaikinLubensky}), here run in reverse: We start from the physical currents $S^z_\ell$ and reconstruct the phase field. The same mapping onto an $XY$ model was used for $S = 1$ in Refs.~\cite{Pandey2025_S1, Watanabe2026} and for $S = 3/2$ in Ref.~\cite{Pandey2026_S32}; the derivation below generalizes this mapping to arbitrary $S$ while keeping track of the exact link weights.

\subsection{Exact dual representation}

From Eq.~\eqref{eq:partition_exact}, taking the exact limit $v \to 0$, the partition function is restricted to the minimally frustrated manifold where $Q_{\bm{r}} = 0$ identically:
\begin{equation}
    Z_{v=0} = \sum_{\{S_\ell^z\}} \left( \prod_{\bm{r}} \delta_{Q_{\bm{r}}, 0} \right) \prod_\ell w^{(S_\ell^z)^2}.
\end{equation}

Each Kronecker delta is resolved by a Lagrange-multiplier phase $\theta_{\bm{r}} \in [0,2\pi)$:
\begin{equation}
    \delta_{Q_{\bm{r}}, 0} = \int_0^{2\pi} \frac{d\theta_{\bm{r}}}{2\pi} e^{-i \theta_{\bm{r}} Q_{\bm{r}}}.
\end{equation}
Substituting into $Z_{v=0}$ and exchanging the order of summation and integration gives
\begin{equation}
    Z_{v=0} = \sum_{\{S_\ell^z\}} \int \mathcal{D}\theta \exp\left( -i \sum_{\bm{r}} \theta_{\bm{r}} Q_{\bm{r}} \right) w^{\sum_\ell (S_\ell^z)^2},
\end{equation}
with $\mathcal{D}\theta \coloneqq \prod_{\bm{r}} \frac{d\theta_{\bm{r}}}{2\pi}$.

A discrete summation by parts converts the vertex sum in the exponent into a sum over oriented links $\ell = \langle \bm{r}_A, \bm{r}_B \rangle$:
\begin{align}
    \sum_{\bm{r}} \theta_{\bm{r}} Q_{\bm{r}} &= \sum_{\bm{r}_A} \theta_{\bm{r}_A} \Big( \sum_{\ell \in \bm{r}_A} S_\ell^z \Big) - \sum_{\bm{r}_B} \theta_{\bm{r}_B} \Big( \sum_{\ell \in \bm{r}_B} S_\ell^z \Big) \nonumber \\
    &= - \sum_{\ell=\langle \bm{r}_A, \bm{r}_B \rangle} (\theta_{\bm{r}_B} - \theta_{\bm{r}_A}) S_\ell^z \nonumber \\
    &= - \sum_\ell (\nabla\theta_\ell) S_\ell^z,
\end{align}
where we have defined the phase difference along the link as $\nabla\theta_\ell \coloneqq \theta_{\bm{r}_B} - \theta_{\bm{r}_A}$. 

This transformation decouples the global constraint, allowing the summation over $S_\ell^z \in \{-S, -S+1, \dots, S\}$ to factorize into a product of independent local sums:
\begin{equation}
    Z_{v=0} = \int \mathcal{D}\theta \prod_\ell W_S(\nabla\theta_\ell),
\end{equation}
where we define the exact local link weight function:
\begin{equation}
    W_S(\nabla\theta_\ell)\coloneqq \sum_{m=-S}^S w^{m^2} e^{i m \nabla\theta_\ell}.
\end{equation}
The macroscopic statistical mechanics of this exact dual continuous representation is thus governed by the effective action $S_{\text{eff}}[\theta] \coloneqq -\sum_\ell \ln W_S(\nabla\theta_\ell)$.

\subsection{Integer spins: Protection of 3D $XY$ universality}

For integer spins ($S \in \mathbb{Z}$), the absolute minimum of the energy uniquely defines the trivial nonmagnetic vacuum state $m=0$, which carries a weight of $w^0 = 1$. The elementary thermal excitations are the fundamental defects with $m = \pm 1$, carrying a suppressed weight $w \ll 1$. Higher-spin states, such as $m = \pm 2$, carry penalized weights $\mathcal{O}(w^4)$.

For integer $S \ge 2$, the exact link weight $W_S(\nabla\theta_\ell)$ expands as
\begin{equation}
    W_S(\nabla\theta_\ell) = 1 + 2w \cos(\nabla\theta_\ell) + 2w^4 \cos(2\nabla\theta_\ell) + \dots.
\end{equation}
Expanding $\ln W_S = \ln(1+x)$ with $x = W_S - 1$, we obtain the effective action:
\begin{align}
    S_{\text{eff}}[\theta] = \sum_\ell \Big[ &-(2w + 2w^3) \cos(\nabla\theta_\ell) + w^2 \cos(2\nabla\theta_\ell) \nonumber \\
    &- \tfrac{2}{3}w^3\cos(3\nabla\theta_\ell) + \tfrac{1}{2}w^4\cos(4\nabla\theta_\ell) \Big] \nonumber \\
    &+ \mathcal{O}(w^5).
    \label{eq:Seff_integer}
\end{align}

The dominant term, $-(2w+2w^3) \sum_\ell \cos(\nabla\theta_\ell)$, is independent of $S$ (it holds for all integer $S \ge 1$; see below) and maps the small-$w$ regime to the classical 3D $XY$ model \cite{Jose1977} with an effective coupling $K_{\text{eff}} = 2w + 2w^3 + \mathcal{O}(w^5)$. The critical point $w_{c1}$ is determined by $K_{\text{eff}}(w_{c1}) = K_c^{XY}$, where $K_c^{XY}$ is the 3D $XY$ critical coupling on the diamond lattice. Our Monte Carlo simulations of the classical $XY$ model on the diamond lattice (the ferromagnetic and antiferromagnetic models share the same $T_c$ on this bipartite lattice) yield $T_c^{XY} = 1.30036(1)$ \cite{WatanabeH2026}, corresponding to $K_c^{XY} = 1/T_c^{XY} \approx 0.769$. Solving $2w + 2w^3 = K_c^{XY}$ to $\mathcal{O}(w^3)$ gives $w_{c1} \approx 0.344$, in good agreement with the numerical result $w_{c1} = 0.3609(1)$ obtained for $S = 1$ by Pandey and Damle \cite{Pandey2025_S1}.

All non-$XY$ harmonics $\cos(p\nabla\theta_\ell)$ ($p \ge 2$) are strongly irrelevant operators at the 3D $XY$ fixed point \cite{Jose1977, Nahum2011}, so the small-$w$ regime belongs to the 3D $XY$ universality class for all integer $S$. Two remarks clarify the logic here. First, the 3D $XY$ universality is ultimately a consequence of the exact $U(1)$ invariance of the dual action, which encodes the real-valued divergence-free constraint; the harmonic expansion serves to identify the two phases that the transition connects (an $XY$ paramagnet, corresponding to the trivial phase, and an $XY$ ``ferromagnet,'' corresponding to the Coulomb liquid) and to locate the transition quantitatively. Second, by 3D $XY$ universality we refer to the critical fixed point governing the transition at $w_{c1}$. The ordered phase of the dual $XY$ model is not merely an auxiliary construct: Its spin stiffness maps onto the stiffness of the emergent gauge field of the Coulomb liquid, so the dual description remains physically informative on both sides of $w_{c1}$, although the truncation to the leading harmonic is quantitatively controlled only for small $w$.

The leading $XY$ coupling $-(2w+2w^3)\cos(\nabla\theta_\ell)$ is completely independent of $S$ for all integer $S \ge 1$; only the irrelevant higher harmonics ($p \ge 2$) are sensitive to the available spin states. The critical point $w_{c1}$ is therefore universal across all integer spins. For $S=1$, the absence of $m = \pm 2$ states renders $W_1(\nabla\theta_\ell) = 1 + 2w\cos(\nabla\theta_\ell)$ exact, so the higher-harmonic coefficients differ from Eq.~\eqref{eq:Seff_integer} at subleading orders (e.g., $\cos(2\nabla\theta_\ell)$ acquires an additional $+2w^4$ correction). These differences are confined to the irrelevant sector and do not affect the universal $XY$ critical behavior.

\subsection{Half-integer spins: Absence of phase transitions}

The physical conclusion of this subsection can be stated simply: As $w \to 0$, a half-integer-$S$ model reduces to the ordinary $S=1/2$ spin ice, because the minimal doublet $S^z = \pm 1/2$ is the only surviving degree of freedom. The small-$w$ phase is therefore the $U(1)$ Coulomb liquid, and no thermodynamic phase transition can occur at small $w$. The purpose of the analysis below is to make this statement quantitative---in particular, to show that the leading corrections in $w$ do not generate any relevant (confining) perturbation but merely renormalize the stiffness of the Coulomb liquid.

For half-integer spins ($S = 1/2, 3/2, 5/2, \dots$), the nonmagnetic integer state $m=0$ is strictly forbidden. The lowest-energy states available to each link are the physical doublets $m = \pm 1/2$, both carrying an identical Boltzmann weight $w^{1/4}$. The higher-spin thermal states (e.g., $m = \pm 3/2$) carry weights $w^{9/4}$, which are heavily suppressed by a factor of $w^2 \ll 1$.

For $S \ge 3/2$, evaluating the exact link weight function yields the analytical expansion:
\begin{align}
   & W_S(\nabla\theta_\ell) \notag\\
    &= 2w^{1/4} \cos\Big(\frac{\nabla\theta_\ell}{2}\Big) + 2w^{9/4} \cos\Big(\frac{3\nabla\theta_\ell}{2}\Big) + \mathcal{O}(w^{25/4}).
\end{align}
Factoring out $2w^{1/4}\cos(\nabla\theta_\ell/2)$, expanding the logarithm, and applying the identity $\cos(3x)/\cos x = 2\cos(2x)-1$ with $x=\nabla\theta_\ell/2$, we obtain the effective action (up to constants):
\begin{align}
    S_{\text{eff}}[\theta]
    = &- \sum_\ell \ln \cos\Big(\frac{\nabla\theta_\ell}{2}\Big) \nonumber \\
    &- 2w^2 \sum_\ell \cos(\nabla\theta_\ell)
    + \mathcal{O}(w^4).
    \label{eq:Seff_halfinteger}
\end{align}
The physical tuning parameter $w$ factors out of the leading action term as an overall global normalization constant. While this shifts the global energy scale, it does not lift the topological degeneracy of the system. The leading, $w$-independent effective action, $- \sum_\ell \ln \cos\Big(\frac{\nabla\theta_\ell}{2}\Big)$, is identical to the gauge theory formulation of the classical $S=1/2$ spin ice \cite{Hermele2004, Henley2005}. This action describes the degenerate manifold of two-in, two-out configurations, which hosts the $U(1)$ Coulomb liquid phase characterized by algebraic dipolar correlations and unconfined integer fluxes ($\bm{P} \in \mathbb{Z}^3$).

The subleading $\mathcal{O}(w^2)$ correction is a global $U(1)$-symmetric $XY$ rotor coupling $-2w^2 \cos(\nabla\theta_\ell)$---the same functional form as the standard 3D $XY$ model action. In the dual framework, this term strictly preserves the continuous $U(1)$ symmetry and merely renormalizes the phase stiffness (corresponding to the photon stiffness of the emergent $U(1)$ gauge field) without introducing any symmetry-breaking relevant operator (such as a monopole potential $\cos\theta_{\bm{r}}$) that could generate a confining gap. More broadly, the 3D $U(1)$ Coulomb phase is intrinsically robust against arbitrary small local perturbations \cite{Hermele2004, Henley2010}, and the $\mathcal{O}(w^2)$ perturbation does not generate any relevant operator that could confine the gauge field. Notably, this $\mathcal{O}(w^2)$ photon stiffness vanishes identically in the pure $S=1/2$ limit, where $W_{1/2} = 2w^{1/4}\cos(\nabla\theta_\ell/2)$ contains no higher harmonics. It is the admixture of higher spin states ($|m| \ge 3/2$), available only for $S \ge 3/2$, that generates the phase stiffness through virtual fluctuations. Consequently, the duality mapping confirms and quantifies the simple picture stated at the beginning of this subsection: For any half-integer $S$, the small-$w$ regime is the $S=1/2$ Coulomb liquid with a weakly renormalized stiffness, and no thermodynamic phase transition occurs.

A technical remark: The link weight $W_S(\alpha)$ of any half-integer $S$ is antiperiodic, $W_S(\alpha+2\pi) = -W_S(\alpha)$---the dual manifestation of the half-integer spin, reminiscent of the $(-1)$ phase acquired by a spin-$1/2$ wave function under a $2\pi$ rotation. The dual integrand nonetheless remains single valued under $\theta_{\bm{r}} \to \theta_{\bm{r}} + 2\pi$, because each such shift flips the sign of exactly $z = 4$ incident link weights and $(-1)^4 = +1$; the even coordination number of the diamond lattice is what makes the half-angle representation consistent, and the partition function is strictly real and positive. Alternatively, all half-angles can be avoided from the outset by treating $2S^z_\ell \in \{\pm 1, \pm 3, \dots\}$ as the integer-valued conserved current and coupling the dual phase field to it; the resulting link weights involve only integer harmonics $\cos(p\,\nabla\theta_\ell)$ and lead to identical conclusions. We have kept the half-angle convention to maintain an explicit connection to the physical spin variables.

\section{Large-$w$ regime: Graphical mappings and the nature of the deconfinement transitions}
\label{sec:large_w}

In the large-$w$ limit ($\mu \to -\infty$), the system uniquely resides in the highly confined vacuum state where all spins take their maximum absolute values, $S_\ell^z = S \sigma_\ell$, with background polarities $\sigma_\ell \in \{-1, 1\}$. Defect strings, characterized by a reduced spin amplitude $|S^z| < S$, form graphs constrained by the local ice rule. In this section, working again at $v = 0$, we derive graphical (loop-gas) representations of the partition function and use them to determine the nature of the deconfinement transition. The outcome is a sharp dichotomy: For $S \le 3/2$ the loop gas coincides with the high-temperature expansion of a \emph{discrete} spin model (Ising for $S = 1$, three-state Potts for $S = 3/2$), whereas for $S \ge 2$ it coincides with that of the \emph{continuous} 3D $XY$ model.

\subsection{Defect graphs, ice-rule compatibility, and configurational entropy}
\label{sec:defect_graphs_compatibility}

\subsubsection{Ice-rule compatibility criterion}

To analyze the macroscopic defect statistics, we map the physical spin variables into a discrete effective flux. For each link $\ell$, we define the shifted nonnegative variable $\phi_\ell \coloneqq S_\ell^z + S \in \{0, 1, \dots, 2S\}$. In this representation, the maximal-amplitude vacuum states correspond to $\phi_\ell \in \{0, 2S\} \equiv 0 \pmod{2S}$. Consequently, any link hosting an elementary defect ($|S^z| < S$) carries a nonzero $\mathbb{Z}_{2S}$ flux $\phi_\ell \in \{1, 2, \dots, 2S-1\}$.

At each vertex, $\sum_{\ell \in \bm{r}} \phi_\ell = \sum_{\ell \in \bm{r}} S_\ell^z + 4S$. The ice rule ($\nabla \cdot \bm{S}_{\bm{r}} = 0$) strictly fixes the first sum to zero on both sublattices, giving the \emph{exact} constraint $\sum_{\ell \in \bm{r}} \phi_\ell = 4S$. Taking this equation modulo $2S$ yields a weaker, purely local conservation law:
\begin{equation}
    \sum_{\ell \in \bm{r}} \phi_\ell \equiv 0 \pmod{2S}.
    \label{eq:Z2S_conservation}
\end{equation}
Because $4S \equiv 0 \pmod{2S}$, the ice rule implies Eq.~\eqref{eq:Z2S_conservation}, but the converse need not hold: The modular condition admits solutions with $\sum_{\ell \in \bm{r}} \phi_\ell = 2S, 6S, \dots$ that violate $Q_{\bm{r}}=0$.

Given a configuration of defect links (those with $\phi_\ell \in \{1, \dots, 2S{-}1\}$) satisfying Eq.~\eqref{eq:Z2S_conservation} at every vertex, a natural question arises: Can one always choose the values of the remaining vacuum links ($\phi_\ell \in \{0, 2S\}$) so that the strict ice-rule constraint $\sum_{\ell \in \bm{r}} \phi_\ell = 4S$ holds at every vertex? Such a choice is called an \emph{ice-rule-compatible vacuum assignment}.

The answer reveals a topological dichotomy.

\emph{Ice-rule compatibility criterion.}---Consider any $\mathbb{Z}_{2S}$-conserving defect configuration on the diamond lattice and any single vertex $\bm{r}$ of it. The modular conservation law alone guarantees---for every such configuration and every such vertex---that the non-defect links of $\bm{r}$ can be assigned vacuum values $\phi_\ell \in \{0, 2S\}$ such that the ice rule $\sum_{\ell \in \bm{r}} \phi_\ell = 4S$ holds at $\bm{r}$ if and only if $S \le 3/2$. For $S \ge 2$ the assignment still exists whenever fewer than four links of $\bm{r}$ carry defects, and can fail only at fully excited vertices.

\emph{Proof.}---Let $k$ denote the number of defect bonds ($\phi_\ell \in \{1,\dots,2S{-}1\}$) at a vertex, and write $\Sigma_{\mathrm{def}}$ for their partial sum. The remaining $4-k$ vacuum bonds each carry $\phi_\ell \in \{0, 2S\}$; if $n_+$ of them take the value $2S$, the ice rule $\sum_{\ell \in \bm{r}} \phi_\ell = 4S$ becomes
\begin{equation}
  n_+ = \frac{4S - \Sigma_{\mathrm{def}}}{2S} = 2 - \frac{\Sigma_{\mathrm{def}}}{2S},
  \label{eq:nplus_constraint}
\end{equation}
which must be solved with $n_+ \in \{0, 1, \dots, 4{-}k\}$.

(i) $k < 4$ (at least one vacuum bond exists, so $n_+$ is adjustable). Case by case:
$k = 0$: Trivially $\Sigma_{\mathrm{def}} = 0$ and $n_+ = 2$.
$k = 1$: $\Sigma_{\mathrm{def}} \in [1,2S{-}1]$ is never a multiple of $2S$, so $k=1$ is forbidden by $\mathbb{Z}_{2S}$ conservation itself.
$k = 2$: $\Sigma_{\mathrm{def}} \in [2,4S{-}2]$; the unique multiple of $2S$ is $2S$ itself, giving $n_+=1 \le 2$, which is satisfiable.
$k = 3$: $\Sigma_{\mathrm{def}}/(2S) \in \{1,2\}$ with $n_+ \in \{1,0\}$, both in $[0,1]$, again satisfiable.
Thus every vertex with $k < 4$ admits an ice-rule-compatible vacuum assignment for all $S$.

(ii) $k = 4$ (all bonds excited; no adjustable vacuum bonds). Since $\phi_\ell \in [1,2S{-}1]$, we have $\Sigma_{\mathrm{def}} \in [4,\, 4(2S{-}1)]$. With no vacuum bonds available, Eq.~\eqref{eq:nplus_constraint} has no free variable, and the strict ice rule demands $\Sigma_{\mathrm{def}} = 4S$. The modular conservation law $\Sigma_{\mathrm{def}} \equiv 0 \pmod{2S}$ guarantees this if and only if $4S$ is the unique multiple of $2S$ within the kinematically allowed interval $[4,\, 4(2S{-}1)]$ (Table~\ref{tab:ice_rule_compatibility}).
For $S \le 3/2$, $4S$ is the unique multiple, so $\mathbb{Z}_{2S}$ conservation guarantees $Q_{\bm{r}}=0$. For $S \ge 2$, extra multiples of $2S$ appear in the interval, satisfying $\mathbb{Z}_{2S}$ conservation but giving $Q_{\bm{r}} \ne 0$---a situation we refer to as \emph{ice-rule incompatibility}. (We deliberately avoid the term ``monopole'' here: The physical, thermally excited monopoles of Sec.~\ref{sec:finite_T} are absent in the present $v = 0$ setting, and the incompatibility is a purely arithmetic property of the modular conservation law.)

\begin{table}[t]
\centering
\caption{Ice-rule compatibility for a fully excited vertex ($k=4$). The interval $[4, 4(2S{-}1)]$ lists the allowed range of $\sum_{\ell \in \bm{r}} \phi_\ell$. The ice rule requires $\sum_{\ell \in \bm{r}} \phi_\ell = 4S$; for $S \ge 2$, additional multiples of $2S$ appear in the interval, so the modular condition no longer guarantees the ice rule (ice-rule incompatibility).}
\label{tab:ice_rule_compatibility}
\begin{tabular}{cccc}
  \toprule
  Spin $S$ & Interval & \parbox[c]{2.8cm}{\centering Multiples of $2S$\\in the interval} & Compatible? \\
  \midrule
  $1$       & $[4, 4]$          & $4$              & \checkmark \\
  $3/2$     & $[4, 8]$          & $6$              & \checkmark \\
  \midrule
  $2$       & $[4, 12]$         & $4, 8, 12$       & \ding{55} \\
  $\ge 5/2$ & $[4, 4(2S{-}1)]$  & $2S, 4S, \dots$  & \ding{55} \\
  \bottomrule
\end{tabular}
\end{table}

Combining (i) and (ii) completes the proof. \hfill$\square$

\begin{figure}
\centering
\includegraphics[width=0.8\columnwidth]{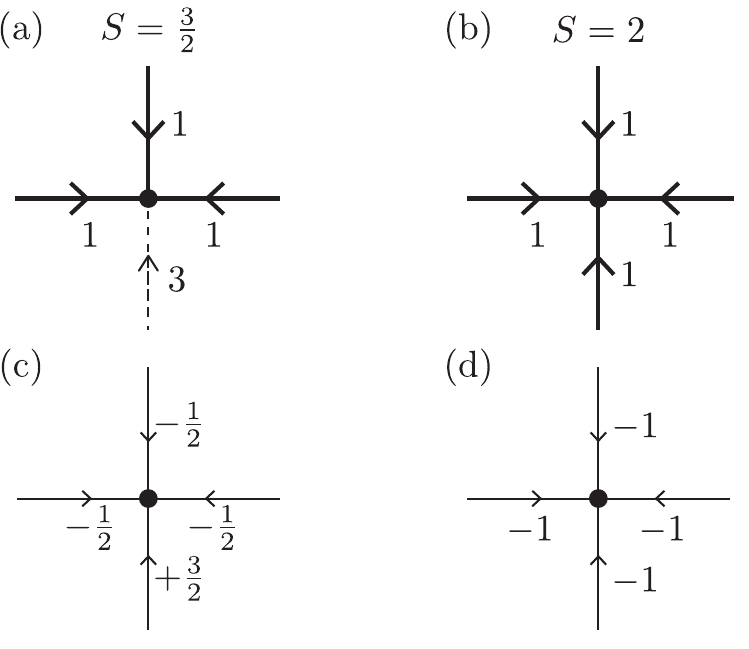}
\caption{Ice-rule compatibility at a single $z = 4$ diamond-lattice vertex; arrows indicate the bipartite A$\to$B link orientation along which $\phi_\ell$ is defined. (a)~$S = 3/2$: three fundamental defects ($\phi = 1$) annihilate at a single vertex, and the remaining vacuum link at $\phi = 2S$ completes the ice rule $\sum_{\ell \in \bm{r}} \phi_\ell = 4S$. (b)~$S = 2$: four fundamental defects give $\sum_{\ell \in \bm{r}} \phi_\ell = 4 = 2S \neq 4S$---the modular condition is satisfied but the ice rule is violated. (c), (d)~The spin configurations $S^z_\ell$ corresponding to (a) and (b), with $\sum_{\ell \in \bm{r}} S^z_\ell = 0$ and $-4$, respectively. The four bonds are drawn as a planar cross for clarity; on the diamond lattice they point toward the corners of a tetrahedron.}
\label{fig:annihilation}
\end{figure}

This criterion establishes a sharp dichotomy at the vertex level. For $S \le 3/2$, the modular conservation law guarantees vertex-by-vertex compatibility with the ice rule. One caveat must be recorded: Vertex-level compatibility does not by itself guarantee a \emph{globally} consistent vacuum assignment, because vacuum links are shared between vertices. For $S = 1$, this mechanism does not arise (junctions are absent, and no vacuum value is ever forced). For $S = 3/2$, however, a cubic junction fixes the value of its unique vacuum link ($\phi = 2S$ when all three defect currents point in, $\phi = 0$ when they point out), so rare arrangements in which the forced vacuum links of two like-oriented junctions meet at a common pass-through vertex admit no consistent assignment and are excluded from the physical ensemble. This exclusion acts as a short-range, $\mathbb{Z}_3$-symmetric hard-core constraint between nearby junctions; it first arises at second order in the junction density, and its role is accounted for in Sec.~\ref{sec:S32_Potts}. The minimal excluded configuration, together with an exact finite-lattice verification, is recorded in Sec.~V of Supplemental Material~\cite{SMarxiv}. For $S \ge 2$, compatibility fails already at the single-vertex level. As a concrete illustration, consider $S=2$ [Figs.~\ref{fig:annihilation}(b) and~\ref{fig:annihilation}(d)]: A vertex with $(\phi_1, \phi_2, \phi_3, \phi_4) = (1,1,1,1)$ gives $\Sigma_{\mathrm{def}} = 4 = 2S$, while $(3,3,3,3)$ gives $\Sigma_{\mathrm{def}} = 12 = 6S$. Both satisfy $\mathbb{Z}_{4}$ conservation but violate the ice rule $\Sigma_{\mathrm{def}} = 4S = 8$, producing charges $Q_{\bm{r}} = \pm 4$. Consequently, for $S \ge 2$, any loop model built on $\mathbb{Z}_{2S}$ conservation alone would include configurations that the spin-ice model forbids. Together with the mismatch of defect fugacities quantified in Sec.~\ref{sec:S_ge_2}, this obstructs a mapping to a discrete $\mathbb{Z}_{2S}$ spin or gauge model \cite{Wegner1971, Fradkin1979, Kogut1979}.

\subsubsection{Configurational entropy}
\label{sec:entropy_dichotomy}

The Boltzmann weight of a single defect link relative to the vacuum background changes by a factor of $w^{(S-1)^2 - S^2} = w^{1-2S}$. To evaluate the full loop-gas partition function, we must additionally account for the macroscopic configurational entropy of the background spins.

We employ Pauling's independent-vertex approximation \cite{Pauling1935}. In the vacuum, each vertex has $\binom{4}{2}=6$ valid ice-rule configurations out of $2^4=16$ unconstrained binary assignments, yielding a Pauling fraction $p_0 = 3/8$. We now compute, following Ref.~\cite{Watanabe2026}, the ratio $W(C)/W(\emptyset)$, where $W(C)$ denotes the number of ice-rule-satisfying background spin configurations in the presence of a defect loop $C$ of length $|C|$, and $W(\emptyset)$ is the corresponding count in the defect-free vacuum.

For $S \ge 3/2$, the defect retains a nonzero amplitude [$S^z_\ell = \pm(S-1) \neq 0$], so each defect link carries an internal arrow (the sign of $S^z$), making the loops directed. For such a directed loop with a fixed orientation, the two defect links at each loop vertex are completely determined---their $\phi$ values are fixed by the loop's flux charge---leaving only the two background links as free variables. Each background link carries $\phi \in \{0, 2S\}$, and the ice rule $\sum_{\ell \in \bm{r}} \phi_\ell = 4S$ constrains the pair: If the defect links contribute $\Sigma_{\mathrm{def}}$ to the vertex sum, the two background links must sum to $4S - \Sigma_{\mathrm{def}} = 2S$, which is uniquely satisfied by one link carrying $0$ and the other $2S$. There are exactly two such assignments (either ordering), so out of $2^2 = 4$ background configurations, exactly two satisfy the ice rule, giving a vertex fraction $p_{\text{loop}} = 1/2$. This holds uniformly for all $S$. In addition, each defect link replaces a two-state vacuum link ($S_\ell^z = \pm S$) with a single fixed state, contributing a link-degeneracy factor of $1/2$ per defect link. Combining both factors, the Pauling entropy reduction per loop link is
\begin{equation}
    \frac{W(C)}{W(\emptyset)} \approx \left( \frac{1}{2} \times \frac{p_{\text{loop}}}{p_0} \right)^{|C|} = \left( \frac{1}{2} \times \frac{4}{3} \right)^{|C|} = \left(\frac{2}{3}\right)^{|C|}.
\end{equation}
Together with the bare Boltzmann weight $w^{1-2S}$ per defect link, the effective fundamental string fugacity evaluates to (see Secs.~II and~III of Supplemental Material~\cite{SMarxiv} for detailed derivations)
\begin{equation}
    \tilde{x}_1 = \frac{2}{3} w^{-(2S-1)}.
    \label{eq:fundamental_fugacity}
\end{equation}

The internal degrees of freedom $n$ hosted by each continuous loop depend on $S$. For $S=1$, the defect is $S_\ell^z = 0$, yielding undirected loops ($n=1$). For $S \ge 3/2$, as noted above, the nonzero defect amplitude preserves the sign of the displacement, yielding chiral, directed loops that host $n=2$ distinct physical states.

\begin{figure*}[t]
\centering
\includegraphics[width=0.8\textwidth]{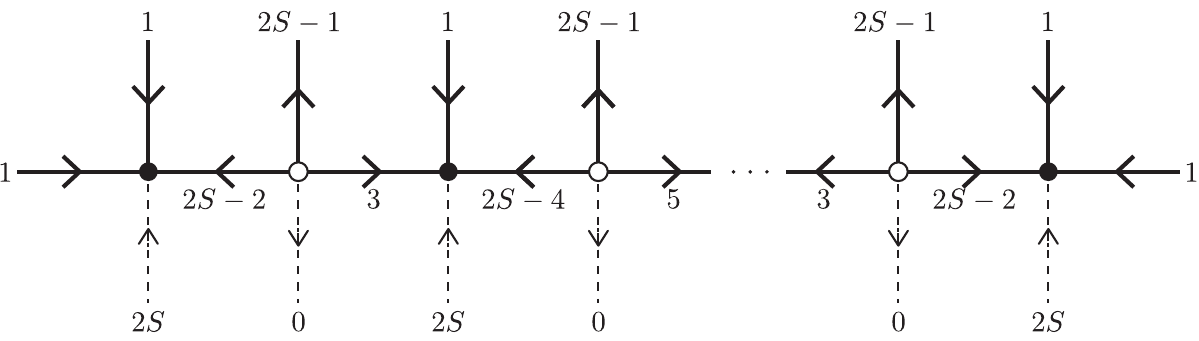}
\caption{The simplest (linear-chain) fusion channel for $S \ge 2$: at each of the $2S - 2$ fusion vertices, an incoming fundamental defect ($\phi = 1$) merges with the accumulated intermediate defect, incrementing the effective charge by one. Arrows indicate the bipartite A$\to$B link orientation along which $\phi_\ell$ is defined; open and filled circles denote A- and B-sublattice vertices. The displayed link labels alternate between $\phi$ and $2S - \phi$ along the chain because the vertices alternate between the two sublattices. The half-odd-integer case is shown; for integer $S$ the same structure applies with uniform link orientations. The chain is drawn as a straight line for clarity; on the diamond lattice the backbone follows a zigzag path.}
\label{fig:hierarchical_fusion}
\end{figure*}

\subsection{Mapping to the three-state Potts model for $S=3/2$}
\label{sec:S32_Potts}

For the specific case of $S=3/2$, our ice-rule compatibility criterion guarantees that the geometric $z=4$ lattice constraints and the $\mathbb{Z}_3$ flux conservation are matched. Up to the short-range junction-junction exclusion recorded in Sec.~\ref{sec:defect_graphs_compatibility}, every $\mathbb{Z}_3$-conserving defect graph is realizable within the physical ice-rule manifold, and the loop expansion below is that of a discrete three-state spin model.

The fundamental defects $S_\ell^z = \pm 1/2$ naturally act as conjugate strings carrying fluxes $\phi=1$ and $\phi=2 \equiv -1 \pmod 3$. The strict $\mathbb{Z}_3$ conservation geometrically permits degree-2 pass-throughs, degree-4 crossings, and uniquely, degree-3 junctions where three fundamental strings converge [$\sum_{\ell \in \bm{r}} \phi_\ell = 1+1+1 = 3 \equiv 0 \pmod 3$]. This geometry accommodates the simultaneous local annihilation of exactly three topological strings at a single vertex [Figs.~\ref{fig:annihilation}(a) and~\ref{fig:annihilation}(c)].

The vertex weights are determined by the ice-rule state counts at each vertex degree. In the defect-free vacuum (degree $d_0 = 0$), each vertex has $\binom{4}{2} = 6$ valid configurations out of $2^4 = 16$ possibilities, giving a local state sum $z(0) = 6$. At a degree-2 (pass-through) vertex, the two defect links are fixed by the loop direction, and the two remaining vacuum links must satisfy the ice rule, giving $z_{\mathrm{dir}}(2)/z(0) = 2/6 = 1/3$. At a degree-3 (cubic junction) vertex, the $\mathbb{Z}_3$ conservation law already forces all three defect currents to share one orientation ($|\Sigma_{\mathrm{def}}| = 3$, i.e., all ingoing or all outgoing); the ice rule additionally fixes the value of the single remaining vacuum link, which halves the allowed configurations relative to the degree-2 baseline, giving $z_{\mathrm{dir}}(3)/z(0) = 1/6$. At a degree-4 (crossing) vertex, all four bonds are occupied by defects (no vacuum links remain); the directed ice rule then uniquely fixes the defect configuration, giving $z_{\mathrm{dir}}(4)/z(0) = 1/6$. The vertex fugacities, normalized relative to the degree-2 weight, are thus
$\lambda_3 = (1/6)/(1/3)^{3/2} = \sqrt{3}/2 \approx 0.866$ (penalty at cubic junctions) and $\lambda_4 = (1/6)/(1/3)^2 = 3/2$ [enhancement at crossings; the denominator is $(1/3)^2$ because a crossing replaces two independent degree-2 vertices]. Utilizing the standard topological graph identity $\sum d \cdot V_d = 2|G|$, the complete monopole-free partition function takes the form
\begin{equation}
    Z_{v=0} = Z_{\text{vac}}\sum_{G} 2^{N_{\text{comp}}(G)} \tilde{x}_1^{|G|} \lambda_3^{V_3(G)} \lambda_4^{V_4(G)},
\end{equation}
where $Z_{\text{vac}}$ is the partition function of the vacuum sector, in which every link carries maximum amplitude $|S_\ell^z| = S$ subject to the ice rule. Each vacuum configuration contributes a Boltzmann weight $w^{2NS^2}$, and the Pauling approximation (Sec.~\ref{sec:entropy_dichotomy}) yields an ice-rule degeneracy $(3/2)^{N}$, so that $Z_{\text{vac}} \approx w^{2NS^2}(3/2)^{N}$. The sum runs over all graphs $G$ on the diamond lattice whose edges carry directed $\mathbb{Z}_3$ fluxes satisfying the conservation law $\sum_{\ell \in \bm{r}} \phi_\ell \equiv 0 \pmod{3}$ at every vertex $\bm{r}$ [Eq.~\eqref{eq:Z2S_conservation}], excluding the short-range junction-junction configurations identified in Sec.~\ref{sec:defect_graphs_compatibility}; $|G|$ is the total number of edges, $V_3(G)$ and $V_4(G)$ are the numbers of degree-3 and degree-4 vertices, and $N_{\text{comp}}(G)$ is the number of closed defect strands, each carrying two chiral orientations (at a degree-4 crossing, the four incident defect links pair into two through-going strands, so two loops crossing at a vertex count as two strands).

This graph ensemble is isomorphic to the high-temperature graphical expansion of the classical three-state Potts model on the diamond lattice \cite{Savit1980, Wu1982} (see Secs.~III and~V of Supplemental Material~\cite{SMarxiv} for detailed derivations),
\begin{equation}
    Z_{3\text{-Potts}} = \sum_{\{\sigma_{\bm{r}} \in \mathbb{Z}_3\}} \exp\!\Bigl(K \sum_{\langle \bm{r}\bm{r}'\rangle} \delta_{\sigma_{\bm{r}},\sigma_{\bm{r}'}}\Bigr).
    \label{eq:Potts}
\end{equation}
The standard Fourier decomposition on $\mathbb{Z}_3$ expands each bond weight as $e^{K\delta_{\sigma,\sigma'}} = a_0[1 + t(\omega^{\sigma-\sigma'} + \omega^{-(\sigma-\sigma')})]$ with $\omega = e^{2\pi i/3}$, $a_0 = (e^K+2)/3$, and $t = (e^K-1)/(e^K+2)$. Summing over the spin variables, the $\mathbb{Z}_3$ flux conservation at every vertex generates exactly the same graph ensemble as above, yielding
\begin{equation}
    Z_{3\text{-Potts}} = a_0^{2N}\cdot 3^N \sum_{G} 2^{N_{\text{comp}}(G)}\, t^{|G|}.
    \label{eq:Potts_HT}
\end{equation}
The identification with the spin-ice expansion is established by the correspondence $t = \tilde{x}_1$; the local geometric weights $\lambda_3$ and $\lambda_4$ act merely as short-distance renormalizations of the bare coupling and do not alter the universality class.

We spell out the status of this mapping. The correspondence between the two graph ensembles is fixed by the $\mathbb{Z}_3$ conservation law and the vertex-level compatibility criterion, up to the explicit short-range exclusion recorded in Sec.~\ref{sec:defect_graphs_compatibility}: Configurations in which the forced vacuum links of two like-oriented junctions collide occur in the Potts ensemble but not in the spin-ice ensemble. The vertex weights, by contrast, are computed within Pauling's independent-vertex approximation: Correlations of the background ice manifold induce additional interactions between defect strings beyond the independent-vertex level. In particular, a cubic junction, which forces its single background link to $\phi = 2S$, acts as a localized perturbation (an effective charge) for the background height field, so junctions experience entropic, dipolar-type interactions mediated by the background Coulomb phase.

These interactions are short ranged or screened at finite defect density---including, at shortest distance, the strict junction-junction exclusion above, which acts as a hard core---and, crucially, they respect the $\mathbb{Z}_3$ symmetry of the loop gas; they therefore renormalize the couplings of the effective Potts description but do not affect the symmetry classification on which the Landau argument below rests. Their quantitative smallness is supported a posteriori by the accuracy of the same approximation elsewhere: It reproduces the $S=1$ transition point $w_{c2} = 1.8904(1)$ of Ref.~\cite{Pandey2025_S1} to within $0.3\%$ \cite{Watanabe2026} and the $S=3/2$ transition point of Ref.~\cite{Pandey2026_S32} to within $\sim 3.0\%$ (below).

Finally, for $q = 3$ the Potts and clock descriptions are equivalent. Since the Kronecker delta on $\mathbb{Z}_3$ satisfies $\delta_{\sigma,\sigma'} = \tfrac{1}{3}[1+2\cos(2\pi(\sigma-\sigma')/3)]$, the three-state Potts model is equivalent to the three-state clock model---the identity $\cos(2\pi/3) = \cos(4\pi/3)$ ensures that all nonidentical spin pairs receive equal weight, making the clock interaction identical to the Potts Kronecker delta. This equivalence holds for $q \le 3$ but fails for $q \ge 4$, where different angular separations yield distinct weights (cf.\ Sec.~\ref{sec:S_ge_2}).

In the continuum Landau-Ginzburg-Wilson effective action, the $\mathbb{Z}_3$ defect-string fusion generates a symmetry-allowed cubic invariant ($\sim \Phi^3$). Within Landau mean-field theory, a cubic invariant drives the transition first order, and in three dimensions this expectation is borne out: Indeed, extensive Monte Carlo studies have established that the three-state Potts model in three dimensions undergoes a first-order transition \cite{BloteSwendsen1979, Wu1982}. The graphical mapping derived above thus provides a theoretical explanation for the \emph{first-order phase transition} observed numerically by Pandey \textit{et al.} for $S=3/2$ \cite{Pandey2026_S32}.

We note that Ref.~\cite{Pandey2026_S32} itself provided a rationale for the first-order character by a different, symmetry-based route: Through the mapping to an $XY$ partition function and the observation that the large-$w$ phase exhibits long-range order of $e^{3i\theta}$ but not of $e^{i\theta}$, the transition was identified there with that of a three-state spin system. The graphical derivation given here reaches the same conclusion from a complementary, microscopic starting point, and in addition yields a quantitative estimate of the transition point, as follows.

The location of the first-order transition can be estimated analytically via the Bethe-Peierls (cavity) approximation \cite{Baxter1982}. The idea is to replace the diamond lattice ($z = 4$) by the Bethe lattice (Cayley tree) with the same coordination number, on which the partition function of the $q=3$ Potts model can be evaluated exactly via a single-variable cavity recursion. On this tree, each site has $b = z - 1 = 3$ ``children''; the cavity message---the ratio $r \coloneqq P(\sigma=\sigma_0)/P(\sigma = \sigma_1)$ ($\sigma_1 \neq \sigma_0$) of the probability that a boundary spin matches a reference state $\sigma_0$ to that of any single non-reference state $\sigma_1$ (the $q-1$ non-reference states are equivalent by the $S_q$ symmetry of the Potts model)---satisfies the self-consistency equation,
\begin{equation}
    r = \left[\frac{e^K r + 2}{r + e^K + 1}\right]^b,
\end{equation}
where $K$ is the Potts coupling and $b = 3$.

The disordered solution $r = 1$ is always present; for the $q = 3$ Potts model, a second solution $r^* > 1$ (the ordered phase) appears discontinuously---a hallmark of a first-order transition driven by the cubic $\Phi^3$ invariant. Equating the Bethe free energies of the two solutions (see Sec.~VIII of Supplemental Material~\cite{SMarxiv}) determines the coexistence point \emph{in closed form}: $e^{K_c} = 1 + \sqrt{2}$, i.e., $t_c = (3\sqrt{2}-2)/7 \approx 0.3204$, where $t = (e^K - 1)/(e^K + 2)$ is the Potts high-temperature parameter identified with the spin-ice fugacity via $t = \tilde{x}_1 = \frac{2}{3}w^{-2}$. Converting back to the spin-ice fugacity gives the closed-form estimate
\begin{equation}
    w_c^2 = \sqrt{2} + \tfrac{2}{3} \approx 2.081, \qquad w_c \approx 1.443
\end{equation}
(note that Ref.~\cite{Pandey2026_S32} uses $w$ to denote what corresponds to $w^2$ in our convention). This overestimates the Monte Carlo value $w^2 \approx 2.02$ \cite{Pandey2026_S32} by only $\sim 3.0\%$, a level of accuracy expected for the Bethe approximation on a $z = 4$ lattice: At a first-order transition the correlation length remains finite, so the absence of loops on the Bethe lattice---its only structural deficiency---has a minor quantitative effect (see Sec.~VIII of Supplemental Material~\cite{SMarxiv} for the full derivation including free-energy evaluation).

\subsection{$S \ge 2$: 3D $XY$ universality}
\label{sec:S_ge_2}

\subsubsection{Breakdown of the clock mapping and continuous $U(1)$ baseline}

We begin by recalling a fact already used in Sec.~\ref{sec:small_w}: The ice rule is a \emph{real-valued} constraint---the lattice divergence of $S^z$ vanishes exactly, not merely modulo an integer---so the microscopic model possesses an exact $U(1)$ structure for every $S$. This is precisely what underlies the dual $XY$ representations here and in Refs.~\cite{Pandey2025_S1, Watanabe2026, Pandey2026_S32}. The question addressed in this subsection is therefore \emph{not} how a $U(1)$ symmetry emerges, but the converse: whether the large-$w$ defect ensemble reduces this microscopic $U(1)$ structure to a discrete one, as it does for $S = 1$ (Ising) and $S = 3/2$ (Potts). We show that for $S \ge 2$ it does not: No discrete clock model reproduces the loop gas, and the transition is governed by the microscopic $U(1)$ structure, i.e., it is of the 3D $XY$ type.

For $S \ge 2$, the situation changes qualitatively relative to $S \le 3/2$. As shown by the ice-rule compatibility criterion (Sec.~\ref{sec:defect_graphs_compatibility}), the $\mathbb{Z}_{2S}$ Gauss law no longer coincides with the physical ice rule (ice-rule incompatibility). Furthermore, the hierarchy of defect tensions creates a bond-fugacity mismatch. The natural comparison model is the $\mathbb{Z}_{2S}$ clock model \cite{Jose1977} (rather than the Potts model), because the defect charges $\phi = 1, 2, \ldots, 2S{-}1$ naturally correspond to the clock-model current harmonics, which distinguish different angular separations---a structure absent in the Potts model that treats all nonidentical states equally. In a standard $\mathbb{Z}_{2S}$ clock model, higher-current harmonics yield polynomial fugacity decay ($t_m \sim K^m/m!$ at high temperature), whereas in the spin-ice model, each defect level carries an independent chemical potential with fugacities $x_\phi \coloneqq w^{-\phi(2S-\phi)}$ ($\phi = 1, \dots, 2S{-}1$) that decay exponentially.

A key consequence is the arithmetic prohibition of the cubic junction. For $S=3/2$ ($2S=3$), three fundamental strings satisfy $1+1+1 = 3 \equiv 0\pmod{3}$, permitting a cubic vertex that maps to the symmetry-breaking $\Phi^3$ invariant driving the first-order Potts transition (Sec.~\ref{sec:S32_Potts}). For $S \ge 2$ ($2S \ge 4$), however, $1+1+1 = 3 \not\equiv 0\pmod{2S}$, so a cubic junction composed solely of fundamental strings is strictly forbidden by the $\mathbb{Z}_{2S}$ conservation law itself. This arithmetic obstruction inherently suppresses the $\Phi^3$ invariant that would otherwise drive a first-order transition.

A second, purely geometric, obstruction emerges in the annihilation of fundamental strings. In a $\mathbb{Z}_{2S}$ clock model, the modular conservation law $\sum_{\ell \in \bm{r}} n_\ell \equiv 0\pmod{2S}$ allows up to $z = 4$ fundamental currents to merge at a single vertex whenever $2S \le z$; in particular, for $S = 2$ ($q = 4$), all four fundamentals annihilate locally with no intermediate bonds and unit penalty (Table~\ref{tab:annihilation}). Even for $S > 2$, a short cascade of only $\lceil S{-}2 \rceil$ intermediate bonds carrying odd currents $n = 3, 5, \ldots$ suffices (see Sec.~VI of Supplemental Material~\cite{SMarxiv}).

In the spin-ice model, by contrast, the exact ice rule $\sum_{\ell \in \bm{r}} \phi_\ell = 4S$ is more restrictive than the modular condition: It forbids simultaneous merging even for $2S = 4$ [Figs.~\ref{fig:annihilation}(b) and~\ref{fig:annihilation}(d)] and, more generally, forbids any vertex from absorbing more than two \emph{fundamental} strings (with three or four fundamentals at a vertex, $\sum_{\ell \in \bm{r}} \phi_\ell \le 3 + 2S < 4S$ for $S \ge 2$). The $2S$ fundamentals therefore fuse pairwise into composite strands, which fuse further through a fusion tree containing at least $2S-3$ intermediate bonds with charges $2 \le \phi \le 2S-2$; the linear chain, in which one fundamental is absorbed at a time ($\phi = 2, 3, \dots, 2S-2$; Fig.~\ref{fig:hierarchical_fusion}), is the simplest such tree.

These combined obstructions---the ice-rule incompatibility, the exponential bond-fugacity mismatch, and the extended cascade geometry---mean that, unlike for $S \le 3/2$, the large-$w$ loop gas is \emph{not} the high-temperature expansion of any discrete $\mathbb{Z}_{2S}$ clock model. The only processes that could distinguish the discrete $\mathbb{Z}_{2S}$ conservation law from the full $U(1)$ conservation law are the fusion cascades themselves, whose statistical weight we now quantify.

Quantitatively, each intermediate bond incurs a fugacity $x_\phi = w^{-\phi(2S-\phi)}$, so the total statistical cost of a fusion channel---the \emph{bridge penalty}---is the product of its intermediate fugacities (see Supplemental Material~\cite{SMarxiv}, Sec.~II, for a self-contained derivation). For the linear chain,
\begin{align}
    \mathcal{P}_{\text{bridge}} &= \prod_{\phi=2}^{2S-2} x_\phi = w^{-\sum_{\phi=2}^{2S-2} \phi(2S-\phi)}\notag\\
    & = w^{-(2S-3)(2S-1)(2S+4)/6}.
    \label{eq:cascade_suppression}
\end{align}

\begin{figure}[t]
\centering
\includegraphics[width=0.8\columnwidth]{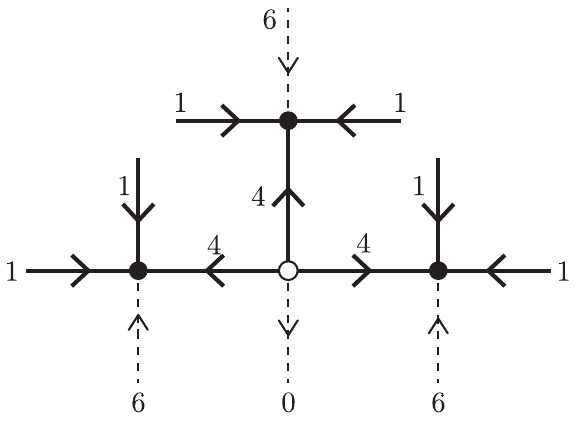}
\caption{The cheapest annihilation channel for $S = 3$ ($2S = 6$): six fundamental strings ($\phi_\ell = 1$) fuse pairwise at three vertices with link values $(1, 1, 4, 6)$, and the three resulting charge-2 intermediate bonds (displayed link value $\phi_\ell = 4 = 2S - 2$) annihilate at a common vertex with link values $(4, 4, 4, 0)$, whose vacuum link absorbs the total charge $2S$. Arrows indicate the bipartite A$\to$B link orientation along which $\phi_\ell$ is defined; solid and dashed lines denote defect and vacuum links; open and filled circles denote A- and B-sublattice vertices. An equivalent configuration is obtained by the global flip $\phi_\ell \to 2S - \phi_\ell$.}
\label{fig:cascade2}
\end{figure}

Other fusion trees carry slightly different exponents [for $S = 3$, the cheapest channel fuses the fundamentals pairwise into three charge-2 bonds of fugacity $x_2 = w^{-8}$ each, costing $w^{-24}$ instead of the chain's $w^{-25}$---a minimum shared by the pairwise-doubling channel $2 + 2 \to 4$ terminated by $(4,1,1)$; see Fig.~\ref{fig:cascade2}], but the following bound is channel independent. Each pairwise fusion produces exactly one intermediate bond, so a channel that starts from the $2S$ fundamentals and terminates at a vertex absorbing $k \in \{2, 3\}$ strands contains $2S - k$ fusion outputs; of these, at most one---a conjugate fundamental with $\phi = 2S-1$, which can only be consumed in the pair termination $(2S-1, 1)$---falls outside the counted range $2 \le \phi \le 2S-2$. Every channel therefore contains at least $2S-3$ counted intermediate bonds and, since the tension exponent $\phi(2S-\phi)$ is minimized at the edges of this range, each of them costs at least $w^{-2(2S-2)}$. Hence the penalty of \emph{any} channel is bounded by
\begin{equation}
    \mathcal{P}_{\text{bridge}} \;\le\; w^{-4(S-1)(2S-3)}.
    \label{eq:bridge_bound}
\end{equation}

For $S = 2$, $\mathcal{P}_{\text{bridge}} = w^{-4}$; for $S = 5/2$, $\mathcal{P}_{\text{bridge}} = w^{-12}$; and the suppression strengthens rapidly with $S$, the penalty decaying at least as fast as $e^{-c\,S^2}$ at fixed $w > 1$. We also note that the exponent $4(S-1)(2S-3)$ vanishes precisely for $S \le 3/2$---the minimal number of intermediate bonds in any annihilation channel is $\max(0,\, 2S-3)$---recovering the \emph{unsuppressed} pair annihilation of the $S = 1$ loops and the unsuppressed cubic junction at $S = 3/2$. The exponential protection of the $U(1)$ structure thus sets in exactly at $S = 2$, encoding the $S \le 3/2$ vs $S \ge 2$ dichotomy of Sec.~\ref{sec:defect_graphs_compatibility} in a single formula.

In the $\mathbb{Z}_{2S}$ clock model, by contrast, the weaker modular conservation $\sum_{\ell \in \bm{r}} n_\ell \equiv 0 \pmod{2S}$ (rather than the exact ice rule) permits a qualitatively shorter cascade. For $2S \le z = 4$ (i.e., $S \le 2$), all $2S$ fundamental currents annihilate at a single vertex ($\underbrace{1+\cdots+1}_{2S} = 2S \equiv 0$), so no intermediate bond is needed and the penalty is unity. For $S > 2$, a cascade is required but is shorter than in the spin-ice model. For example, at $S = 3$ ($q = 6$), three fundamental currents ($n = 1$ each) merge at the first vertex to produce an intermediate bond with current $n = 3$ (using three of the four links); at the second vertex, this intermediate bond meets three more fundamentals, and $3+1+1+1 = 6 \equiv 0 \pmod{6}$ terminates the cascade with just one intermediate bond (carrying $n = 3$, with fugacity $t_3$), compared to $2S - 3 = 3$ intermediate bonds in the spin-ice model. In general, the clock-model cascade requires only $\lceil S - 2 \rceil$ intermediate bonds with polynomially suppressed fugacities (Sec.~IV of Supplemental Material~\cite{SMarxiv}). A comparison of the two cascade costs is given in Table~\ref{tab:annihilation}.

\begin{table}[t]
\centering
\caption{Number of intermediate bonds and total cascade penalty for annihilating $2S$ fundamental strings on the diamond lattice ($z = 4$). Each row shows the number of intermediate bonds connecting the fusion vertices and the product of their fugacities (the ``penalty''); a penalty of 1 means no intermediate bond is needed. For $S \le 3/2$ the two models agree and no cascade is needed. For $S \ge 2$ they diverge: the clock model requires at most $\lceil S{-}2 \rceil$ bonds with polynomially suppressed fugacities $t_n$ (Sec.~IV of Supplemental Material~\cite{SMarxiv}), while the spin-ice model requires $2S-3$ bonds with exponentially suppressed fugacities $x_\phi = w^{-\phi(2S-\phi)}$; see the text for details.}
\label{tab:annihilation}
\renewcommand{\arraystretch}{1.3}
\begin{tabular}{c cc cc}
\toprule
& \multicolumn{2}{c}{\textbf{$\mathbb{Z}_{2S}$ clock}} & \multicolumn{2}{c}{\textbf{Spin ice}} \\
\cmidrule(lr){2-3}\cmidrule(lr){4-5}
Spin $S$ & Bonds & Penalty & Bonds & Penalty \\
\midrule
$1$    & $0$ & $1$             & $0$ & $1$ \\
$3/2$  & $0$ & $1$             & $0$ & $1$ \\
\midrule
$2$    & $0$ & $1$             & $1$ & $w^{-4}$ \\
$5/2$  & $1$ & $t_3$           & $2$ & $w^{-12}$ \\
$3$    & $1$ & $t_3$           & $3$ & $w^{-24}$ \\
$7/2$  & $2$ & $t_3 t_5$       & $4$ & $w^{-42}$ \\
\bottomrule
\end{tabular}
\end{table}

Because the cascade is exponentially suppressed by the spin stiffness, at leading order the active macroscopic graph ensemble is stripped of degree-3 branching junctions and collapses to fundamental strings forming closed directed loops and degree-4 loop crossings. This restricted ensemble is topologically isomorphic to the high-temperature graphical expansion of the continuous classical 3D $XY$ model \cite{Nahum2011}, differing only in short-distance vertex weights.

\subsubsection{Decomposition of the partition function and 3D $XY$ universality}

To turn this continuous $U(1)$ baseline into a controlled analytical argument for the 3D $XY$ universality class for $S \ge 2$, we construct a decomposition of the vacuum-normalized partition function $\tilde{Z}_{\text{ice}} \coloneqq Z_{v=0}/Z_{\text{vac}}$, where $Z_{\text{vac}}$ is the partition function restricted to the vacuum sector in which every link carries the maximal amplitude $|S_\ell^z| = S$ subject to the ice rule (i.e., the $\mathbb{Z}_{2S}$-confined phase with no defects). The ratio $\tilde{Z}_{\text{ice}}$ thus measures the total statistical weight of all defect configurations relative to the defect-free background. We decompose $\tilde{Z}_{\text{ice}}$ into a continuous $U(1)$ sector and an exponentially suppressed discrete $\mathbb{Z}_{2S}$ correction.

To separate the $U(1)$-conserving sector from the cascade corrections, we note that at crossing vertices (degree 4), the $S \ge 2$ spin ice carries a constant Pauling weight $\lambda_4 = (1/6)/(1/3)^2 = 3/2$ relative to two independent passing strings. We therefore define a \emph{decorated $XY$ model}, denoted $\tilde{Z}_{XY}^{(3/2)}$, which employs the ice-model bond fugacities $\tilde{x}_{|n|}$ for currents $|n| \le 2S-1$ and exact $U(1)$ current conservation at every vertex (the same Gauss law as the ice rule), while additionally enhancing every crossing vertex by a factor $\beta = 3/2$.

Since $\lambda_4 = 3/2$ is a constant independent of $S$ and $w$, the crossing decoration acts as a short-range quartic composite operator that flows to zero under the renormalization group (RG). Because $\tilde{Z}_{XY}^{(3/2)}$ employs the same bond fugacities $\tilde{x}_{|n|}$ as the ice model, every non-cascade graph of the spin-ice loop gas appears in both partition functions with the same weight, the weights being evaluated---here as throughout this section---within the independent-vertex approximation.

There are two remaining differences to keep track of. First, the spin-ice configurations not captured by the decorated $XY$ model are the fusion cascades analyzed above, whose total contribution $\delta Z_{\text{cascade}}$ is controlled by the bridge penalty $\mathcal{P}_{\text{bridge}}$ [Eq.~\eqref{eq:cascade_suppression}] and is exponentially small in $1/T$. Second, the decorated $XY$ ensemble contains some local current configurations with no spin-ice counterpart [e.g., a vertex with currents $(+1,+1,+1,-3)$, which conserves the $U(1)$ charge but is incompatible with the ice rule, which permits at most two positive-deviation links at any vertex]; excluding them is a $U(1)$-symmetric, short-range modification of vertex weights, on the same footing as the crossing decoration, and is treated as such in the RG argument below. With this bookkeeping (see Supplemental Material~\cite{SMarxiv}, Sec.~VII), the partition function decomposes as
\begin{equation}
    \tilde{Z}_{\text{ice}} = \tilde{Z}_{XY}^{(3/2)} + \delta Z_{\text{cascade}}.
    \label{eq:sandwich}
\end{equation}

The physical content of this decomposition is as follows. The decorated $XY$ model $\tilde{Z}_{XY}^{(3/2)}$ captures the $U(1)$-conserving sector of the spin-ice loop gas, and shares the 3D $XY$ universality class (as shown below). The cascade contribution $\delta Z_{\text{cascade}}$ is the only term sensitive to the discreteness of the $\mathbb{Z}_{2S}$ conservation law, and its bare weight is exponentially suppressed by the lattice geometry. Combined with RG arguments, this decomposition places $\tilde{Z}_{\text{ice}}$ in the 3D $XY$ universality class, as follows.

First, the crossing decoration operator $\beta^{V_4}$ introduced in the decorated $XY$ model preserves the continuous $U(1)$ current conservation at every vertex and therefore corresponds to a $U(1)$-symmetric irrelevant perturbation. At the 3D $XY$ critical fixed point, the leading correction-to-scaling exponent is $\omega \approx 0.79$ \cite{Hasenbusch2019}, so the scaling dimension of this operator is $\Delta = 3 + \omega \approx 3.79 > 3$. Likewise, the differences between the ice-model bond fugacities $\tilde{x}_{|n|}$ and the Bessel ratios $\tau_{|n|}$ are $U(1)$-symmetric short-range perturbations involving higher-current operators ($|n| \ge 2$), which are even more strongly irrelevant at the $XY$ fixed point. Being irrelevant, the RG flow guarantees that the decorated $XY$ model flows to the 3D $XY$ universality class.

Second, the cascade correction $\delta Z_{\text{cascade}}$ is the only mechanism that distinguishes the discrete $\mathbb{Z}_{2S}$ conservation law from the continuous $U(1)$ conservation law; in the coarse-grained $XY$ description it is represented by the leading discrete anisotropy $\cos(2S\theta)$. Two logically independent facts now combine. (i)~\emph{RG irrelevance}: At the 3D $XY$ fixed point, the anisotropy $\cos(p\theta)$ has scaling dimension $\Delta_p > 3$ for all $p \ge 4$ ($\Delta_4 \approx 3.11$ from Monte Carlo RG studies \cite{Lou2007, Shao2020}, and $\Delta_p$ increases monotonically with $p$; here $p = 2S \ge 4$). It is therefore a (dangerously) irrelevant operator \cite{Jose1977, Lou2007}: It cannot alter the critical universality class for any bare coupling within the basin of attraction of the $XY$ fixed point---in particular, for any sufficiently weak bare coupling. (A sufficiently strong bare anisotropy can escape this basin: The hard $\mathbb{Z}_4$ clock model, the infinite-anisotropy limit at $p = 4$, decomposes into two decoupled Ising models and orders with Ising rather than $XY$ exponents; this is precisely why the following point matters.) (ii)~\emph{Bare smallness}: The decomposition shows that the bare weight of this operator is not merely small but exponentially small [$\mathcal{O}(e^{-c/T})$, Eqs.~\eqref{eq:cascade_suppression} and~\eqref{eq:bridge_bound}], placing the system deep inside the basin of attraction. It also controls the crossover scale: The length scale beyond which the asymptotic $XY$ behavior becomes visible is short, so no remnant of the discreteness should be observable even in moderate-size simulations. Together, (i) and (ii) place the deconfinement transitions of the $S \ge 2$ pyrochlore spin ice in the 3D $XY$ universality class, with an early onset of the asymptotic critical behavior.

\subsubsection{Quantitative estimate of $w_c$ via $O(2)$ loop-gas mapping}

Because the discrete anisotropy is highly suppressed and irrelevant for $S \ge 2$, the system asymptotes to a continuous $O(2)$ symmetry near criticality. We can thus quantitatively estimate the deconfinement transition point by mapping the dilute directed-loop gas to the high-temperature graphical expansion of the classical 3D $XY$ model on the diamond lattice \cite{Nahum2011}. The transition occurs when the fundamental string fugacity $\tilde{x}_1 = \frac{2}{3}w^{-(2S-1)}$ reaches the critical fugacity $\tau_1 \coloneqq I_1(K_c^{XY}) / I_0(K_c^{XY})$, where $K_c^{XY}$ is the 3D $XY$ critical coupling on the diamond lattice. Equating $\tilde{x}_1 = \tau_1$, we obtain the general scaling formula:
\begin{equation}
    w_c(S) = \left(\frac{2}{3\,\tau_1}\right)^{\!\frac{1}{2S-1}}, \qquad S \ge 2. \label{eq:scaling}
\end{equation}
From our Monte Carlo determination \cite{WatanabeH2026} $K_c^{XY} \approx 0.769$ [i.e., $T_c^{XY} = 1.30036(1)$], we obtain $\tau_1 \approx 0.359$ and $2/(3\tau_1) \approx 1.859$, giving $w_c(S) \approx (1.859)^{1/(2S-1)}$.
As $S \to \infty$, the critical fugacity asymptotes to $w_c \to 1$ (i.e., $\mu \to 0$), consistent with the classical continuous-spin limit: as $S \to \infty$ the discrete spin variable $S^z_\ell \in \{-S, \ldots, S\}$ approaches a continuous variable, the $\mathbb{Z}_{2S}$ flux quantization becomes infinitely fine, and the discrete $\mathbb{Z}_{2S}$-confined phase vanishes; any weak easy-axis anisotropy, i.e., $w$ slightly below $w_c(S)$, within the present model then suffices to stabilize the continuous $U(1)$ Coulomb liquid.
The geometric suppression of the $\mathbb{Z}_{2S}$ fusion cascade, established by the decomposition above, ensures that the continuous $O(2)$ approximation underlying this formula remains accurate for all $S \ge 2$. The complete classification of phases, flux quantization, and critical phenomena obtained in the monopole-free limit is collected in Tables~\ref{tab:integer_spins} and~\ref{tab:halfinteger_spins} and illustrated schematically in Fig.~\ref{fig:phase_diagram}.

\section{Finite-temperature effects: Magnetic monopoles, string severing, and crossovers}
\label{sec:finite_T}

The analysis in the previous sections relied on the strict enforcement of the local ice rule, $Q_{\bm{r}} = 0$. This monopole-free limit is realized in the limit $T/J \to 0$ taken at fixed $w$, i.e., at fixed $\mu/T$ (Sec.~\ref{sec:model_and_review}\,C). At any realistic finite temperature $T > 0$, the finite exchange $J$ permits the thermal excitation of magnetic monopoles \cite{Castelnovo2008, Ryzhkin2005}, characterized by a small but strictly nonzero fugacity $v \coloneqq \exp(-J/2T) \ll 1$.

In this section, we return to the exact partition function generated in Eq.~\eqref{eq:partition_exact}. It is well established on general grounds that three-dimensional topological order is fragile against thermal fluctuations \cite{Nussinov2008, Castelnovo2008b, Zhou2025}. Building on this principle, we show concretely that the thermal excitation of monopoles explicitly breaks the microscopic $U(1)$ invariance encoded in the divergence-free constraint---entering the dual description of the small-$w$ regime as a symmetry-breaking field---and acts as a topological severing of defect strings in the large-$w$ regime \cite{Watanabe2026,Watanabe2026ice}, destroying the macroscopic topological invariants and rounding the phase transitions into crossovers for all $S$ (with the exception of $S=3/2$).

\subsection{Small-$w$ regime: Emergence of a symmetry-breaking sine-Gordon field}

The construction in this subsection is standard \cite{Jose1977, Savit1980}, and we keep the derivation brief for self-containedness; the conclusion that thermal monopoles act as an effective external field in the dual $XY$ language was reached for $S = 1$ in Ref.~\cite{Watanabe2026}. We note, however, that this dual viewpoint is itself recent: Thermal monopoles in spin ice have traditionally been treated through the continuum electrostatic analogy, with screening inserted at the Debye level \cite{Ryzhkin2005, Castelnovo2008, Henley2010}. In the exact dual representation below, by contrast, a monopole of charge $Q_{\bm{r}}$ appears as the local vertex operator $e^{i Q_{\bm{r}} \theta_{\bm{r}}}$, so the full monopole gas enters the partition function---for every $S$ at once---as an explicit $U(1)$ symmetry-breaking site weight, from which the screening physics follows rather than being assumed. To properly capture the effect of thermal monopoles in the small-$w$ limit (Sec.~\ref{sec:small_w}), we relax the constraint in the partition function by reinstating the sum over local divergences. The partition function in the dual phase field representation becomes
\begin{align}
    Z =& \sum_{\{Q_{\bm{r}}\}} \sum_{\{S_\ell^z\}} \int \mathcal{D}\theta\notag\\
    & \exp\left( -i \sum_{\bm{r}} \theta_{\bm{r}} (\nabla \cdot \bm{S}_{\bm{r}} - Q_{\bm{r}}) \right) v^{\sum_{\bm{r}} Q_{\bm{r}}^2} w^{\sum_\ell (S_\ell^z)^2}.
\end{align}

Following the exact discrete integration by parts introduced previously, the exact spin sector decouples and maps to the local link weights $W_S(\nabla\theta_\ell)$ as before. The new addition is the unconstrained independent summation over the local discrete monopole charges $Q_{\bm{r}}$ at every single site, which we define as the local site action $M(\theta_{\bm{r}})$:
\begin{equation}
    M(\theta_{\bm{r}}) = \sum_{Q_{\bm{r}} \in \mathbb{Z}} v^{Q_{\bm{r}}^2} e^{i Q_{\bm{r}} \theta_{\bm{r}}}.
\end{equation}

Because each diamond lattice vertex connects exactly four links, the local divergence $Q_{\bm{r}}$ is the algebraic sum of four spin variables. Even for half-integer spins, the sum of four half-integers is an integer, so $Q_{\bm{r}} \in \mathbb{Z}$ for all $S$.

In the low-temperature regime $v \ll 1$, this sum is dominated by the fundamental monopole excitations $Q_{\bm{r}} = \pm 1$. Truncating at leading order,
\begin{align}
    M(\theta_{\bm{r}}) &\approx 1 + v e^{i \theta_{\bm{r}}} + v e^{-i \theta_{\bm{r}}} + \mathcal{O}(v^4) \nonumber \\
    &= 1 + 2v \cos(\theta_{\bm{r}}) \approx \exp(2v \cos \theta_{\bm{r}}).
\end{align}

Taking the logarithm, we find that thermal monopoles generate a local sine-Gordon potential at every lattice vertex. For integer spins ($S \ge 1$), the effective action evaluates to
\begin{equation}
    S_{\text{eff}}[\theta] \approx -2w\sum_\ell \cos(\nabla\theta_\ell) - 2v \sum_{\bm{r}} \cos(\theta_{\bm{r}}).
\end{equation}

The generated term $-2v \cos(\theta_{\bm{r}})$ acts as a uniform external magnetic field of strength $h \propto v$ applied along the $\theta=0$ direction of the emergent 3D $XY$ model. This field explicitly breaks the continuous $U(1)$ symmetry and is a relevant perturbation that gaps the Goldstone modes and removes the thermodynamic singularity. Consequently, the 3D $XY$ phase transition is rounded into a smooth crossover at any finite temperature $T > 0$ \cite{Watanabe2026, Watanabe2026ice}.

For half-integer spins ($S \ge 1/2$), where the unperturbed system resides in the $U(1)$ Coulomb liquid phase, the effective action acquires the same term. This confirms the standard Debye-H\"uckel screening mechanism \cite{Castelnovo2011, Ryzhkin2005}: The monopole plasma introduces a screening length that exponentially cuts off the algebraic dipolar correlations, smoothly converting the Coulomb liquid into a trivial paramagnet without a phase transition.

\subsection{Large-$w$ regime: Generalized directed loop gas and topological string severing}

\begin{figure}[t]
\centering
\includegraphics[width=\columnwidth]{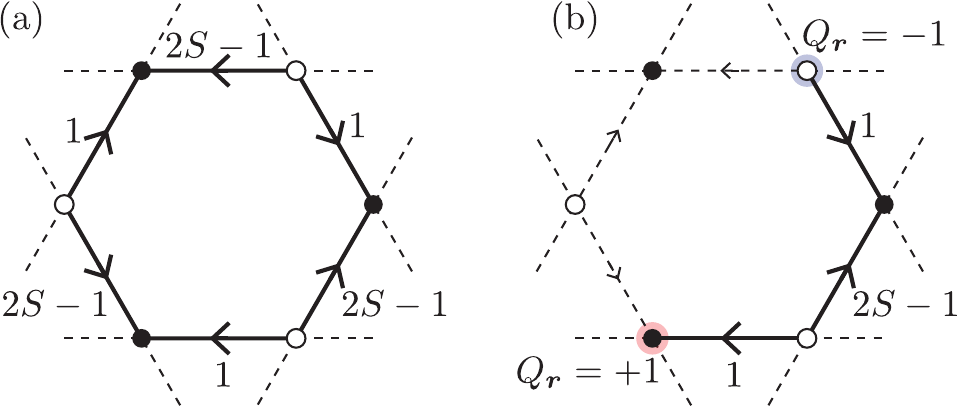}
\caption{Defect-string topology on the diamond lattice. Solid lines denote defect links carrying $\phi_\ell = 1$ or $2S - 1$ (alternating with the sublattice convention); dashed lines denote vacuum links ($\phi_\ell = 0$ or $2S$); arrows indicate the bipartite A$\to$B link orientation along which $\phi_\ell$ is defined; open and filled circles denote A- and B-sublattice vertices. (a)~At $v = 0$, defect strings form closed directed loops with $Q_{\bm{r}} = 0$ at every vertex. (b)~At $v > 0$, thermal monopoles ($|Q_{\bm{r}}| = 1$) act as string end points, severing closed loops into open segments. The hexagonal loop is drawn schematically in two dimensions; on the actual lattice it is a chair-shaped six-membered ring.}
\label{fig:string_severing}
\end{figure}

We now symmetrically generalize the large-$w$ directed loop-gas expansion to finite temperatures. The geometric consequence of thermally activating local monopoles ($Q_{\bm{r}} \neq 0$) is that the defect strings are no longer topologically forced to form closed continuous loops. The directed strings can now terminate, creating open end points $\partial G$ that correspond geometrically to the locations of the thermal monopoles \cite{Morris2009, Jaubert2009} (Fig.~\ref{fig:string_severing}).

The loop-gas partition function now expands over all directed graphs $G$ on the diamond lattice, which include both closed loops and open string segments terminated by monopole end points. For $S \ge 3/2$, the end point vertex weight is computed as follows. At a monopole end point, one of the four links carries a fundamental defect ($\phi = 1$, fixed), while the remaining three links are in the vacuum sector ($\phi \in \{0, 2S\}$). The monopole condition $\sum_{\ell \in \bm{r}} \phi_\ell = 4S \pm 1$ (i.e., $|Q_{\bm{r}}| = 1$) constrains how many of the three vacuum links carry $\phi = 2S$ versus $\phi = 0$. In the defect-free vacuum, each vertex has $\binom{4}{2} = 6$ valid ice-rule configurations (those with exactly two links at $\phi = 2S$ and two at $\phi = 0$). Of these 6 configurations, exactly $\hat{z}(1) = 3$ are compatible with one link being promoted to a defect while maintaining $|Q_{\bm{r}}| = 1$ (see Sec.~II of Supplemental Material~\cite{SMarxiv}). The end point vertex weight is therefore $W_1 = \hat{z}(1)/6 = 1/2$. By the same normalization convention used for the junction and crossing weights ($\lambda_d = W_d / W_2^{d/2}$), the end point correction factor is $\lambda_1 = W_1/\sqrt{W_2} = (1/2)/\sqrt{1/3} = \sqrt{3}/2$. Including the monopole fugacity $v$, each end point contributes a factor $v\lambda_1 = \sqrt{3}\,v/2$, giving the generalized dilute loop gas partition function:
\begin{equation}
    \frac{Z_{v > 0}}{Z_{\text{vac}}} = \sum_{\{G\}} 2^{N_{\text{comp}}(G)}\, \tilde{x}_1^{\,|G|} \left( \frac{\sqrt{3}\,v}{2} \right)^{|\partial G|},
    \label{eq:Z_large_w_v}
\end{equation}
where $|G|$ is the total number of edges, $|\partial G|$ is the number of monopole end points (string termination sites with $|Q_{\bm{r}}| = 1$), and $N_{\text{comp}}(G)$ is the number of strands---both closed loops and open strings---each carrying two chiral orientations on the directed diamond lattice.
This graphical expansion maps exactly to the high-temperature expansion of a classical $O(2)$ (i.e., 3D $XY$) spin model in the presence of a uniform external magnetic field $h$. In the standard $O(2)$ high-temperature expansion, an external field $h$ assigns a weight $h/2$ to each string end point. Equating $h/2 = \sqrt{3}\,v/2$ immediately yields the effective field strength:
\begin{equation}
    h_{\text{eff}} \approx \sqrt{3}\, v.
    \label{eq:h_eff}
\end{equation}

This result extends to $S=1$ ($n=1$, undirected loops mapping to the 3D Ising model). Unlike directed loops at $S \ge 3/2$, the $S=1$ defect $S^z_\ell = 0$ is directionless: $\phi = 1$ is its own conjugate under $\phi \to 2S - \phi$, so a single defect link can terminate at either a $Q = +1$ or $Q = -1$ monopole. This doubles the valid vacuum configurations at each end point from $\hat{z}(1) = 3$ to $3+3 = 6$, and hence the end point weight from $W_1 = 1/2$ to $W_1 = 1$ (see Ref.~\cite{Watanabe2026} for the detailed $S=1$ derivation). The resulting end point factor $\sqrt{3}v$ (doubled relative to the directed case $\sqrt{3}v/2$) precisely compensates for the halved magnetic coupling in the Ising convention ($\exp(h \sigma) \propto 1 + \tanh(h) \sigma$), yielding the same universal effective field: $h_{\text{eff}} \approx \sqrt{3}v$.

Thus, irrespective of $S$, thermal monopoles act as a uniform external magnetic field applied to the emergent gauge variables. This field severs the defect strings, allowing the confined topological loops to terminate in the bulk. The macroscopic flux quantization laws established in Sec.~\ref{sec:completeness} are thereby destroyed, rendering all topological sectors adiabatically connected \cite{Watanabe2026, Chern2014}.

\subsection{Order-dependent fate of topological transitions}

The thermodynamic consequence of this string-severing field depends on the order of the monopole-free phase transition. On the one hand, for continuous transitions ($S=1$ and $S \ge 2$), any finite symmetry-breaking field conjugate to the order parameter destroys a second-order critical point \cite{Polyakov1977, Castelnovo2008b}; the continuous deconfinement transitions are therefore rounded into smooth crossovers for any $v > 0$. On the other hand, for the first-order transition ($S=3/2$), the situation is qualitatively different: Because the transition involves a discontinuous jump between two macroscopic minima separated by a free-energy barrier (driven by the relevant cubic invariant), a coexistence line is expected to persist under a weak symmetry-breaking field.

Therefore, the first-order transition unique to $S=3/2$ is predicted to survive at finite temperatures as a genuine thermodynamic phase boundary [Fig.~\ref{fig:phase_diagram}(b)]. The situation is analogous to the liquid-gas transition: At zero external field ($v = 0$), the free energy has two degenerate minima (ordered and disordered phases) separated by a barrier. A weak symmetry-breaking field ($v > 0$) tilts the free-energy landscape but cannot eliminate the barrier, so the coexistence line should persist. As the monopole fugacity $v$ increases, the free-energy barrier shrinks and vanishes at a finite critical monopole density $v_c$, where the two minima merge into one. At this point, the first-order line terminates at a critical end point, analogous to the liquid-gas critical point and expected to belong to the 3D Ising universality class.

To estimate $v_c$, we extend the Bethe-Peierls analysis of Sec.~\ref{sec:S32_Potts} by including the monopole-induced symmetry-breaking field $h$, which enters as $s = (e^h-1)/(e^h+2) = \sqrt{3}v/2$ [cf.\ Eq.~\eqref{eq:Z_large_w_v}]. The external field modifies the cavity recursion to $r = e^h [(e^K r + 2)/(r + e^K + 1)]^b$, where the prefactor $e^h$ biases the cavity message toward the field-favored state. The critical end point---the point at which the free-energy barrier just vanishes---is determined by three simultaneous conditions on the iterative map $g(r)$: (i)~$g(r) = r$ (self-consistency of the cavity message), (ii)~$g'(r) = 1$ (the ordered and disordered fixed points have merged, i.e., marginal stability), and (iii)~$g''(r) = 0$ (the inflection point of $g(r) - r$ vanishes, signifying the disappearance of the intervening barrier). These three equations for the three unknowns $(K_c, h_c, r_c)$ can be solved in closed form, yielding $e^{K_c} = (\sqrt{33}-1)/2$ and $h_c \approx 0.010$, corresponding to $v_c \approx 0.004$ or $T_c/J \approx 0.09$ (see Sec.~VIII of Supplemental Material~\cite{SMarxiv} for the complete algebraic derivation).

The small value of $h_c$ reflects the weakness of the first-order transition at $q=3$: Since $q = 3$ is the smallest integer for which the Potts model exhibits a first-order transition in three dimensions, the free-energy barrier is intrinsically shallow and easily destroyed by a weak perturbation. Indeed, the Bethe-Peierls approximation systematically overestimates the barrier height because it neglects the long-wavelength fluctuations that further erode the shallow free-energy landscape. Monte Carlo simulations of the 3D three-state Potts model in an external ordering field on the simple cubic lattice \cite{KarschStickan2000, WadaKitazawaKanaya2025} have established that the first-order coexistence line terminates at a critical end point in the 3D Ising universality class at a critical field $h_c = 7.75(10) \times 10^{-4}$---more than an order of magnitude smaller than the Bethe-Peierls estimate $h_c \approx 0.010$. On the diamond lattice relevant to the present pyrochlore problem, where the lower coordination number ($z = 4$ vs $z = 6$) amplifies thermal fluctuations, the true critical field is expected to be smaller still.

This critical end point provides a distinctive experimental signature: In candidate $S=3/2$ pyrochlore materials, the specific heat and susceptibility should exhibit Ising-like critical scaling at an isolated temperature, offering a direct thermodynamic probe of the underlying string-fusion physics. Notably, unlike the liquid-gas-type critical end points in conventional $S=1/2$ spin ice, which require a finely tuned external magnetic field \cite{Castelnovo2008}, this critical end point occurs at strictly zero external field. The symmetry-breaking field is intrinsically generated by the thermal monopole plasma, making the $S=3/2$ pyrochlore spin ice a rare example of a system that spontaneously generates its own topological critical end point through thermal fluctuations alone.

\section{Monte Carlo simulations at finite temperature}
\label{sec:MC}

To test the analytical predictions of the preceding sections at finite temperatures, we perform classical Monte Carlo simulations of the Hamiltonian Eq.~\eqref{eq:model} on finite pyrochlore lattices comprising $2N = 16L^3$ spins (where $N = 8L^3$ is the number of diamond-lattice vertices) and periodic boundary conditions. Throughout this section we fix $J = 1$ as the energy unit.

\subsection{Simulation method}

A standard Monte Carlo sweep consists of $2N$ single-spin heat-bath updates. After $10^6$ thermalization sweeps, measurements are accumulated over $5 \times 10^6$ sweeps.

The primary thermodynamic observable is the specific heat per spin,
\begin{equation}
    c \coloneqq \frac{1}{2N}\frac{\langle E^2 \rangle - \langle E \rangle^2}{T^2},
    \label{eq:specific_heat}
\end{equation}
whose peaks trace the crossover scales. To unambiguously distinguish collective topological phenomena from noncooperative local single-ion physics, we systematically compute the thermal occupancy fraction of each spin amplitude,
\begin{equation}
    n_s \coloneqq \frac{1}{2N}\sum_\ell \langle \delta_{|S^z_\ell|, s} \rangle,
    \label{eq:occupancy}
\end{equation}
i.e., the fraction of links carrying amplitude $|S^z_\ell| = s$, averaged over all $2N$ links and Monte Carlo configurations.

\subsection{Integer spins ($S = 1, 2, 3$)}

\begin{figure*}
\centering
\includegraphics[width=0.8\textwidth]{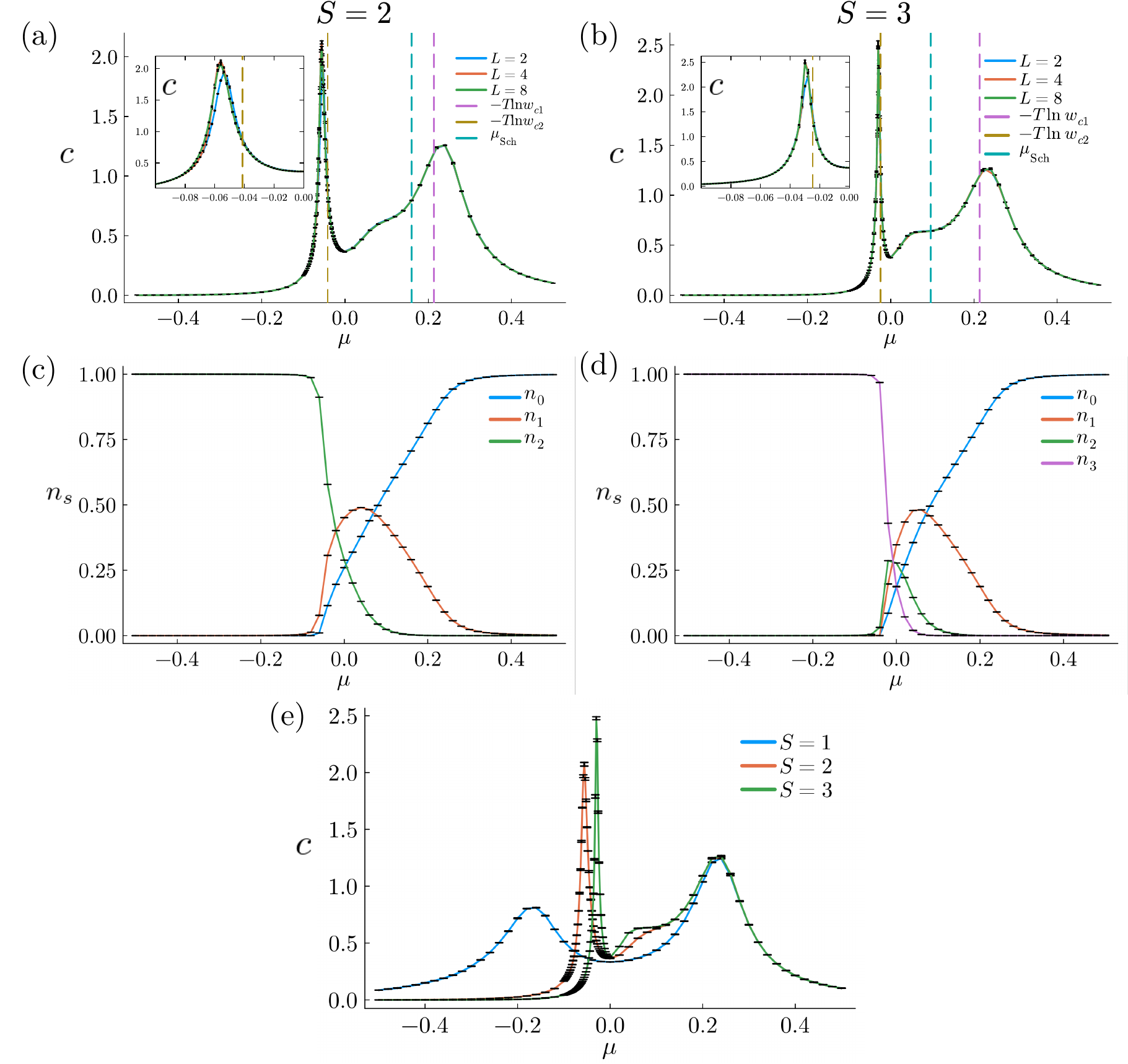}
\caption{Monte Carlo results for integer spins at $T = 0.2$ with $L = 2$, 4, 8; the exchange coupling is fixed to $J = 1$, and $\mu$ and $T$ are quoted in units of $J$. (a),(b)~Specific heat per spin $c$ for $S = 2$ and $S = 3$; insets enlarge the $\mu < 0$ peak region. Vertical dashed lines mark the predicted crossover scales $-T \ln w_{c1}$ (purple) and $-T \ln w_{c2}$ (dark yellow) and the Schottky position $\mu_{\text{Schottky}} = 2.4\,T/(2S-1)$ (cyan). (c),(d)~Thermal occupancy fractions $n_s = \langle \delta_{|S^z_\ell|, s} \rangle$ for $S = 2$ and $S = 3$ ($L = 8$). (e)~Specific heat across $S = 1$, 2, and 3 ($L = 8$).}\label{fig:integer}
\end{figure*}

We compute the specific heat as a function of $\mu$ at low temperatures ($T = 0.2$) for $S = 1$, 2, and~3 with system sizes $L = 2$, 4, and~8; Figs.~\ref{fig:integer}(a) and~\ref{fig:integer}(b) show the results for $S=2$ and $S=3$, respectively. In all three cases, $c$ exhibits prominent peaks whose positions are in reasonable agreement with the analytically predicted crossover scales $\mu = -T \ln w_{c1}$ and $\mu = -T \ln w_{c2}$, where $w_{c1}$ and $w_{c2}$ are the monopole-free critical fugacities estimated in Secs.~\ref{sec:small_w} and~\ref{sec:large_w}, respectively. Notably, the peak heights do not diverge with increasing $L$, confirming that the monopole-free phase transitions are rounded into smooth crossovers at finite temperature, as predicted by the monopole-screening analysis of Sec.~\ref{sec:finite_T}. For $S = 1$, the results are fully consistent with the detailed numerical study of Ref.~\cite{Watanabe2026}, which demonstrated that the specific-heat peaks saturate to finite values in the thermodynamic limit and that the $\mathbb{Z}_2$ deconfinement transition (which maps onto the ice-VII to ice-X transformation in high-pressure water ice) is a continuous crossover driven by Debye-H\"uckel monopole screening.

For $S \ge 2$, in addition to the cooperative crossover peaks, the specific heat exhibits a broad secondary shoulder near $\mu \approx 0.1$, which is absent for $S=1$ (because $S=1$ hosts only two amplitudes, $|S^z|=0$ and $1$, so no intermediate depopulation step exists). As established by the completeness argument of Sec.~\ref{sec:completeness}, this feature does not correspond to an intermediate thermodynamic phase. Rather, it is a noncooperative \emph{Schottky anomaly} driven by the single-ion anisotropy. The single-ion energy splitting between the two highest-amplitude states, $|S^z_\ell| = S$ and $|S^z_\ell| = S-1$, is $\Delta E = \mu[S^2 - (S-1)^2] = \mu(2S-1)$. For $\mu > 0$, the maximal-amplitude state $|S^z_\ell| = S$ is the excited (higher-energy) state of this two-level system. Approximating this as an independent two-level system whose excited state $|S^z_\ell| = S$ is thermally depopulated as $\mu$ increases, the Schottky specific heat peaks at $\Delta E/T \approx 2.4$, yielding an estimated peak position $\mu_{\text{Schottky}} \approx 2.4T/(2S-1)$. This analytical estimate accurately captures the position and the $1/(2S-1)$ scaling of the secondary humps observed in the Monte Carlo data: As $S$ increases, the hump shifts closer to $\mu = 0$ [Fig.~\ref{fig:integer}(e)]. Furthermore, the explicitly computed microscopic spin occupancies $n_s$ [Figs.~\ref{fig:integer}(c) and~\ref{fig:integer}(d)] reveal that this shoulder coincides precisely with the continuous thermal depopulation from the maximal-amplitude state $|S^z_\ell| = S$ ($n_S \to 0$) to the lower spin amplitudes ($n_{S-1} \to 1$). The absence of finite-size scaling for this shoulder, combined with the occupancy data, clearly decouples the local single-ion entropy release from the collective topological phenomena.

\subsection{Half-integer spins ($S = 3/2, 5/2, 7/2$)}

For half-integer spins, the analytical framework of Sec.~\ref{sec:small_w} predicts no transition in the small-$w$ regime because the half-integer link weights map directly onto the $S=1/2$ Coulomb liquid. The only nontrivial topological crossover is therefore the large-$w$ deconfinement crossover, whose monopole-free scale is set by $w_c \approx 1.44$ for $S=3/2$ (our closed-form Bethe-Peierls estimate, Sec.~\ref{sec:S32_Potts}; the transition point measured in Ref.~\cite{Pandey2026_S32} is $w_c \approx 1.42$) and $w_c \approx (1.859)^{1/(2S-1)}$ for $S \ge 5/2$ [Eq.~\eqref{eq:scaling}].

\begin{figure*}
\centering
\includegraphics[width=0.8\textwidth]{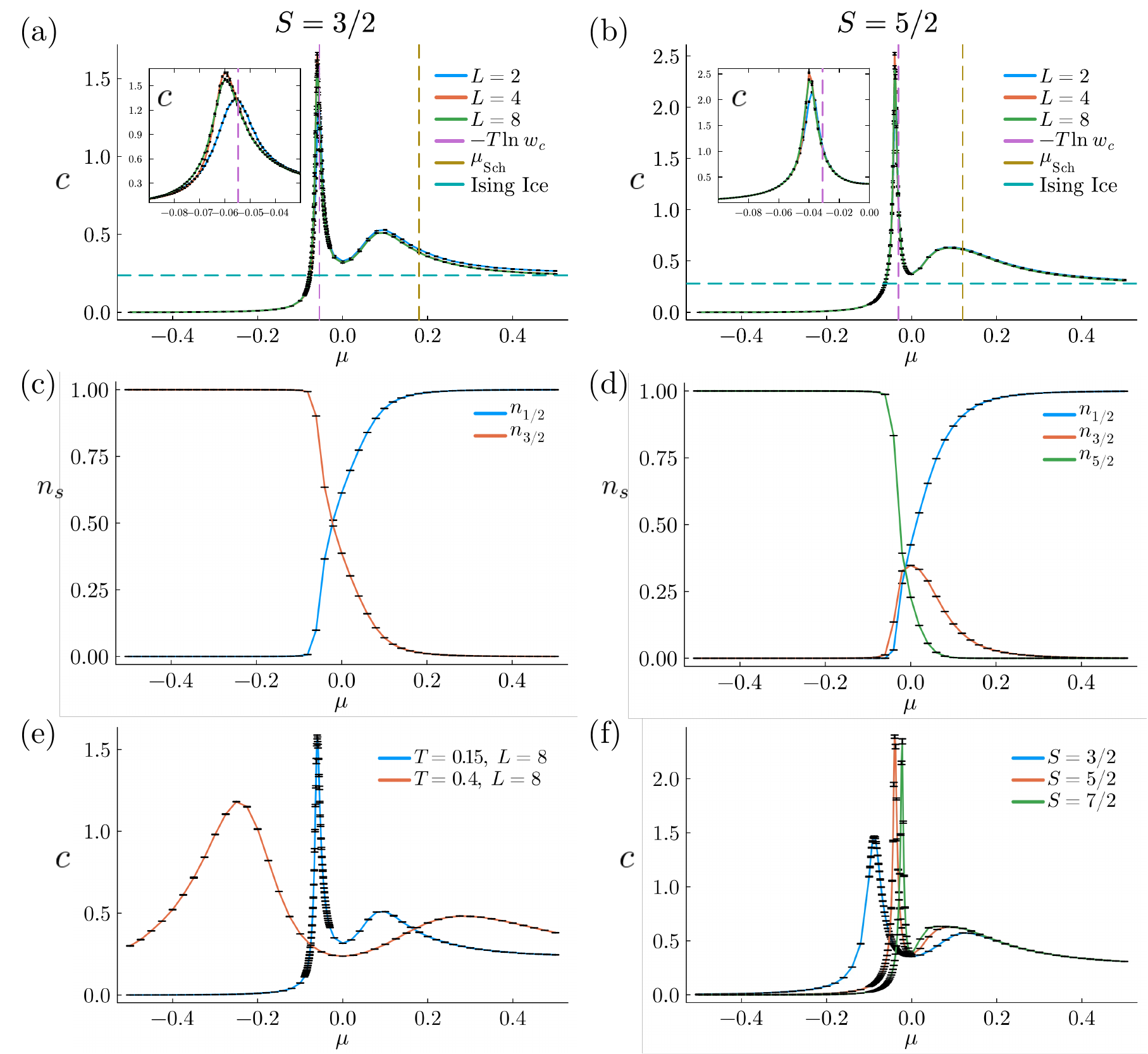}
\caption{Monte Carlo results for half-integer spins; the exchange coupling is fixed to $J = 1$, and $\mu$ and $T$ are quoted in units of $J$. (a)~Specific heat per spin $c$ for $S = 3/2$ at $T = 0.15$ and (b)~for $S = 5/2$ at $T = 0.2$, with $L = 2$, 4, 8; insets enlarge the peak region. Vertical dashed lines mark the predicted crossover scale $-T \ln w_c$ (purple; for $S = 3/2$, $w_c \approx 1.44$ from Table~\ref{tab:halfinteger_spins}) and the Schottky position $\mu_{\text{Schottky}}$ (dark yellow); the horizontal dashed line (cyan) marks the specific heat of the effective $S = 1/2$ Ising spin ice at $T_{\text{eff}} = 4T$. (c),(d)~Thermal occupancy fractions $n_s$ for $S = 3/2$ and $S = 5/2$ ($L = 8$). (e)~The $S = 3/2$ specific heat at $T = 0.15$ and $0.4$ ($L = 8$). (f)~Comparison across $S = 3/2$, $5/2$, and $7/2$ at $T = 0.2$ ($L = 8$).}\label{fig:halfinteger}
\end{figure*}

The Monte Carlo specific heat at $T = 0.15$ ($S = 3/2$) and $T = 0.2$ ($S = 5/2$) confirms this picture [Figs.~\ref{fig:halfinteger}(a) and~\ref{fig:halfinteger}(b)]: A sharp dominant peak associated with the topological crossover appears near $\mu = -T \ln w_c$, and its height saturates with increasing $L$; the corresponding peak for $S = 7/2$ at $T = 0.2$ is shown in Fig.~\ref{fig:halfinteger}(f) ($L = 8$). In the large-positive-$\mu$ (small-$w$) limit, where the single-ion anisotropy $\mu (S^z)^2$ confines every spin to its minimal doublet $S^z = \pm 1/2$, the model reduces to an effective $S = 1/2$ Ising spin ice whose exchange coupling is rescaled by $(1/2)^2 = 1/4$, yielding an effective temperature $T_{\text{eff}} = 4T$. The horizontal dashed lines in Figs.~\ref{fig:halfinteger}(a) and~\ref{fig:halfinteger}(b) show the specific heat of this effective $S = 1/2$ model computed independently, and the Monte Carlo data converge to these values at large $\mu$ for all half-integer spins, providing direct numerical confirmation of the effective Hamiltonian mapping. Furthermore, similar to the integer-spin case, a broader secondary hump is clearly visible on the large-$\mu$ side for all $S \ge 3/2$. As confirmed by the fractional occupancies $n_s$ [Figs.~\ref{fig:halfinteger}(c) and~\ref{fig:halfinteger}(d)], which track the thermal depopulation cascade, this is the identical noncooperative Schottky anomaly driven by $\Delta E \sim T$. As predicted by our scaling formula $\mu_{\text{Schottky}} \approx 2.4T/(2S-1)$, this Schottky peak gradually shifts toward $\mu = 0$ and merges with the topological crossover peak as $S$ increases [Fig.~\ref{fig:halfinteger}(f)].

For $S = 3/2$, the monopole-free limit ($T/J \to 0$) predicts a first-order transition driven by the symmetry-allowed $\mathbb{Z}_3$ Potts cubic invariant. At the finite temperatures accessible to our simulations, however, the specific-heat peak shows no sign of thermodynamic divergence with $L$, and the energy Binder cumulant exhibits no negative dip characteristic of a first-order transition. This implies that the first-order coexistence line has either terminated at a critical end point below the simulated temperatures or that exponentially larger system sizes are needed to resolve a weakly first-order nature. Indeed, a comparison of the specific heat at $T=0.15$ and $T=0.4$ [Fig.~\ref{fig:halfinteger}(e)] reveals that while the peak sharpens upon cooling, it remains a crossover. Quantitatively, at $T = 0.15$, the monopole fugacity $v = e^{-J/(2T)} \approx 0.036$ is an order of magnitude larger than the critical end point value $v_c \approx 0.004$ estimated from the Bethe-Peierls analysis in Sec.~\ref{sec:finite_T} (see Sec.~VIII of Supplemental Material~\cite{SMarxiv})---and likely 2 orders of magnitude larger than the true lattice value, given that the Bethe-Peierls approximation overestimates $h_c$ by more than a factor of 10 relative to the exact Monte Carlo result for the 3D three-state Potts model \cite{KarschStickan2000}. The simulation therefore unambiguously places the system well inside the crossover regime, confirming that thermal monopoles have completely washed out the macroscopic phase coexistence. We stress the corresponding limitation: Because all accessible temperatures lie above the estimated critical end point, the present simulations test the crossover regime but do not directly verify the predicted survival of the first-order line at lower temperatures.

Prompted by this limitation, we have additionally performed parallel-tempering simulations of the physical ($v > 0$) model at smaller anisotropies, for which the predicted transition temperature lies \emph{below} the estimated critical end point; the question probed is not the existence of the first-order transition---established at $v = 0$ in Ref.~\cite{Pandey2026_S32}---but its predicted survival at $v > 0$. These runs show signatures consistent with a weakly first-order transition---a specific-heat peak that sharpens and grows with system size at the predicted location, a negative minimum of the energy Binder cumulant, and an incipient bimodal energy distribution---and the signatures become clearer as $\mu \to 0^-$, where the monopole fugacity at the transition is smaller. The finite-size hallmarks that would make the identification definitive are, however, not yet resolved at the accessible sizes; a conclusive test requires more sophisticated sampling and is left for future work.

We close this section by commenting on the prospects for simulating the monopole-free limit directly. For $S = 1/2$ spin ice, it is standard to sample the ice-rule manifold at $v = 0$ using nonlocal loop updates \cite{Melko2001}, and worm algorithms of exactly this type were used in Refs.~\cite{Pandey2025_S1, Pandey2026_S32} to map out the $v = 0$ phase diagram at $S = 1$---including finite-size-scaling analyses consistent with the 3D $XY$ and 3D Ising universality classes of the two transitions---and to establish the first-order transition at $S = 3/2$. The same strategy generalizes to arbitrary $S$: A worm-type update that shifts $S^z_\ell$ by $\pm 1$ along a closed loop (respecting $|S^z_\ell| \le S$) is ergodic within the ice-rule manifold. For $S \ge 2$, where our theory predicts 3D $XY$ criticality at the couplings $w_c(S)$ of Eq.~\eqref{eq:scaling} and an $S$-independent small-$w$ transition at $w_{c1}$ for integer $S$, no such simulations have yet been performed; they would provide a direct test of these predictions and are deferred to a separate study. The finite-temperature fate of the $S = 3/2$ coexistence line near the predicted critical end point ($T/J \approx 0.09$, where $v \approx v_c$) could likewise be probed by loop-based sampling supplemented with monopole insertions.

\section{Experimental outlook}
\label{sec:outlook}

Before concluding, we discuss experimental contexts for the model, Eq.~\eqref{eq:model}, and for our predictions.

\subsection{Higher-spin pyrochlore magnets}
The single-ion term in Eq.~\eqref{eq:model} is not a theoretical idealization: The magnetic site of every pyrochlore magnet has $D_{3d}$ point symmetry with its trigonal axis along the local $\langle 111 \rangle$ direction, so for any ion with $S \ge 1$, the leading symmetry-allowed single-ion anisotropy is precisely $\mu (S^z)^2$ with $S^z$ along the local axis. Anisotropies comparable to the exchange coupling are documented on both sides of $\mu = 0$: Electron-paramagnetic-resonance measurements for the $S = 7/2$ gadolinium pyrochlores find an easy-plane single-ion term ($\mu > 0$) of about $0.75\,J$ in $\mathrm{Gd_2Ti_2O_7}$ \cite{Glazkov2005}, while for the $S = 3/2$ pyrochlore $\mathrm{Cd_2Os_2O_7}$, an easy-axis term ($\mu < 0$) with $|\mu| \approx J$ has been obtained by ab initio quantum-chemistry calculations and adopted in the resonant inelastic x-ray scattering analysis of the magnetic excitations \cite{Bogdanov2013, Calder2016}. Higher-spin pyrochlore magnets now exist across the full range of spin values---most systematically in the fluoride family, which realizes effective $S = 1/2$ ($\mathrm{NaCaCo_2F_7}$ \cite{Ross2017}), $S = 1$ ($\mathrm{NaCaNi_2F_7}$ \cite{KrizanCava2015}), $S = 2$ ($\mathrm{NaCaFe_2F_7}$, $\mathrm{NaSrFe_2F_7}$), and $S = 5/2$ ($\mathrm{NaSrMn_2F_7}$) \cite{Sanders2017} on the same lattice \cite{ReigPlessisHallas2021}. We stress that none of these compounds is currently placed at the transitions studied here: The Ni and Mn members of the fluoride family are nearly Heisenberg-like (for $\mathrm{NaCaNi_2F_7}$, polarized neutron scattering bounds the single-ion anisotropy to a small fraction of the exchange \cite{Plumb2019}), the single-ion anisotropy of the $S = 2$ iron members has not yet been determined experimentally, and materials with dominant easy-axis exchange typically sit deep inside one phase. Rather, the present model provides the organizing phase diagram for this growing materials landscape---with the canonical spin ices $\mathrm{Ho_2Ti_2O_7}$ and $\mathrm{Dy_2Ti_2O_7}$ occupying the extreme easy-axis limit $w \to \infty$, in which the crystal field isolates the largest-$|m_J|$ doublet---and identifies which single-ion parameters, tunable by chemical substitution, strain, or pressure, would drive a material across the crossover ridges---and, for $S = 3/2$, across the first-order line---classified here.

\subsection{High-pressure ice}
Outside magnetism, the $S = 1$ member of Eq.~\eqref{eq:model} describes the hydrogen-bond network of high-pressure water ice, with $S^z_\ell = \pm 1$ representing the two off-center proton positions on a hydrogen bond, $S^z_\ell = 0$ the bond-centered position, $\mu$ the pressure-tuned centering energy, and ionic defects ($\mathrm{H_3O^+}$/$\mathrm{OH^-}$) playing the role of monopoles. In this language, proton-disordered ice VII corresponds to the $\mathbb{Z}_2$-confined Coulomb phase at large $w$, bond-symmetrized ice X to the trivial paramagnet at small $w$, and the monopole-induced rounding derived in Sec.~\ref{sec:finite_T} explains why the ice VII--X transformation is observed as a broad, probe-dependent crossover rather than a sharp transition; a companion paper develops this application in detail \cite{Watanabe2026ice}. Notably, the competition $\mu \sim J$ is realized there directly by the proton coordinate, without fine-tuning, and the isotope effect ($\mathrm{H_2O}$ versus $\mathrm{D_2O}$) shifts the crossover pressure by about $10\,\mathrm{GPa}$ \cite{Goncharov1996}, making the model parameters experimentally scannable.

\subsection{Engineered platforms}
Blume-Capel ($S = 1$) degrees of freedom have recently been realized in an artificial nanomagnet array with lithographically tunable anisotropy \cite{Berchialla2026}, and programmable platforms---superconducting-flux-qubit annealers realizing artificial spin ice \cite{King2021} and proposed three-dimensional Rydberg-atom arrays on the pyrochlore lattice \cite{Shah2025}---offer settings in which both fugacities $w$ and $v$ can, in principle, be controlled independently. The quantitative phase boundaries of Tables~\ref{tab:integer_spins} and~\ref{tab:halfinteger_spins} and the predicted zero-field critical end point provide concrete targets for such platforms.

\subsection{Measurable signatures}
Three predictions of this work are directly testable. First, for $S = 3/2$ realizations, the first-order line should terminate at a critical end point at strictly zero external field, at which the specific heat and susceptibility exhibit 3D Ising critical scaling at an isolated temperature $T \lesssim J/10$ (Sec.~\ref{sec:finite_T}). Second, the crossover ridges of the specific heat trace the straight lines $\mu_c = -T \ln w_c$ with the $S$-dependent slopes of Tables~\ref{tab:integer_spins} and~\ref{tab:halfinteger_spins} (for $S = 1$ and $S = 3/2$, the corresponding experimental temperature scales were already emphasized in Refs.~\cite{Pandey2025_S1, Pandey2026_S32}). Third, the Schottky anomaly at $\mu_{\text{Schottky}} \approx 2.4\,T/(2S-1)$ provides a practical protocol for separating single-ion features from the collective crossovers in candidate higher-spin pyrochlore magnets (Sec.~\ref{sec:MC}).

\section{Conclusion and outlook}
\label{sec:conclusion}

In this work, we have developed a self-contained theoretical framework that classifies the macroscopic topological phases, emergent gauge theories, and critical phenomena of classical spin-$S$ pyrochlore magnets for arbitrary $S$. The framework combines thermodynamic energy convexity, the microscopic lattice geometry ($z=4$), exact graphical mappings, and RG analysis.

In the monopole-free limit, we established the completeness of the phase diagram: Uniform backgrounds of intermediate spin amplitude are excluded by the convexity of the single-ion energy, and partial-deconfinement scenarios are excluded by a free-energy comparison of the competing defect strings. A spin parity dichotomy emerges in the small-$w$ limit: half-integer spins reduce to the divergence-free Coulomb fluid, whereas integer spins map onto the 3D $XY$ model. In the large-$w$ limit, we proved geometrically that the macroscopic polarization flux is quantized to multiples of $2S$, establishing an emergent $\mathbb{Z}_{2S}$-confined Coulomb phase. The resulting classification in this limit is summarized in Tables~\ref{tab:integer_spins} and~\ref{tab:halfinteger_spins}.

The nature of the deconfinement transitions reveals a further dichotomy. While $S=1$ defects map to the 3D Ising class, all $S \ge 3/2$ defects generate chiral, directed loops. The ice-rule compatibility criterion shows that the $\mathbb{Z}_{2S}$ Gauss law is compatible with the physical ice rule at the vertex level if and only if $S \le 3/2$. Exploiting this geometric property, we mapped the $S=3/2$ defect gas to the three-state Potts model, whose symmetry-allowed cubic invariant drives a first-order transition.

For $S \ge 2$, this discrete clock mapping breaks down: The modular $\mathbb{Z}_{2S}$ conservation law is no longer equivalent to the physical ice rule. The system is instead forced into hierarchical fusion processes with exponentially suppressed intermediate defects. Using a decomposition of the partition function, we showed that the macroscopic graph ensemble reduces to the continuous $U(1)$ 3D $XY$ model at leading order. The residual fusion cascade acts as a dangerously irrelevant operator at the 3D $XY$ fixed point; its exponentially small bare weight places the system deep inside the basin of attraction of that fixed point, where its irrelevance fixes the universality class and guarantees an early onset of the asymptotic critical behavior, placing all $S \ge 2$ deconfinement transitions in the 3D $XY$ universality class.

Finally, by relaxing the strict ice rules, we showed that thermally excited monopoles act as an explicit symmetry-breaking field $h_{\text{eff}} \approx \sqrt{3}v$ in the emergent gauge theory. This field severs the topological defect strings and renders all topological sectors adiabatically connected. The consequences are order dependent: The continuous 3D Ising ($S=1$) and 3D $XY$ ($S \ge 2$) transitions are rounded into crossovers for any $v > 0$, while the first-order $S=3/2$ transition is predicted to survive at finite temperatures, protected by a macroscopic free-energy barrier, and to terminate at a critical end point. Classical Monte Carlo simulations for integer ($S = 1$--$3$) and half-integer ($S = 3/2$--$7/2$) spins confirm this crossover picture at the accessible temperatures: The specific-heat peaks saturate to finite values with increasing system size, and the crossover scales are in quantitative agreement with the analytically predicted fugacities. We emphasize, however, an important limitation of the numerical part of this work: The accessible temperatures lie above the estimated critical end point of the $S = 3/2$ coexistence line ($v \approx 0.036$ at $T = 0.15$, versus $v_c \approx 0.004$ from the Bethe-Peierls estimate), so the predicted survival of the $S = 3/2$ first-order transition---and, more generally, the universality classes of the monopole-free transitions---are not directly verified by the present simulations. Direct tests would require sampling the $v = 0$ manifold with loop algorithms, as discussed in Sec.~\ref{sec:MC}, and constitute an important direction for future work. Preliminary parallel-tempering simulations at smaller $|\mu|$, reaching below the estimated end point, already show signatures consistent with the predicted survival (Sec.~\ref{sec:MC}), though not yet its definitive demonstration. The spin-amplitude occupancies $n_s$ further identify a noncooperative Schottky anomaly for $S \ge 2$, clearly separating local single-ion physics from collective topological phenomena. This clear decoupling, governed by the analytically predicted $1/(2S-1)$ scaling of the Schottky peak position, provides a practical diagnostic for experimentalists seeking to disentangle broad specific-heat anomalies in candidate higher-spin pyrochlore magnets from the signatures of collective topological phenomena.

Extending this framework to the quantum regime is a natural direction for future work. The interplay between quantum fluctuations, emergent discrete $\mathbb{Z}_{2S}$ gauge fields, and the geometric properties unique to $S=3/2$ may give rise to novel fractionalized quantum spin liquids and higher-spin topological phases \cite{Kitaev2006, Savary2012, Shannon2012, Savary2017, GingrasMcClarty2014, Benton2016}. More broadly, the mechanism identified here---in which the microscopic real-valued constraint keeps the model $U(1)$-symmetric while the lattice geometry exponentially suppresses the processes that would single out a discrete subgroup---may apply to other frustrated lattice models where discrete gauge symmetries compete with continuous gauge symmetries, including dimer models on nonbipartite lattices and higher-rank tensor gauge theories.

\begin{acknowledgments}
The work of S.W. is supported by JST SPRING, Grant No.~JPMJSP2108.
The work of Y.M. is supported by JSPS KAKENHI Grant No.~JP25H01247.
The work of H.W. is supported by JSPS KAKENHI Grant No.~JP24K00541.
The authors used Claude (Anthropic) to assist with analytical derivations, the development and review of the numerical simulation code, literature verification, and manuscript preparation; all output was independently verified by the authors.
\end{acknowledgments}

\section*{Data Availability}

The data that support the findings of this study, including the Monte Carlo simulation data shown in Figs.~\ref{fig:integer} and~\ref{fig:halfinteger}, are available from the corresponding author upon reasonable request.

\bibliography{ref_spinS}

\clearpage
\setcounter{section}{0}
\setcounter{subsection}{0}
\setcounter{subsubsection}{0}
\makeatletter
\renewcommand{\theHsection}{SM.\arabic{section}}
\renewcommand{\theHsubsection}{SM.\arabic{section}.\arabic{subsection}}
\renewcommand{\theHsubsubsection}{SM.\arabic{section}.\arabic{subsection}.\arabic{subsubsection}}
\makeatother
\onecolumngrid
\begin{center}
{\large\bfseries Supplemental Material for ``Topological Phase Transitions and Their Thermodynamic Fate in Arbitrary-$S$ Classical Pyrochlore Spin Ice''}\\[1.5ex]
Sena Watanabe, Yukitoshi Motome, and Haruki Watanabe
\end{center}
\vspace{1ex}
\twocolumngrid


\section{Phase diagram in the $(w, T/J)$ plane}
\label{app:phase-w}

\begin{figure}[t]
\centering
\includegraphics[width=0.8\columnwidth]{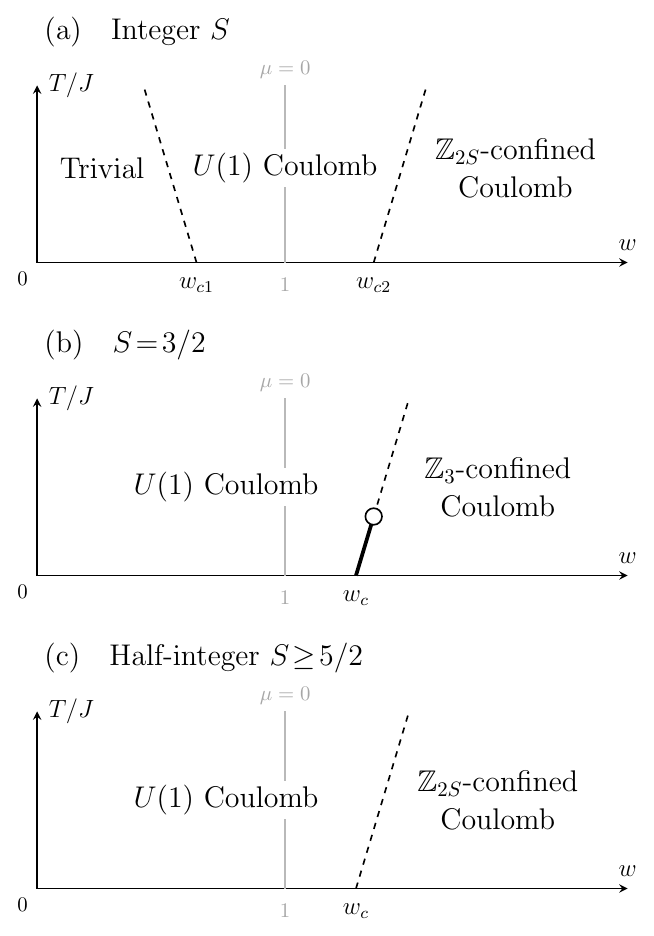}
\caption{Schematic phase diagram in the $(w, T/J)$
plane, where $w \coloneqq e^{-\mu/T}$, for (a)~integer $S \ge 1$,
(b)~$S = 3/2$, and (c)~half-integer $S \ge 5/2$; the phases
realized on the lower axis are labeled in each panel, and the
vertical gray line marks $w = 1$ ($\mu = 0$). Dashed lines denote
crossover ridges (no thermodynamic singularity at $T/J > 0$); the
solid line in (b) denotes the first-order coexistence line,
terminating at a critical endpoint (open circle). The lower
boundary $T/J = 0$ is approached at fixed $w$, i.e., with
$\mu/T = -\ln w$ held fixed, and realizes the monopole-free
($v = 0$) constrained ensemble---which retains an extensive
configurational entropy at any $w > 0$ and hosts sharp,
entropy-driven transitions at the critical fugacities $w_c$---rather
than a zero-temperature limit at fixed couplings.}
\label{fig:SM_phase_w}
\end{figure}

Figure~\ref{fig:SM_phase_w} shows the schematic phase diagram of
Fig.~1 of the main text replotted in the $(w, T/J)$ plane, where
$w = e^{-\mu/T}$ is the single-ion fugacity. For $T > 0$ the map
$(\mu, T) \mapsto (w, v) = (e^{-\mu/T}, e^{-J/2T})$ is a smooth
bijection, so the two representations carry identical information;
the merit of the $(w, T/J)$ plane is that the monopole-free
($v = 0$) transitions appear as separate points on the lower axis,
at the critical fugacities $w_c$ of Tables~I and~II of the main
text. In particular, the two transitions of the integer-$S$ case,
at $w_{c1}$ and $w_{c2}$, are separately visible, with the $U(1)$
Coulomb regime as the interval between them.

The lower boundary $T/J = 0$ at fixed $w$ corresponds to
approaching the origin of the $(\mu, T)$ plane along the fixed
direction $\mu/T = -\ln w$: the $w$ axis resolves the origin of
Fig.~1 of the main text into a full axis, each point of which
represents one ray $\mu = -T \ln w$. This is the parametrization
in which the $T/J \to 0$ phase diagrams of
Refs.~\cite{Pandey2025_S1, Pandey2026_S32} were obtained, and in
which the constrained ($v = 0$) ensemble hosts the sharp,
entropy-driven transitions analyzed in the main text. At any
$T/J > 0$ the monopole fugacity is nonzero, and, exactly as in the
main text, the continuous transitions are rounded into crossovers
(dashed lines), while the $S = 3/2$ first-order line (solid)
survives up to its critical endpoint (open circle).

\section{Large-$w$ Expansion of the Spin-Ice Model ($S \ge 2$)}
\label{app:spin-ice-HTE}

In this section we present a self-contained derivation of the
large-$w$ expansion of the spin-$S$ pyrochlore spin-ice model for
$S \ge 2$.
The exceptional case $S = 3/2$ is treated separately in
Sec.~\ref{app:spin-ice-S32}, and the $S = 1$ case in the
Supplemental Material of Ref.~\cite{Watanabe2026}.
Each bond of the pyrochlore lattice carries a bounded spin
$S^z_\ell \in \{-S,\dots,S\}$, and the coordination number $z=4$ of
the diamond lattice imposes geometric constraints that are absent
in the clock model and will be analyzed in detail below.

\subsection{Notation and conventions}
\label{app:ice-def}

We adopt the model and notation introduced in
Sec.~II of the main text.
For ease of reference within this section, we restate the key
equations with local labels.
The Hamiltonian [Eq.~(1) of the main text] reads
\begin{equation}\label{eq:app-ice-H}
    H = \frac{J}{2}\sum_{\bm{r}}
    \Bigl(\sum_{\ell \in \bm{r}} S^z_\ell\Bigr)^{\!2}
    + \mu \sum_\ell (S^z_\ell)^2,
\end{equation}
and the exact partition function [Eq.~(3) of the main text], expressed
in terms of the fugacities $w \coloneqq e^{-\mu/T}$ and
$v \coloneqq e^{-J/(2T)}$, is
\begin{equation}\label{eq:app-ice-Z}
    Z_{\mathrm{ice}} = \sum_{\{Q_{\bm{r}}\}}
    \sum_{\{S^z_\ell\}}
    \Bigl(\prod_{\bm{r}}
      \delta_{\nabla\!\cdot\!\bm{S}_{\bm{r}},\,Q_{\bm{r}}}\Bigr)\,
    v^{\sum_{\bm{r}} Q_{\bm{r}}^2}\;
    w^{\sum_\ell (S^z_\ell)^2},
\end{equation}
with the sublattice-dependent lattice divergence
\begin{equation}\label{eq:app-ice-div}
    \nabla\!\cdot\!\bm{S}_{\bm{r}} \coloneqq
    \begin{cases}
        +\sum_{\ell \in \bm{r}} S^z_\ell
        & \text{if } \bm{r} \in A, \\[4pt]
        -\sum_{\ell \in \bm{r}} S^z_\ell
        & \text{if } \bm{r} \in B.
    \end{cases}
\end{equation}

\subsection{Monopole-free limit and loop-gas expansion}
\label{app:ice-monopole-free}

We now specialize to the monopole-free limit $v = 0$
($J/T \to \infty$), setting $Q_{\bm{r}} = 0$ at every site in
Eq.~\eqref{eq:app-ice-Z}:
\begin{equation}\label{eq:app-ice-Z-v0}
    Z_{\mathrm{ice}} \coloneqq Z_{\mathrm{ice}}\big|_{v=0}
    = \sum_{\{S^z_\ell\}}
    \Bigl(\prod_{\bm{r}}
      \delta_{\nabla\!\cdot\!\bm{S}_{\bm{r}},\,0}\Bigr)\,
    \prod_\ell w^{(S^z_\ell)^2}.
\end{equation}
From this point onward, all \emph{algebraic} manipulations within the
$v=0$ sector are exact; the only approximation entering below is
Pauling's independent-vertex evaluation of the ice-rule degeneracy,
which is flagged explicitly wherever it is used.

\subsubsection{Vacuum partition function}
In the large-$w$ limit ($\mu \to -\infty$), the dominant spin
configuration on each link is the fully polarized state
$S^z_\ell = \pm S$, with Boltzmann weight $w^{S^2}$ per link.
We call this the \emph{vacuum}: every link carries $|S^z_\ell| = S$,
and the remaining freedom is the sign
$\sigma_\ell \coloneqq \mathrm{sgn}(S^z_\ell) \in \{+1,-1\}$.
The ice-rule constraint $\nabla\!\cdot\!\bm{S}_{\bm{r}} = 0$
[Eq.~\eqref{eq:app-ice-div}] then reads
$S\sum_{\ell\in\bm{r}} e_{\bm{r}}\,\sigma_\ell = 0$
at every diamond-lattice site, which reduces to the two-in, two-out
rule: exactly two of the four links at each vertex carry $\sigma_\ell = +1$
and two carry $\sigma_\ell = -1$ (in the sublattice convention).
The vacuum partition function is therefore
\begin{equation}\label{eq:app-ice-Zvac}
    Z_{\mathrm{ice,vac}}
    = w^{|E|\,S^2}\,\Omega_{\mathrm{ice}},
\end{equation}
where $\Omega_{\mathrm{ice}}$ is the number of two-in, two-out sign
configurations on the diamond lattice.

We evaluate $\Omega_{\mathrm{ice}}$ in the \emph{Pauling
independent-vertex approximation}.
Each vertex has $2^4 = 16$ possible sign patterns on its four links,
of which $\binom{4}{2} = 6$ satisfy the two-in, two-out rule.
Treating the vertex constraints as independent, the fraction of the
$2^{|E|}$ total sign configurations that is globally valid is
$(6/16)^N$, giving
\begin{equation}\label{eq:app-ice-Pauling}
    \Omega_{\mathrm{ice}}
    \approx 2^{|E|}\!\left(\frac{6}{16}\right)^{\!N}
    = 2^{2N}\!\left(\frac{3}{8}\right)^{\!N}
    = \left(\frac{3}{2}\right)^{\!N}\!,
\end{equation}
and hence
\begin{equation}\label{eq:app-ice-Zvac-Pauling}
    Z_{\mathrm{ice,vac}}
    = w^{2N S^2}\!\left(\frac{3}{2}\right)^{\!N}\!.
\end{equation}
This is the Pauling entropy $S_P = Nk_B\ln(3/2)$ of the spin-ice
manifold.

\subsubsection{Defect expansion and closed loops}

Configurations with $|S^z_\ell| < S$ on some links represent
excitations above the vacuum.
We define the \emph{defect number} on each link as
\begin{equation}\label{eq:app-ice-defect}
    d_\ell \coloneqq S - |S^z_\ell|
    \in \{0, 1, \dots, S\},
\end{equation}
so that $d_\ell = 0$ is the vacuum and $d_\ell \ge 1$ is an excited
state.
This unsigned defect number is related to the shifted flux variable
$\phi_\ell \coloneqq S^z_\ell + S \in \{0, 1, \dots, 2S\}$
used in the main text by $d_\ell = \min(\phi_\ell,\, 2S - \phi_\ell)$.
The Boltzmann weight of a link with defect number $d$ relative to the
vacuum is
\begin{equation}\label{eq:app-ice-xd}
    \frac{w^{(S-d)^2}}{w^{S^2}}
    = w^{-d(2S-d)},
\end{equation}
which is exponentially small for $w \gg 1$.
In particular, the lightest excitation $d=1$ carries the suppression
factor $w^{-(2S-1)}$, and all higher defects $d \ge 2$ are
exponentially heavier.

The monopole-free constraint $Q_{\bm{r}} = 0$ imposes a strong
geometric restriction: links excited to $d_\ell = 1$
(i.e.\ $|S^z_\ell| = S-1$) must form \emph{closed loops} on the
diamond lattice.
In the dilute regime $w \gg 1$, the dominant contribution comes from
non-overlapping closed loops of fundamental ($d = 1$) defects.
We now derive the effective fugacity per link of such a loop.

\subsubsection{Loop-gas representation}

Consider a configuration in which a subset $G \subseteq E$ of links
is excited to $d_\ell = 1$ while all remaining links carry $d_\ell = 0$.
For $S \ge 3/2$, every link has a well-defined sign
$\sigma_\ell \coloneqq \mathrm{sgn}(S^z_\ell) \in \{+1,-1\}$
(since $S^z_\ell = \pm(S-1) \ne 0$ on defect links).
Let $d_0(\bm{r})$ denote the number of excited links incident
to vertex $\bm{r}$.
At each vertex, the ice rule
$\nabla\!\cdot\!\bm{S}_{\bm{r}} = 0$ constrains the signs via
\begin{equation}\label{eq:app-ice-vertex-general}
    (S-1)\,\Sigma_{\mathrm{def}} + S\,\Sigma_{\mathrm{vac}} = 0,
\end{equation}
where
$\Sigma_{\mathrm{def}} \coloneqq \sum_{\ell\in\bm{r},\,\ell\in G}\sigma_\ell$ and
$\Sigma_{\mathrm{vac}} \coloneqq \sum_{\ell\in\bm{r},\,\ell\notin G}\sigma_\ell$
are the partial sums of defect-link and vacuum-link signs.
We define the \emph{local state sum} $z(d_0)$ as the number of
sign configurations $\{\sigma_\ell\} \in \{+1,-1\}^4$ satisfying
Eq.~\eqref{eq:app-ice-vertex-general};
by the bipartite symmetry of the diamond lattice, $z(d_0)$ depends
only on $d_0$.

\paragraph{Evaluation of $z(d_0)$.}
Since $S-1$ and $S$ are coprime and since $\Sigma_{\mathrm{def}}$ and $\Sigma_{\mathrm{vac}}$ are bounded
integers, Eq.~\eqref{eq:app-ice-vertex-general} is very restrictive.

\emph{$d_0 = 0$ (vacuum):}
The constraint reduces to $\Sigma_{\mathrm{vac}} = 0$, giving
$z(0) = \binom{4}{2} = 6$---the Pauling ice-rule count.

\emph{Odd $d_0$ (monopole):}
For $d_0 = 1$, the constraint gives
$\Sigma_{\mathrm{vac}} = -(S-1)\sigma_1/S$, so $|\Sigma_{\mathrm{vac}}| = (S-1)/S < 1$;
but $|\Sigma_{\mathrm{vac}}| \ge 1$ for any three-sign sum, so $z(1) = 0$.
For $d_0 = 3$, one similarly requires
$|\sigma_4| = (S-1)|\Sigma_{\mathrm{def}}|/S$ with $|\Sigma_{\mathrm{def}}| \in \{1,3\}$;
since $S \ge 2$, the right-hand side never equals~1, giving $z(3) = 0$.
Physically, odd $d_0$ means the vertex is a string endpoint
($Q_{\bm{r}} \ne 0$), forbidden at $v = 0$.

\emph{$d_0 = 2$ (loop vertex):}
The constraint gives $\Sigma_{\mathrm{vac}} = -(S-1)\Sigma_{\mathrm{def}}/S$.
Since $S-1$ and $S$ are coprime for every integer or half-integer
$S \ge 3/2$, the coprimality forces $S \mid \Sigma_{\mathrm{def}}$.
Combined with the bound $|\Sigma_{\mathrm{def}}| \le 2 < 2S$ (which holds for all
$S \ge 3/2$), this leaves only $\Sigma_{\mathrm{def}} = 0$, and hence $\Sigma_{\mathrm{vac}} = 0$, for $S \neq 2$.
For the boundary case $S = 2$, the condition $2 \mid \Sigma_{\mathrm{def}}$ with $|\Sigma_{\mathrm{def}}| \le 2$ permits $\Sigma_{\mathrm{def}} = \pm 2$; however, $\Sigma_{\mathrm{def}} = \pm 2$ gives $\Sigma_{\mathrm{vac}} = -(S-1)\Sigma_{\mathrm{def}}/S = \mp 1$, which is odd, whereas $\Sigma_{\mathrm{vac}}$ is the sum of two vacuum-link signs $(\pm 1 \pm 1) \in \{-2, 0, 2\}$ and must be even. This parity contradiction eliminates $\Sigma_{\mathrm{def}} = \pm 2$, leaving $\Sigma_{\mathrm{def}} = 0$ and $\Sigma_{\mathrm{vac}} = 0$ for all $S \ge 3/2$.
$\Sigma_{\mathrm{def}} = 0$ means the two defect-link signs are opposite at every
loop vertex.
This constraint propagates along a closed loop:
the requirement that consecutive defect-link signs alternate
determines \emph{all} defect-link signs from a single global
binary choice---the loop direction.
These two directions will be accounted for by a factor
$n = 2$ in the loop-gas formula below.
With the direction fixed, only the vacuum-link signs remain free;
$\Sigma = 0$ then requires one $+1$ and one $-1$ among the two
vacuum links ($\binom{2}{1} = 2$ ways):
\begin{equation}\label{eq:app-ice-z2}
    z(2) = 2.
\end{equation}

\emph{$d_0 = 4$ (self-intersection):}
All four defect-link signs are fixed by the loop directions,
so whether the ice rule is satisfied depends on the specific
intersection geometry.
In the dilute-gas limit (non-overlapping loops), self-intersections
do not arise.

\paragraph{Effective fugacity $\tilde{x}_1$.}
We evaluate the partition-function weight of a graph $G$ in the
Pauling independent-vertex approximation.
With the loop direction fixed, the free sign variables live only on
the $2N - |G|$ vacuum links.
The total number of free configurations is $2^{2N-|G|}$,
and the fraction satisfying the ice rule at each vertex is
$z(d_0)/2^{4-d_0}$.
Using $\sum_{\bm{r}}(4 - d_0(\bm{r})) = 4N - 2|G|$,
the Pauling estimate of the number of valid sign configurations
(for a given direction) is
\begin{align}\label{eq:app-ice-Omega-Pauling}
    \Omega(G)
    &\approx 2^{2N - |G|}
      \prod_{\bm{r}}
      \frac{z\bigl(d_0(\bm{r})\bigr)}{2^{4-d_0(\bm{r})}}
      \notag\\
    &= 2^{|G| - 2N}
      \prod_{\bm{r}} z\bigl(d_0(\bm{r})\bigr).
\end{align}
Each excited link contributes $w^{(S-1)^2}$ and each vacuum link
$w^{S^2}$, so the total weight of $G$ is
\begin{align}\label{eq:app-ice-contrib-G}
    \mathrm{contrib}(G)
    &= w^{S^2(2N-|G|)+(S-1)^2|G|}
      \;\Omega(G)
      \notag\\
    &= w^{2NS^2}\,w^{-(2S-1)|G|}
      \;\Omega(G).
\end{align}
Dividing by the vacuum contribution
$Z_{\mathrm{ice,vac}}
= w^{2NS^2}\cdot 2^{-2N}\,z(0)^N$,
the prefactors cancel and we obtain
\begin{equation}\label{eq:app-ice-WG-ratio}
    \frac{\mathrm{contrib}(G)}
         {Z_{\mathrm{ice,vac}}}
    = w^{-(2S-1)|G|} \cdot 2^{|G|}
      \prod_{\bm{r}:\,d_0 > 0}
      \frac{z\bigl(d_0(\bm{r})\bigr)}{z(0)}.
\end{equation}
For closed, non-overlapping fundamental loops,
$d_0(\bm{r}) \in \{0,2\}$ at every vertex.
A single loop of length $|C|$ traverses $|C|$ vertices with
$d_0 = 2$, giving (for one direction)
\begin{align}\label{eq:app-ice-loop-weight}
    \frac{\mathrm{contrib}(C)}{Z_{\mathrm{ice,vac}}}
    &= w^{-(2S-1)|C|} \cdot 2^{|C|}
      \left(\frac{z(2)}{z(0)}\right)^{|C|}
      \notag\\
    &= \left(\frac{2}{3}\,w^{-(2S-1)}\right)^{|C|}\!,
\end{align}
where we used $z(2)/z(0) = 2/6 = 1/3$.
The effective per-link fugacity is therefore
\begin{equation}\label{eq:app-ice-x1}
    \boxed{
    \tilde{x}_1 = \frac{2}{3}\,w^{-(2S-1)}.
    }
\end{equation}

\paragraph{Loop-gas formula.}
Including the direction degeneracy $n = 2$ per loop, the
monopole-free partition function in the dilute-gas limit
(non-overlapping loops with $d_\ell \le 1$ on every link) is
\begin{equation}\label{eq:app-ice-loopgas}
    Z_{\mathrm{ice}}
    = Z_{\mathrm{ice,vac}}
    \sum_{\{C\}} n^{N_{\mathrm{loop}}(C)}\,\tilde{x}_1^{|C|},
\end{equation}
where $Z_{\mathrm{ice,vac}}$ is given
by~\eqref{eq:app-ice-Zvac-Pauling} and the sum runs over all
non-overlapping closed loop graphs $C$ on the diamond lattice.
Eq.~\eqref{eq:app-ice-loopgas} retains only fundamental ($d = 1$)
defects on non-overlapping loops.
Configurations involving heavier defects $d \ge 2$
---including the bridge process in which $2S$ fundamental loops
annihilate via $2S - 2$ vertices and $2S - 3$ intermediate bonds
carrying effective charges $\phi = 2, 3, \dots, 2S-2$
(see Sec.~\ref{app:ice-hierarchical})---enter as subleading
corrections of relative order $w^{-(2S-3)}$ or higher.
The effective fugacity $\tilde{x}_1$ is unaffected by these higher-order
processes; what changes is the topology of graphs that contribute to
the partition function at a given order in $w^{-1}$.

\subsection{Hierarchical fusion: $(2S{-}2)$-vertex bridge process}
\label{app:ice-hierarchical}

The loop-gas formula~\eqref{eq:app-ice-loopgas} involves only
fundamental ($d = 1$) defects.
Higher defects $d \ge 2$ become relevant when $2S$ fundamental loops
annihilate; the cost of this process is the subject of this section.

\subsubsection{Higher-defect fugacity and geometric obstruction}

The Boltzmann weight of a defect-$d$ bond relative to the
fundamental bond is
\begin{equation}\label{eq:app-ice-xd-ratio}
    \frac{x_d}{x_1}
    = w^{-(2S-d)d + (2S-1)}
    = w^{-(d-1)(2S-d-1)}.
\end{equation}
For $d = 2$, this gives $x_2/x_1 = w^{-(2S-3)}$, which is
\emph{exponentially} suppressed for $S \ge 2$.

Can all $2S$ fundamental strings annihilate at a single vertex?
Each diamond-lattice vertex has exactly two outward background links
($\sigma'_{\ell,\bm{r}} = +1$) and two inward background links
($\sigma'_{\ell,\bm{r}} = -1$).
Single-vertex annihilation requires
$k$ incoming fundamental defects ($d=1$) on the outward background
links and $m$ incoming anti-fundamental defects
($d=2S-1$) on the inward background links, with
$k + m = 2S$ and $k = m(2S-1)$
(divergence-free condition: $k \cdot 1 - m \cdot (2S-1) = 0$),
giving $m = 1$, $k = 2S - 1 \ge 3$;
since the diamond lattice provides only $k \le 2$ outward links,
single-vertex annihilation is \emph{forbidden by the exact ice rule} for
$S \ge 2$.
(Note that this obstruction is specific to the exact conservation
$\sum_{\ell \in \bm{r}} \phi_\ell = 4S$; the weaker modular conservation
$\sum_{\ell \in \bm{r}} n_\ell \equiv 0\pmod{2S}$ of the $\mathbb{Z}_{2S}$ clock model
does permit single-vertex annihilation for $2S \le z = 4$; see
Sec.~\ref{app:corr-Sge2-subleading} for a detailed comparison.)
The annihilation in the spin-ice model must therefore proceed through a
\emph{spatially extended bridge} consisting of multiple vertices.

\subsubsection{Sequential fusion cascade}

Consider $2S$ fundamental loops ($\phi = 1$) that are to annihilate.
Since at most two fundamental strings can be fused per vertex
(using the two outward background links), the process requires a
chain of $2S - 2$ diamond-lattice vertices
$\bm{r}_1, \bm{r}_2, \dots, \bm{r}_{2S-2}$,
connected by $2S - 3$ intermediate bonds
(see Fig.~3 of the main text).

The hierarchical fusion proceeds as follows.

\begin{enumerate}
\item \emph{Vertex $\bm{r}_1$}: Two incoming fundamental strings
  ($\phi = 1$) arrive on the two outward background links.
  They fuse into a single outgoing intermediate string with
  effective charge $\phi = 2$ on one inward background link.
  The divergence-free condition is satisfied:
  $2 \times 1 - 1 \times 2 = 0$.

\item \emph{Vertex $\bm{r}_j$} ($2 \le j \le 2S - 3$):
  One incoming intermediate string with $\phi = j$ arrives on an
  outward background link, together with one additional incoming
  fundamental string ($\phi = 1$) on the other outward link.
  They fuse into a single outgoing intermediate string with
  $\phi = j + 1$ on one inward link.
  The divergence-free condition reads:
  $1 \times j + 1 \times 1 - 1 \times (j+1) = 0$.

\item \emph{Vertex $\bm{r}_{2S-2}$} (final fusion):
  One incoming intermediate string with $\phi = 2S - 2$ arrives
  together with one additional fundamental string ($\phi = 1$).
  They produce $\phi = 2S - 1$.
  Since $\phi = 2S - 1 \equiv -1 \pmod{2S}$, this final intermediate
  string carries the dual charge and is absorbed as the
  last remaining anti-string.
  The total charge absorbed is
  $(2S - 2) \times 1 + 1 \times 1 + 1 \times 1 = 2S \equiv 0
  \pmod{2S}$.
\end{enumerate}

At each step, the vertex uses at most $k = 2$ outward links and
$m = 1$ inward link, always respecting the geometric bound $k \le 2$,
$m \le 2$.

\subsubsection{Energetic cost of the bridge}

The bridge introduces $2S - 3$ intermediate bonds carrying
effective charges $\phi = 2, 3, \dots, 2S - 2$.
Each intermediate bond with charge~$\phi$ carries the
fugacity $x_\phi = w^{-\phi(2S-\phi)}$ [Eq.~\eqref{eq:app-ice-xd}].
The total bridge penalty---the product of all intermediate-bond
fugacities---is
\begin{equation}\label{eq:app-ice-bridge-total}
    \mathcal{P}_{\mathrm{bridge}}
    = \prod_{\phi=2}^{2S-2} x_\phi
    = w^{-\sum_{\phi=2}^{2S-2}\phi(2S-\phi)}.
\end{equation}
Evaluating the sum in the exponent by the standard identities
$\sum_{d=1}^{n} d = n(n+1)/2$ and $\sum_{d=1}^{n} d^2 = n(n+1)(2n+1)/6$
gives
\begin{equation}\label{eq:app-ice-bridge-summary}
    \mathcal{P}_{\mathrm{bridge}}
    = w^{-(2S-3)(2S-1)(2S+4)/6}.
\end{equation}

Let us verify this for small values.
For $S = 2$ ($2S = 4$),
the bridge uses $2S - 2 = 2$ vertices and $2S - 3 = 1$ intermediate
bond (carrying $d = 2$);
the exponent is $d(2S-d) = 2 \times 2 = 4$,
so $\mathcal{P}_{\mathrm{bridge}} = w^{-4}$.
For $S = 5/2$ ($2S = 5$),
the bridge uses $3$ vertices and $2$ intermediate bonds
($d = 2$ and $d = 3$);
the exponents are $2 \times 3 = 6$ and $3 \times 2 = 6$,
giving $\mathcal{P}_{\mathrm{bridge}} = w^{-12}$.
For $S = 3$ ($2S = 6$),
the bridge uses $4$ vertices and $3$ intermediate bonds
($d = 2, 3, 4$);
the exponents are $2 \times 4 = 8$, $3 \times 3 = 9$,
$4 \times 2 = 8$,
giving $\mathcal{P}_{\mathrm{bridge}} = w^{-25}$.
All three cases agree with Eq.~\eqref{eq:app-ice-bridge-summary},
confirming the closed-form result.
The bridge penalty is \emph{exponentially} suppressed in $w$
for all $S \ge 2$.

\subsection{Inclusion of thermal monopoles}
\label{app:ice-monopoles}

The results of
Secs.~\ref{app:ice-monopole-free} and~\ref{app:ice-hierarchical}
were derived in the monopole-free limit $v = 0$,
where only configurations with $Q_{\bm{r}} = 0$ at every diamond site
contribute to the partition function~\eqref{eq:app-ice-Z}.
We now relax this constraint and include thermal monopoles ($v > 0$).

At finite $v$, the monopole penalty $v^{Q_{\bm{r}}^2}$ allows
configurations with nonzero monopole charge $Q_{\bm{r}} \ne 0$.
Geometrically, monopoles act as \emph{endpoints} of defect strings:
the defect graph~$G$ is no longer restricted to closed loops but may
contain open strings whose endpoints carry monopole charge.

\subsubsection{Endpoint local state sum}

Consider a vertex $\bm{r}$ that serves as the endpoint of a single
fundamental defect string, so that $d_0(\bm{r}) = 1$.
In the direction-fixed Pauling approximation
(Sec.~\ref{app:ice-monopole-free}), the sign of the defect link is
determined by the string direction, and the three remaining vacuum
links carry independent signs $s_\ell = \pm 1$, each corresponding
to $S^z_\ell = \pm S$.

Without loss of generality, suppose the string exits $\bm{r}$ on
an outward background link with $S^z = +(S-1)$.
The monopole charge~\eqref{eq:app-ice-div} is then
\begin{equation}\label{eq:app-ice-Qendpoint}
    Q_{\bm{r}} = (S{-}1) + S(s_1 - s_2 - s_3),
\end{equation}
where $s_1$ is the sign of the remaining outward vacuum link and
$s_2, s_3$ are the signs of the two inward vacuum links.
The sum $s_1 - s_2 - s_3$ takes values in $\{-3, -1, 1, 3\}$ with
degeneracies $1, 3, 3, 1$ respectively, giving
\begin{equation}\label{eq:app-ice-Qvalues}
    Q_{\bm{r}} \in
    \bigl\{-(2S{+}1),\; -1,\; 2S{-}1,\; 4S{-}1\bigr\}.
\end{equation}
The minimum monopole charge is $|Q_{\bm{r}}| = 1$,
achieved by the three vacuum-sign configurations with
$s_1 - s_2 - s_3 = -1$.
The local state sum at the endpoint is therefore
\begin{equation}\label{eq:app-ice-zv1}
    z_v(1) = 3v + O\!\left(v^{(2S-1)^2}\right),
\end{equation}
where the subleading term comes from the next smallest charge
$|Q_{\bm{r}}| = 2S - 1$, which carries the Boltzmann weight
$v^{(2S-1)^2}$.
An analogous calculation for the opposite string direction or for
a defect on an inward background link yields the same result.

\subsubsection{Generalized loop-gas formula}

We extend the Pauling analysis of Sec.~\ref{app:ice-monopole-free}
to general defect graphs~$G$ that may contain both closed loops and
open strings.
As before, for each connected component we fix the direction,
which determines all defect-link signs; the $2N - |G|$ vacuum-link
signs remain as independent random variables.
The local state sums at degree-0 and degree-2 vertices receive only
exponentially small corrections,
$z_v(0) = 6 + O(v^{4S^2})$ and $z_v(2) = 2 + O(v^{4S^2})$,
and are unchanged at leading order.

Let $N_{\mathrm{loop}}(G)$ and $N_{\mathrm{string}}(G)$ denote the
number of closed-loop and open-string components of~$G$, respectively,
and $|\partial G| = 2N_{\mathrm{string}}$ the total number of
endpoints.
In the dilute gas (non-overlapping fundamental defects, $d_0 \le 2$
at interior vertices), the graph~$G$ consists of
$V_1 = |\partial G|$ endpoint vertices (degree~1),
$V_2 = |G| - N_{\mathrm{string}}$ interior vertices (degree~2),
and $V_0 = N - V_1 - V_2$ vacuum vertices (degree~0).
The Pauling weight ratio~\eqref{eq:app-ice-WG-ratio} generalizes to
\begin{equation}\label{eq:app-ice-WG-monopole}
\begin{split}
    \frac{\Omega^{\mathrm{dir}}_v(G)}{\Omega(\emptyset)}
    &= 2^{|G|}
      \left(\frac{z(2)}{z(0)}\right)^{\!V_2}
      \left(\frac{z_v(1)}{z(0)}\right)^{\!V_1}\\
    &= 2^{|G|}\,
      \left(\frac{1}{3}\right)^{\!V_2}
      \left(\frac{v}{2}\right)^{\!|\partial G|}\!,
\end{split}
\end{equation}
where the superscript ``dir'' indicates a single fixed direction per
component.

Eliminating $V_2 = |G| - |\partial G|/2$ via the degree-sum
identity $1 \cdot V_1 + 2 \cdot V_2 = 2|G|$
(each edge contributes to the degree of its two endpoints):
\begin{equation}\label{eq:app-ice-WG-monopole2}
    \frac{\Omega^{\mathrm{dir}}_v(G)}{\Omega(\emptyset)}
    = \left(\frac{2}{3}\right)^{\!|G|}
      \left(\frac{3v^2}{4}\right)^{\!|\partial G|/2}\!.
\end{equation}
Each connected component admits two directions that give the same
weight (for open strings, reversing the direction swaps
$Q \to -Q$ at the endpoints, leaving $v^{Q^2}$ unchanged).
Including the direction factor
$2^{N_{\mathrm{loop}} + N_{\mathrm{string}}}$
and the Boltzmann weight $w^{-(2S-1)|G|}$:
\begin{align}\label{eq:app-ice-contrib-monopole}
    \frac{\mathrm{contrib}(G)}{Z_{\mathrm{ice,vac}}}
    &= n^{N_{\mathrm{loop}}}
      \,\tilde{x}_1^{|G|}
      \left(\frac{3v^2}{2}\right)^{\!|\partial G|/2}
      \notag\\
    &= n^{N_{\mathrm{loop}}}
      \,\tilde{x}_1^{|G|}
      \,y^{|\partial G|},
\end{align}
where
\begin{equation}\label{eq:app-ice-y}
    \boxed{
    y \coloneqq v\sqrt{\frac{3}{2}}
    }
\end{equation}
is the monopole (endpoint) fugacity.
Summing over all non-overlapping fundamental defect graphs,
the partition function at finite~$v$ becomes
\begin{equation}\label{eq:app-ice-loopgas-v}
    \boxed{
    Z_{\mathrm{ice}}
    = Z_{\mathrm{ice,vac}}
      \sum_{G}
      n^{N_{\mathrm{loop}}(G)}\,
      \tilde{x}_1^{|G|}\,
      y^{|\partial G|}\,,
    }
\end{equation}
with $n = 2$,
$\tilde{x}_1 = \tfrac{2}{3}\,w^{-(2S-1)}$~\eqref{eq:app-ice-x1},
and $y = v\sqrt{3/2}$~\eqref{eq:app-ice-y}.
At $v = 0$ ($y = 0$), only closed graphs ($|\partial G| = 0$)
contribute, recovering the monopole-free loop-gas
formula~\eqref{eq:app-ice-loopgas}.

\section{Large-$w$ Expansion of the $S = 3/2$ Spin-Ice Model}
\label{app:spin-ice-S32}

In this section we derive the large-$w$ expansion of the
spin-$S$ pyrochlore spin-ice model for $S = 3/2$.
The model definition and the Pauling vacuum
[Secs.~\ref{app:ice-def} and~\ref{app:ice-monopole-free},
Eqs.~\eqref{eq:app-ice-Z}--\eqref{eq:app-ice-Zvac-Pauling}]
carry over unchanged; here we focus on the aspects that are
qualitatively different from the $S \ge 2$ case treated in
Sec.~\ref{app:spin-ice-HTE}.

For $S = 3/2$ ($2S = 3$) the only defect levels are $d = 0$
(vacuum, $|S^z| = 3/2$) and $d = 1$ (fundamental, $|S^z| = 1/2$).
The reversed fundamental $d = 2S - 1 = 2$ corresponds to
$|S^z| = |S - 2| = 1/2$, the same spin magnitude as $d = 1$; its
fugacity ratio is $x_2/x_1 = w^{-(2S-3)} = 1$.
Three fundamental strings carry total $\mathbb{Z}_3$ charge
$3 \times 1 \equiv 0\pmod{3}$, and since each diamond-lattice
vertex provides $k = 2$ outward links, three strings
can annihilate at a \emph{single vertex} without the spatially
extended bridge process required for $S \ge 2$.
These features make the non-overlapping loop gas
qualitatively insufficient.

\subsection{Direction-fixed local state sums}
\label{app:ice-S32-zdir}

As in the $S \ge 2$ case (Sec.~\ref{app:ice-monopole-free}), we
work in the \emph{direction-fixed} Pauling approximation: for each
connected component of the defect graph we choose one of the two
flow values $f \in \{1,2\}$, which fixes all defect-link signs; only
the $2N - |G|$ vacuum-link signs remain as free variables.
The \emph{direction-fixed local state sum} $z_{\mathrm{dir}}(d_0)$
counts the number of vacuum-link sign patterns at a vertex of
degree~$d_0$ that satisfy the ice rule, given the fixed defect-link
signs.

\emph{$d_0 = 0$ (vacuum):}
All four links are vacuum; the ice-rule count is
$z_{\mathrm{dir}}(0) = \binom{4}{2} = 6$.

\emph{$d_0 = 1$ (endpoint):}
The constraint
$(S{-}1)\sigma_1 + S(\sigma_2{+}\sigma_3{+}\sigma_4) = 0$
requires $|\sigma_2{+}\sigma_3{+}\sigma_4| = (S{-}1)/S = 1/3$, but
$|\sigma_2{+}\sigma_3{+}\sigma_4| \ge 1$.
No solution exists: $z_{\mathrm{dir}}(1) = 0$ in the monopole-free
sector.

\emph{$d_0 = 2$ (loop vertex):}
With the direction fixed ($\Sigma_{\mathrm{def}} = \sigma_1 + \sigma_2 = 0$ forced by
coprimality, as in Sec.~\ref{app:ice-monopole-free}), the two
vacuum links must satisfy $\Sigma_{\mathrm{vac}} = \sigma_3 + \sigma_4 = 0$, giving
$z_{\mathrm{dir}}(2) = 2$ (one $+$, one $-$).

\emph{$d_0 = 3$ (cubic vertex):}
Three defect links with $|S^z| = 1/2$ and one vacuum link with
$|S^z| = 3/2$.
The ice rule $\sum_\ell S^z_\ell = 0$ gives
$\Sigma_{\mathrm{def}}/2 + 3\sigma_4/2 = 0$,
where $\Sigma_{\mathrm{def}} = \sigma_1 + \sigma_2 + \sigma_3 \in \{-3,-1,1,3\}$.
The only solutions have $|\Sigma_{\mathrm{def}}| = 3$ (all three defect signs the same),
fixing $\sigma_4 = -\Sigma_{\mathrm{def}}/3 = \mp 1$.
Each flow choice $f$ forces all three defect signs and hence also
$\sigma_4$; therefore
\begin{equation}\label{eq:app-ice-S32-z3dir}
    z_{\mathrm{dir}}(3) = 1.
\end{equation}
Compare the total (direction-summed) count $z_{\mathrm{total}}(3) = 2$
that appears in the odd-$d_0$ paragraph of
Sec.~\ref{app:ice-monopole-free}; the factor of~$2$ between these
two conventions is precisely the direction degeneracy, which is
accounted for by the overall factor $n^{N_{\mathrm{comp}}}$ in the
partition-function formula.

\emph{$d_0 = 4$ (crossing vertex):}
All four links carry defects ($|S^z| = 1/2$) and no vacuum links
remain.
The ice rule $\sum_\ell S^z_\ell = 0$ requires exactly two
positive and two negative defect signs.
At a crossing, two independent loops pass through the vertex;
once the flows on both loops are fixed, all four defect signs are
determined.
The unique sign assignment satisfies the ice rule, giving
\begin{equation}\label{eq:app-ice-S32-z4dir}
    z_{\mathrm{dir}}(4) = 1.
\end{equation}

The Pauling probability at each active vertex is therefore:
\begin{equation}\label{eq:app-ice-S32-Pauling}
    \frac{z_{\mathrm{dir}}(d_0)}{2^{4-d_0}}
    = \begin{cases}
        2/4 = 1/2 & (d_0 = 2), \\[4pt]
        1/2 = 1/2 & (d_0 = 3), \\[4pt]
        1/1 = 1   & (d_0 = 4).
    \end{cases}
\end{equation}
The Pauling probabilities at degree-2 and degree-3 vertices
\emph{coincide}, but the direction-fixed local state sums do not:
\begin{equation}\label{eq:app-ice-S32-zratios}
    \frac{z_{\mathrm{dir}}(2)}{z(0)} = \frac{2}{6} = \frac{1}{3},
    \qquad
    \frac{z_{\mathrm{dir}}(3)}{z(0)} = \frac{1}{6},
    \qquad
    \frac{z_{\mathrm{dir}}(4)}{z(0)} = \frac{1}{6}.
\end{equation}
The difference between $1/3$ and $1/6$ at degree-3 vertices will
produce a nontrivial vertex fugacity for cubic vertices;
the coincidence $z_{\mathrm{dir}}(4)/z(0) = z_{\mathrm{dir}}(3)/z(0)$
will produce a distinct \emph{crossing fugacity} at degree-4
vertices (see below).

\subsection{Direction factor}
\label{app:ice-S32-dir}

The argument of Sec.~\ref{app:ice-monopole-free} shows that
$\Sigma_{\mathrm{def}} = 0$ at every degree-2 vertex forces the $\mathbb{Z}_3$
flow~$f \in \{1,2\}$ to be constant along any path of degree-2
vertices.
At a degree-3 vertex, the constraint $|\Sigma_{\mathrm{def}}| = 3$ requires all
three defect-link signs to be the same, which means all three
branches meeting at the vertex carry the same flow~$f$.
Since reversing the global flow $f \to 3 - f$ maps one valid
configuration to the other, a connected component with any mixture
of degree-2 and degree-3 vertices admits exactly two valid
direction assignments.
The direction factor for a general graph~$G$ is therefore
\begin{equation}\label{eq:app-ice-S32-dir}
    \nu(G) = 2^{N_{\mathrm{comp}}(G)}.
\end{equation}

\subsection{Cubic vertex fugacity and the loop-gas formula}
\label{app:ice-S32-loopgas}

Consider a graph~$G$ with vertex degrees
$d_0(\bm{r}) \in \{0,2,3,4\}$ (no endpoints; the monopole-free case).
Let $V_2$, $V_3$, and $V_4$ denote the numbers of degree-2, degree-3,
and degree-4 (crossing) vertices, respectively.
The degree-sum identity
$2V_2 + 3V_3 + 4V_4 = 2|G|$
(each edge contributes to the degree of its two endpoints)
implies $V_2 = |G| - 3V_3/2 - 2V_4$.
(In particular, $V_3$ is always even for a closed graph, by the
handshaking lemma.)

Following the Pauling analysis of
Sec.~\ref{app:ice-monopole-free}, the direction-fixed weight of
$G$ relative to the vacuum is
\begin{align}\label{eq:app-ice-S32-weight}
    \frac{\Omega^{\mathrm{dir}}(G)}{\Omega(\emptyset)}
    &= 2^{|G|}
      \left(\frac{z_{\mathrm{dir}}(2)}{z(0)}\right)^{\!V_2}
      \left(\frac{z_{\mathrm{dir}}(3)}{z(0)}\right)^{\!V_3}
      \left(\frac{z_{\mathrm{dir}}(4)}{z(0)}\right)^{\!V_4}
      \notag\\
    &= 2^{|G|}
      \left(\frac{1}{3}\right)^{\!V_2}
      \left(\frac{1}{6}\right)^{\!V_3}
      \left(\frac{1}{6}\right)^{\!V_4}\!.
\end{align}
Using $V_2 = |G| - 3V_3/2 - 2V_4$:
\begin{align}\label{eq:app-ice-S32-weight2}
    \frac{\Omega^{\mathrm{dir}}(G)}{\Omega(\emptyset)}
    &= \left(\frac{2}{3}\right)^{\!|G|}
      \cdot \left(\frac{\sqrt{3}}{2}\right)^{\!V_3}
      \cdot \left(\frac{3}{2}\right)^{\!V_4}\!,
\end{align}
where the exponents follow from $|G| - V_2 - V_3 - V_4 = V_3/2 + V_4$
(obtained by substituting the degree-sum identity).
Including the direction factor~\eqref{eq:app-ice-S32-dir} and the
Boltzmann weight $w^{-(2S-1)|G|} = w^{-2|G|}$, the monopole-free
partition function for $S=3/2$ becomes
\begin{equation}\label{eq:app-ice-loopgas-S32}
    \boxed{
    Z_{\mathrm{ice}}
    = Z_{\mathrm{ice,vac}}
      \sum_{G}
      n^{N_{\mathrm{comp}}(G)}\,
      \tilde{x}_1^{|G|}\,
      \lambda_3^{V_3(G)}\,
      \lambda_4^{V_4(G)}
    }
\end{equation}
with $n = 2$, $\tilde{x}_1 = \tfrac{2}{3}\,w^{-2}$, the
\emph{cubic vertex fugacity}
\begin{equation}\label{eq:app-ice-S32-lambda}
    \boxed{
    \lambda_3 \coloneqq \frac{\sqrt{3}}{2} \approx 0.866,
    }
\end{equation}
and the \emph{crossing fugacity}
\begin{equation}\label{eq:app-ice-S32-lambda4}
    \boxed{
    \lambda_4 \coloneqq \frac{3}{2}.
    }
\end{equation}
The sum runs over all $\mathbb{Z}_3$-conserving graphs~$G$ on the
diamond lattice with vertex degrees $d_0(\bm{r}) \le 4$.
This is a strict generalization of the non-overlapping loop
gas~\eqref{eq:app-ice-loopgas}: restricting to $d_0 \le 2$
($V_3 = V_4 = 0$) recovers the standard formula.
The key distinction from the $S \ge 2$ case is that cubic vertices
enter at the \emph{same order} in $\tilde{x}_1$ as pass-through vertices,
but with an additional geometric penalty
$\lambda_3 = \sqrt{3}/2 < 1$ per cubic vertex.
For the simplest graphs with cubic vertices ($V_3 = 2$), this
penalty is $\lambda_3^2 = 3/4$.
Crossing vertices ($d_0 = 4$), by contrast, carry an
\emph{enhancement} $\lambda_4 = 3/2 > 1$: the ice rule at a
crossing is \emph{less} restrictive relative to the vacuum than
at a degree-2 vertex, because all four bonds are defect bonds and
no vacuum-bond sign choices remain.
In the dilute-gas limit $\tilde{x}_1 \to 0$, crossings require at least
two loops to overlap at a single vertex and therefore arise at
total graph order $|G| \ge |C_1| + |C_2|$; they are thus
geometrically subleading relative to simple loops and junctions.

\subsection{Inclusion of thermal monopoles}
\label{app:ice-S32-monopoles}

At finite monopole fugacity $v > 0$, the defect graph may contain
open strings whose endpoints carry monopole charge.
At an endpoint vertex ($d_0 = 1$), the direction-fixed calculation
of Sec.~\ref{app:ice-monopoles} carries over: three of the four
links are vacuum, and the minimum-charge configurations give
$z_v(1) = 3v + O(v^{(2S-1)^2}) = 3v + O(v^4)$.

For $S = 3/2$, the defect graph may now include
degree-1 (endpoint), degree-2 (pass-through), degree-3 (cubic),
and degree-4 (crossing) vertices simultaneously.
The degree-sum identity becomes
$V_1 + 2V_2 + 3V_3 + 4V_4 = 2|G|$,
where $V_1 = |\partial G|$ counts endpoints.
Generalizing Eq.~\eqref{eq:app-ice-S32-weight} to include
endpoint weights:
\begin{equation}\label{eq:app-ice-S32-Omega-v}
    \frac{\Omega^{\mathrm{dir}}_v(G)}{\Omega(\emptyset)}
    = 2^{|G|}
      \left(\frac{1}{3}\right)^{\!V_2}
      \left(\frac{1}{6}\right)^{\!V_3}
      \left(\frac{1}{6}\right)^{\!V_4}
      \left(\frac{v}{2}\right)^{\!V_1}\!,
\end{equation}
where $v/2 = z_v(1)/z(0)$ at leading order.
From the degree-sum identity:
\begin{align}\label{eq:app-ice-S32-Omega-v2}
    \frac{\Omega^{\mathrm{dir}}_v(G)}{\Omega(\emptyset)}
    &= \left(\frac{2}{3}\right)^{\!|G|}
      \left(\frac{v\sqrt{3}}{2}\right)^{\!|\partial G|}
      \left(\frac{\sqrt{3}}{2}\right)^{\!V_3}
      \left(\frac{3}{2}\right)^{\!V_4}\!.
\end{align}
Including the direction factor $2^{N_{\mathrm{comp}}}$ and the
Boltzmann weight $w^{-2|G|}$, and absorbing the direction factor
of open-string components into the endpoint fugacity
$\tilde{y}$ (see below), the full partition function for
$S = 3/2$ at finite $v$ is
\begin{equation}\label{eq:app-ice-S32-loopgas-v}
    \boxed{
    Z_{\mathrm{ice}}
    = Z_{\mathrm{ice,vac}}
      \sum_{G}
      n^{N_{\mathrm{loop}}(G)}\,
      \tilde{x}_1^{|G|}\,
      \lambda_3^{V_3(G)}\,
      \lambda_4^{V_4(G)}\,
      \tilde{y}^{|\partial G|}\,,
    }
\end{equation}
where $N_{\mathrm{loop}}(G)$ counts only the \emph{closed-loop}
components of~$G$.
When every open component of~$G$ has exactly two endpoints
($|\partial G_\alpha| = 2$), the direction factor of open-string
components can be absorbed pairwise into the endpoint fugacity:
$2 = (\sqrt{2})^2$ per pair, giving
\begin{equation}\label{eq:app-ice-S32-ytilde}
    n^{N_{\mathrm{comp}}}
      \left(\frac{v\sqrt{3}}{2}\right)^{\!|\partial G|}
    = n^{N_{\mathrm{loop}}}\,
      \tilde{y}^{\,|\partial G|},
\end{equation}
where the equality holds when $|\partial G_\alpha| = 2$
for all open components,
with
\begin{equation}\label{eq:app-ice-S32-y}
    \tilde{y} \coloneqq v\sqrt{\frac{3}{2}}
    = y,
\end{equation}
the same endpoint fugacity as in the $S \ge 2$
formula~\eqref{eq:app-ice-y}.

A subtlety arises for $S = 3/2$: the cubic vertex allows
connected components with $|\partial G_\alpha| \neq 2$.
By the handshaking lemma, $|\partial G_\alpha|$ and the number of
cubic vertices $V_{3,\alpha}$ in a connected component share the
same parity, so branched topologies can have
$|\partial G_\alpha| = 3, 4, 5, \dots$\footnote{The value $|\partial G_\alpha|=1$ is excluded by the $\mathbb{Z}_3$ flux conservation: each endpoint carries divergence $D=\pm 1$, so $\sum D_{\mathrm{end}} \equiv 0\pmod{3}$ requires at least two (for opposite charges) or three (for equal charges) endpoints.}.
For any component with $|\partial G_\alpha| \neq 2$, the direction
factor~$2$ per component is not equal to
$(\sqrt{2})^{|\partial G_\alpha|}$, and the decomposition
$n^{N_{\mathrm{loop}}} \tilde{y}^{|\partial G|}$ acquires a
correction factor of $2^{1 - |\partial G_\alpha|/2}$ per such
component.
However, all components with $|\partial G_\alpha| \neq 2$ require
at least one cubic vertex \emph{and} at least one monopole
endpoint in the same connected component, and are therefore
suppressed by both $\lambda_3$ and~$v$.
At leading order in $\tilde{x}_1$ and $v$,
Eq.~\eqref{eq:app-ice-S32-loopgas-v} is exact; subleading
corrections from branched components can be tracked systematically
if needed.

At $v = 0$ ($\tilde{y} = 0$), only closed graphs ($|\partial G| = 0$)
contribute and~\eqref{eq:app-ice-S32-loopgas-v} reduces to the
monopole-free formula~\eqref{eq:app-ice-loopgas-S32}.

\section{Exact Partition Function and High-Temperature Expansion of the $\mathbb{Z}_q$ Clock Model}
\label{app:clock-HTE}

In this section we present a self-contained derivation of the
high-temperature expansion (HTE) of the $q$-state clock model on the
diamond lattice.
The resulting graphical expansion---a sum over
$\mathbb{Z}_q$-conserving flow configurations weighted by normalized
Fourier coefficients---provides the natural framework for analyzing
string annihilation costs in the clock model.

\subsection{Definition of the $\mathbb{Z}_q$ clock model}
\label{app:clock-def}

Consider the diamond lattice with $N$ sites and $|E|=2N$ bonds
(coordination number $z=4$). Each site $\bm{r}$ carries a discrete
angular variable
\begin{equation}
    \sigma_{\bm{r}} \in \{0, 1, 2, \dots, q-1\},
\end{equation}
representing angles $\theta_{\bm{r}} = 2\pi\sigma_{\bm{r}}/q$ on the
unit circle. The partition function in the presence of a nearest-neighbor
coupling $K$ and a symmetry-breaking field $h$ is
\begin{align}\label{eq:app-clock-Z}
    Z_{\mathrm{clock}} &\coloneqq \sum_{\{\sigma_{\bm{r}}\}}
    \exp\!\Biggl[
      K\!\sum_{\langle \bm{r},\bm{r}' \rangle}
        \cos\frac{2\pi(\sigma_{\bm{r}} - \sigma_{\bm{r}'})}{q}
        \notag\\
    &\qquad\qquad\qquad
      + h\!\sum_{\bm{r}}
        \cos\frac{2\pi\sigma_{\bm{r}}}{q}
    \Biggr].
\end{align}
The zero-field case is recovered by setting $h = 0$.

Both the bond interaction and the site field share an identical
functional form. We therefore introduce a unified Boltzmann weight
parametrized by a single coupling $\kappa$:
\begin{equation}\label{eq:app-B-kappa}
    B^\kappa(\Delta) \coloneqq \exp\!\left(
      \kappa \cos\frac{2\pi \Delta}{q}
    \right),
    \qquad \Delta \in \mathbb{Z}_q,
\end{equation}
where all arithmetic in $\Delta$ is understood modulo $q$.
The bond weight corresponds to $\kappa = K$ (coupling constant,
$K = \beta J$, so $K \to 0$ is the high-temperature limit),
while the site weight corresponds to $\kappa = h$
(symmetry-breaking field).
Using this notation, the partition function \eqref{eq:app-clock-Z} becomes
\begin{equation}\label{eq:app-clock-Z-factored}
    Z_{\mathrm{clock}} = \sum_{\{\sigma_{\bm{r}}\}}
    \prod_{\bm{r}} B^h(\sigma_{\bm{r}})
    \prod_{\langle \bm{r},\bm{r}' \rangle}
    B^K(\sigma_{\bm{r}} - \sigma_{\bm{r}'}).
\end{equation}

The crucial analytic step is to expand $B^\kappa(\Delta)$ in the discrete
Fourier basis (characters) of $\mathbb{Z}_q$. Defining the primitive
$q$-th root of unity $\omega \coloneqq e^{2\pi i/q}$, the characters
$\{\omega^{m\Delta}\}_{m=0}^{q-1}$ form a complete orthonormal set
under the inner product
$(1/q)\sum_{\Delta=0}^{q-1}(\cdot)\,\overline{(\cdot)}$.
Accordingly, $B^\kappa(\Delta)$ admits the exact expansion
\begin{equation}\label{eq:app-Fourier}
    B^\kappa(\Delta) = \sum_{m=0}^{q-1} \hat{B}^\kappa_m \,\omega^{m\Delta},
\end{equation}
with Fourier coefficients
\begin{equation}\label{eq:app-Fourier-coeff}
    \hat{B}^\kappa_m = \frac{1}{q}\sum_{\Delta=0}^{q-1}
    B^\kappa(\Delta)\,\omega^{-m\Delta}.
\end{equation}
Since this expansion holds for any $\kappa$, it applies to both
the bond weight ($\kappa = K$) and the site weight ($\kappa = h$).

\subsection{Character expansion of the partition function}
\label{app:current-expansion-exact}
We now substitute the character expansion
\eqref{eq:app-Fourier} into the partition function
\eqref{eq:app-clock-Z-factored}. For the bond weights, we set
$\kappa = K$; for the site weights, $\kappa = h$.
Each bond $\ell$ acquires a \emph{bond current index}
$n_\ell \in \{0, 1, \dots, q-1\}$, and each site $\bm{r}$ a
\emph{site character index}
$m_{\bm{r}} \in \{0, 1, \dots, q-1\}$:
\begin{align}
    Z_{\mathrm{clock}}
    &= \sum_{\{\sigma_{\bm{r}}\}}
    \prod_{\bm{r}} \left[
      \sum_{m_{\bm{r}}=0}^{q-1}
      \hat{B}^h_{m_{\bm{r}}}\,\omega^{m_{\bm{r}} \sigma_{\bm{r}}}
    \right] \notag\\
    &\qquad\times
    \prod_\ell \left[
      \sum_{n_\ell=0}^{q-1} \hat{B}^K_{n_\ell}\,
      \omega^{n_\ell(\sigma_{\bm{r}_A} - \sigma_{\bm{r}_B})}
    \right].
\end{align}
Here we have assigned a fixed orientation $\bm{r}_A \to \bm{r}_B$ to
each bond (A-sublattice to B-sublattice on the bipartite diamond
lattice). Exchanging the order of summation over $\{\sigma_{\bm{r}}\}$
and the Fourier indices $\{n_\ell, m_{\bm{r}}\}$, we obtain
\begin{align}
    Z_{\mathrm{clock}}
    &= \sum_{\{n_\ell\}}
    \sum_{\{m_{\bm{r}}\}}
    \left(\prod_\ell \hat{B}^K_{n_\ell}\right)
    \left(\prod_{\bm{r}} \hat{B}^h_{m_{\bm{r}}}\right)
    \notag\\
    &\qquad\times
    \prod_{\bm{r}} \left[
      \sum_{\sigma_{\bm{r}}=0}^{q-1}
      \omega^{\sigma_{\bm{r}} (m_{\bm{r}} + D_{\bm{r}})}
    \right],
\end{align}
where the \emph{$\mathbb{Z}_q$ divergence} at each vertex $\bm{r}$ is
\begin{equation}\label{eq:app-Zq-div}
    D_{\bm{r}} \coloneqq
    \begin{cases}
        +\sum_{\ell \in \bm{r}} n_\ell \pmod{q}
        & \text{if } \bm{r} \in A, \\[4pt]
        -\sum_{\ell \in \bm{r}} n_\ell \pmod{q}
        & \text{if } \bm{r} \in B.
    \end{cases}
\end{equation}
The sign convention ensures that $D_{\bm{r}}$ measures the net
outgoing bond current from $\bm{r}$.

The sum over the site variable $\sigma_{\bm{r}}$ at each vertex yields
the orthogonality relation
\begin{equation}\label{eq:app-ortho}
    \sum_{\sigma_{\bm{r}}=0}^{q-1}
    \omega^{\sigma_{\bm{r}} (m_{\bm{r}} + D_{\bm{r}})}
    = q\,\delta_{m_{\bm{r}} + D_{\bm{r}} \equiv 0 \pmod{q}},
\end{equation}
which fixes $m_{\bm{r}} \equiv -D_{\bm{r}} \pmod{q}$ at every
vertex. In other words, the site character index $m_{\bm{r}}$ is
uniquely determined by the bond current divergence $D_{\bm{r}}$, and
the sum over $\{m_{\bm{r}}\}$ collapses.
Applying Eq.~\eqref{eq:app-ortho} at all $N$ sites, each site
contributes a factor of $q$ (from
$\sum_{\sigma=0}^{q-1}1 = q$), yielding an overall prefactor of $q^N$.
The constraint $m_{\bm{r}} \equiv -D_{\bm{r}}$ replaces the
site Fourier coefficient $\hat{B}^h_{m_{\bm{r}}}$ by
$\hat{B}^h_{-D_{\bm{r}}} = \hat{B}^h_{D_{\bm{r}}}$,
where the last equality follows from the conjugation symmetry
$\hat{B}^\kappa_m = \hat{B}^\kappa_{q-m}$
(all subscripts are understood modulo $q$).

Factoring out $(\hat{B}^K_0)^{|E|}$ from the bond weights and
$(\hat{B}^h_0)^{N}$ from the site weights, and defining the normalized
fugacities
\begin{equation}\label{eq:app-t-def-preview}
    t_{n} \coloneqq \frac{\hat{B}^K_{n}}{\hat{B}^K_0},
    \qquad
    u_{D} \coloneqq \frac{\hat{B}^h_{D}}{\hat{B}^h_0},
\end{equation}
with $t_0 = u_0 = 1$, we arrive at the \emph{exact} result:
\begin{equation}\label{eq:app-HTE}
    \boxed{
    Z_{\mathrm{clock}}
    = q^N\,(\hat{B}^h_0)^N\,(\hat{B}^K_0)^{|E|}
    \sum_{\{n_\ell\}}
    \prod_\ell t_{n_\ell}
    \prod_{\bm{r}} u_{D_{\bm{r}}}.
    }
\end{equation}
Equation~\eqref{eq:app-HTE} is \emph{exact at all temperatures};
no approximation has been made.
The character expansion is a complete discrete Fourier transform,
the exchange of summation order is algebraically exact, and the
orthogonality relation is an identity.
The sum $\sum_{\{n_\ell\}}$ runs over \emph{all} configurations
$n_\ell \in \{0,1,\dots,q-1\}$ without restriction;
the constraint on the flow pattern is encoded in the
vertex fugacities $u_{D_{\bm{r}}}$.

At $h = 0$ one has $\hat{B}^h_0 = 1$ and
$u_{D\ne 0} = 0$, so the vertex fugacities impose a hard constraint
and only $\mathbb{Z}_q$-conserving configurations survive,
i.e.\ those satisfying
\begin{equation}\label{eq:app-conservation}
    D_{\bm{r}} \equiv 0 \pmod{q} \qquad \forall\,\bm{r}.
\end{equation}
Eq.~\eqref{eq:app-HTE} then reduces to
\begin{equation}\label{eq:app-HTE-h0}
    Z_{\mathrm{clock}}\big|_{h=0} = q^N\,(\hat{B}^K_0)^{|E|}
    \sum_{\substack{\{n_\ell\} \\ \mathbb{Z}_q\text{-conserving}}}
    \prod_\ell t_{n_\ell}.
\end{equation}
At finite $h > 0$, open strings whose endpoints carry
$\mathbb{Z}_q$ charges ($D_{\bm{r}} \not\equiv 0$) are admitted, each
penalized by the vertex fugacity $u_{D_{\bm{r}}} < 1$.

\subsection{Closed-form evaluation of the Fourier coefficients}
\label{app:character}
Having established the exact HTE formula~\eqref{eq:app-HTE},
we now derive useful closed-form expressions for the Fourier
coefficients $\hat{B}^\kappa_m$.
The results in this subsection are \emph{exact for arbitrary $\kappa$};
the ``high-temperature'' assumption is not used.

The coefficients $\hat{B}^\kappa_m$ [Eq.~\eqref{eq:app-Fourier-coeff}] admit a closed-form evaluation in terms of modified
Bessel functions. The generating function for the modified Bessel
functions of the first kind reads
\begin{equation}\label{eq:app-Bessel-gen}
    e^{\kappa\cos\theta} = \sum_{n=-\infty}^{\infty} I_n(\kappa)\, e^{in\theta}.
\end{equation}
Substituting $\theta = 2\pi\Delta/q$ into Eq.~\eqref{eq:app-Bessel-gen}
and inserting into Eq.~\eqref{eq:app-Fourier-coeff}, we compute
\begin{align}
    \hat{B}^\kappa_m
    &= \frac{1}{q}\sum_{\Delta=0}^{q-1}
    \left[\sum_{n=-\infty}^{\infty} I_n(\kappa)\, e^{i 2\pi n\Delta/q}\right]
    e^{-i 2\pi m\Delta/q} \nonumber \\
    &= \sum_{n=-\infty}^{\infty} I_n(\kappa)\,
    \underbrace{\frac{1}{q}\sum_{\Delta=0}^{q-1}
    e^{i 2\pi (n-m)\Delta/q}}_{\displaystyle = \delta_{n \equiv m \pmod{q}}}.
    \label{eq:app-alias-step}
\end{align}
The orthogonality relation of the discrete Fourier transform yields a
Kronecker delta that selects $n = m + kq$ for all $k \in \mathbb{Z}$.
This gives the \emph{aliasing formula}:
\begin{equation}\label{eq:app-aliasing}
    \boxed{
    \hat{B}^\kappa_m = \sum_{k=-\infty}^{\infty} I_{m+kq}(\kappa).
    }
\end{equation}
We emphasize that this result is exact at all temperatures and holds
for both $\kappa = K$ (bond) and $\kappa = h$ (site).

\paragraph{Properties.}
For $\kappa > 0$, the coefficients $\hat{B}^\kappa_m$ satisfy two important
properties.

\noindent
(i)~\emph{Conjugation symmetry}: $\hat{B}^\kappa_m = \hat{B}^\kappa_{q-m}$ for
all $m$. This follows from $I_{-n}(\kappa) = I_n(\kappa)$ and the relabeling
$k \to -(k+1)$ in the aliasing sum.

\noindent
(ii)~\emph{Monotone decrease}:
$\hat{B}^\kappa_0 > \hat{B}^\kappa_1 \ge \hat{B}^\kappa_2 \ge \cdots
\ge \hat{B}^\kappa_{\lfloor q/2\rfloor} > 0$.
Each $\hat{B}^\kappa_m$ is a sum of strictly positive terms, with $\hat{B}^\kappa_0$
receiving the dominant $I_0(\kappa)$ contribution; increasing $m$ shifts the
dominant Bessel index further from the origin, reducing the sum.

For the bond weight ($\kappa = K$), the normalized bond fugacities
$t_m \coloneqq \hat{B}^K_m/\hat{B}^K_0$
[already introduced in Eq.~\eqref{eq:app-t-def-preview}]
satisfy $t_0 = 1$, $t_m = t_{q-m}$, and
$1 = t_0 > t_1 \ge t_2 \ge \cdots \ge t_{\lfloor q/2\rfloor} > 0$.
The $\lfloor q/2 \rfloor$ independent fugacities $(t_1, t_2, \dots,
t_{\lfloor q/2\rfloor})$ parametrize the model.
Likewise, the vertex fugacities
$u_D \coloneqq \hat{B}^h_{D}/\hat{B}^h_0$ inherit
the same monotone-decrease property with $K$ replaced by $h$.

\subsubsection{Explicit Fourier coefficients for $q=3$}
\label{app:explicit-coeffs}

For the case most relevant to the main text, we specialize to $q = 3$ and $\kappa = K$.
There is a single independent fugacity $t_1 = t_2$. The bond weights are
\begin{align}
    B^K(0) &= e^{K}, \quad
    B^K(1) = B^K(2) = e^{-K/2},
\end{align}
which gives the Fourier coefficients
\begin{align}
    \hat{B}^K_0 &= \tfrac{1}{3}(e^K + 2e^{-K/2}), \\
    \hat{B}^K_1 = \hat{B}^K_2 &= \tfrac{1}{3}(e^K - e^{-K/2}),
\end{align}
so
\begin{equation}\label{eq:app-t1-q3}
    t_1 = \frac{e^K - e^{-K/2}}{e^K + 2e^{-K/2}}
    = \frac{e^{3K/2} - 1}{e^{3K/2} + 2}.
\end{equation}
In the high-temperature limit $K \to 0$, the standard Bessel asymptotics $I_m(K) \sim (K/2)^m/m!$ applied to the aliasing formula~\eqref{eq:app-aliasing} yield $t_m \sim K^m/m!$, so higher-current fugacities are increasingly suppressed. For general $q \ge 4$, the explicit coefficients follow from the same aliasing formula; we refer to Ref.~\cite{Wu1982} for a comprehensive tabulation.

\subsection{Graph classification, loop gas, and annihilation cost}
\label{app:graph-class}

We now return to the zero-field HTE formula~\eqref{eq:app-HTE-h0} and
explain why it constitutes a ``high-temperature expansion,'' and
how the $\mathbb{Z}_q$-conserving graphs are classified.

\paragraph{Physical meaning of the high-temperature limit.}
Bonds with $n_\ell = 0$ contribute $t_0 = 1$ and are inert. The active
bonds ($n_\ell \ne 0$) form a $\mathbb{Z}_q$-conserving graph $G$,
whose fugacity is $\prod_{\ell \in G} t_{n_\ell}$.
In the high-temperature limit $K \to 0$,
all fugacities $t_m \to 0$
($t_m \sim K^m/m!$), so the dominant contribution to the
partition function comes from the vacuum ($G = \emptyset$), and
configurations with a small number of active bonds $|G|$ are
suppressed as $t_1^{|G|} \sim K^{|G|}$.
\emph{This perturbative series structure in $|G|$ is the essence of
the ``high-temperature expansion''}: one systematically enumerates
contributions from small graphs (short loops) to large graphs (long
loops) in increasing powers of $K$.

\paragraph{Graph classification.}
The $\mathbb{Z}_q$-conserving constraint \eqref{eq:app-conservation}
at each vertex with coordination $z=4$ restricts the admissible
current configurations. Any active vertex must have its four bond
currents $(n_1, n_2, n_3, n_4)$ summing to $0 \pmod{q}$, with the
appropriate sign convention from Eq.~\eqref{eq:app-Zq-div}.
For the simplest closed-loop configurations where each bond carries
the fundamental current $n_\ell = 1$ (fugacity $t_1$), the
conservation law \eqref{eq:app-conservation} requires exactly one
incoming and one outgoing unit current at each active vertex.
These are closed directed loops on the diamond lattice.

\paragraph{Annihilation channels.}
The key feature of the clock model is that higher current
values $n_\ell = 2, 3, \dots, \lfloor q/2\rfloor$ are
available on each bond, with fugacity $t_{n_\ell}$.
The annihilation of $q$ fundamental loops ($n = 1$) on the
diamond lattice is governed by the modular conservation law
$\sum_{\ell \in \bm{r}} n_\ell \equiv 0\pmod{q}$, which is \emph{less restrictive}
than the exact ice rule $\sum_{\ell \in \bm{r}} \phi_\ell = 4S$ of the spin-ice model
(Sec.~\ref{app:ice-hierarchical}).
This leads to qualitatively shorter cascades.

For $q \le z = 4$ (i.e.\ $S \le 2$), all $q$ fundamental strings
can annihilate at a single $z = 4$ vertex:
$\underbrace{1 + 1 + \cdots + 1}_{q} = q \equiv 0\pmod{q}$,
so no intermediate bonds are needed and the annihilation penalty is
trivially unity.
In particular, for $q = 4$ ($S = 2$) the modular conservation allows
single-vertex annihilation---in stark contrast to the spin-ice model,
where the exact ice rule forces a spatially extended cascade even for
$S = 2$ (Table~IV of the main text).

For $q > 4$ ($S > 2$), single-vertex annihilation is impossible
since $q > z$, and a spatially extended cascade is required.
However, because the modular conservation permits up to three
fundamental currents to fuse at the first cascade vertex (using three
of the four links), producing an intermediate bond with current
$n = 3$, and two more fundamentals merge at each subsequent vertex,
incrementing the intermediate current by~$2$,
the intermediate bonds carry odd currents
$n = 3, 5, 7, \ldots$ and the cascade terminates after only
$\lceil (q-4)/2 \rceil = \lceil S - 2 \rceil$ intermediate bonds.
This is roughly half the length of the spin-ice cascade
($2S - 3$ bonds; Sec.~\ref{app:ice-hierarchical}).
A concrete illustration of the $q = 6$ ($S = 3$) cascade
is given in Sec.~V\,C\,1 of the main text.

The crucial point is that each intermediate bond carries
current $n \ge 2$ with weight $t_n$.
For general~$q$, the leading Bessel asymptotics
$t_m \approx (K/2)^m/m!$ gives
\begin{equation}\label{eq:app-clock-ratio-gen}
    \frac{t_m}{t_1}
    \approx \frac{t_1^{m-1}}{m!}
    \quad (K \to 0),
\end{equation}
showing that the per-bond annihilation cost in the clock model is
a polynomial (power-law) suppression in $t_1$.

\section{Correspondence between the $S = 3/2$ Spin-Ice Model
         and the $\mathbb{Z}_3$ Clock Model}
\label{app:correspondence-S32}

The results of the preceding two appendices---the $\mathbb{Z}_q$
clock-model HTE (Sec.~\ref{app:clock-HTE}) and the $S = 3/2$
spin-ice expansion (Sec.~\ref{app:spin-ice-S32})---share
a common graphical structure.
In this section we make the correspondence precise, restricting
throughout to the \emph{monopole-free} sector
($v = 0$ in the spin-ice model, $h = 0$ in the clock model).

\subsection{Same graph ensemble}
\label{app:corr-S32-graphs}

Both partition functions are expressed as sums over
$\mathbb{Z}_3$-conserving graphs on the diamond lattice.
We recall the two formulas:

\paragraph{Clock model ($q = 3$, $h = 0$).}
From Eq.~\eqref{eq:app-HTE-h0} with $q = 3$, each
$\mathbb{Z}_3$-conserving configuration $\{n_\ell\}$
is a set of bonds carrying fundamental current $n_\ell = 1$ or~$2$
(with $n_\ell = 2 \equiv -1 \pmod{3}$ acting as the reversed
current).
At every active vertex, the conservation law
$D_{\bm{r}} \equiv 0 \pmod{3}$ [Eq.~\eqref{eq:app-conservation}]
admits degree-2 (one incoming, one outgoing fundamental),
degree-3 (all three incoming or all three outgoing), and
degree-4 (two incoming and two outgoing, i.e.\ a \emph{crossing})
vertices.
(Degree-4 with all same-direction currents is forbidden because
$4 \times 1 = 4 \not\equiv 0 \pmod{3}$, but the crossing
configuration $(n_1,n_2,n_3,n_4) = (1,1,2,2)$ satisfies
$D = 1 + 1 - 1 - 1 = 0$.)
Since the fugacities satisfy $t_2 = t_1$
[by $\hat{B}^K_m = \hat{B}^K_{q-m}$ for $q = 3$], every active
bond receives the same weight $t_1$ irrespective of the
current direction.
Including the direction factor $\nu(G) = 2^{N_{\mathrm{comp}}}$
(one overall $\mathbb{Z}_3$-flow per component), the zero-field
HTE reads
\begin{equation}\label{eq:app-clock-loopgas-q3}
    \frac{Z_{\mathrm{clock}}\big|_{h=0}}{Z_{\mathrm{clock,vac}}}
    = \sum_{G} 2^{N_{\mathrm{comp}}(G)}\, t_1^{|G|}\,,
\end{equation}
where the sum runs over all $\mathbb{Z}_3$-conserving graphs~$G$ on
the diamond lattice with vertex degrees $d_0 \le 4$.
Note that the clock model assigns no vertex-dependent weight:
junctions ($d_0 = 3$) and crossings ($d_0 = 4$) both enter
with unit fugacity per vertex, just as degree-2 vertices do.

\paragraph{Spin-ice model ($S = 3/2$, $v = 0$).}
Eq.~\eqref{eq:app-ice-loopgas-S32} gives
\begin{equation}\label{eq:app-ice-loopgas-S32-recall}
    \frac{Z_{\mathrm{ice}}}{Z_{\mathrm{ice,vac}}}
    = \sum_{G} 2^{N_{\mathrm{comp}}(G)}\, \tilde{x}_1^{|G|}\,
      \lambda_3^{V_3(G)}\,\lambda_4^{V_4(G)}\,,
\end{equation}
with $\tilde{x}_1 = (2/3)\,w^{-2}$, $\lambda_3 = \sqrt{3}/2$, and
$\lambda_4 = 3/2$.
The sum runs over the same set of $\mathbb{Z}_3$-conserving
graphs on the diamond lattice with vertex degrees $d_0 \le 4$.

The $\mathbb{Z}_3$-conserving constraint is common to both models
because the allowed vertex degrees are identical
($d_0 \in \{0,2,3,4\}$),
and in both cases a connected component admits exactly two global
flow orientations ($f = 1$ or $f = 2$), contributing a direction
factor $2^{N_{\mathrm{comp}}}$.
The graph ensembles therefore coincide term by term, up to the short-range junction--junction exclusion recorded in Sec.~V\,A of the main text: configurations in which the forced vacuum links of two like-oriented cubic junctions meet at a common pass-through vertex appear in the Potts ensemble but not in the spin-ice ensemble. This exclusion is $\mathbb{Z}_3$-symmetric and short ranged, and renormalizes the effective couplings without affecting the symmetry classification; the excluded configuration is made explicit in Sec.~\ref{app:corr-S32-jam} below.


\subsection{Global consistency of the vacuum assignment: the excluded junction configuration}
\label{app:corr-S32-jam}

The vertex-level compatibility criterion of Sec.~V\,A of the main
text guarantees that the non-defect links at each vertex can be
assigned vacuum values $\phi_\ell \in \{0, 2S\}$ consistent with the
ice rule \emph{at that vertex}. Because every vacuum link is shared
by two vertices, however, vertex-level assignability does not by
itself guarantee a globally consistent assignment. Here we record
the minimal configuration for which global consistency fails, which
underlies the junction--junction exclusion invoked above.

A cubic junction ($d_0 = 3$) fixes the value of its unique vacuum
link: if its three defect links carry $S^z = +\tfrac12$, the ice
rule forces the vacuum link to $S^z = -\tfrac32$ (i.e., $\phi = 0$);
if they carry $S^z = -\tfrac12$, it forces $S^z = +\tfrac32$
($\phi = 3$). A pass-through vertex ($d_0 = 2$), by contrast,
requires its two vacuum links to carry opposite values, one
$+\tfrac32$ and one $-\tfrac32$. The minimal excluded motif
[Fig.~\ref{fig:SM_jam}(a)] therefore consists of two
\emph{like-type} junctions whose forced vacuum links meet at a
common pass-through vertex: the central vertex would require
\begin{equation}
    \tfrac12 - \tfrac12 - \tfrac32 - \tfrac32 = -3 \neq 0
\end{equation}
[in flux language, $2 + 1 + 0 + 0 = 3 \neq 4S = 6$], so no vacuum
assignment satisfies the ice rule---even though the modular
$\mathbb{Z}_3$ condition holds at every vertex (central vertex:
$2 + 1 \equiv 0$; junctions: $1 + 1 + 1 \equiv 0 \bmod 3$). If the
two junctions are of opposite types [Fig.~\ref{fig:SM_jam}(b)],
they supply one $+\tfrac32$ and one $-\tfrac32$, and the assignment
is consistent.

We verified this exclusion exactly on a finite lattice. Embedding
the motif in a periodic diamond lattice of $6^3$ primitive cells (432 vertices, 864
links) and deciding the existence of a consistent vacuum assignment
as a maximum-flow feasibility problem yields a maximum flow of
$406$ against the required value of $407$ for the like-type pair,
proving that no assignment exists; reversing the type of one
junction restores $407 = 407$ (assignable), and the configuration
with pass-through strands alone is likewise assignable.

Three ingredients are all necessary for the obstruction: two
like-type junctions, forced vacuum links that share a vertex, and
the shared vertex being a pass-through. If the shared vertex
carries no defects ($d_0 = 0$), its remaining two vacuum links can
be chosen as $+\tfrac32$ each, absorbing the two forced
$-\tfrac32$ values; the constraint then merely propagates to
neighboring vertices and generically resolves. The exclusion
therefore requires two junctions in a specific local geometry: it
is a short-range, $\mathbb{Z}_3$-symmetric hard-core constraint
arising at second order in the junction density, as stated in
Sec.~V\,A of the main text.

\begin{figure*}[t]
\centering
\begin{tikzpicture}[scale=0.9]
\begin{scope}
\node[font=\small] at (0,2.1) {(a) like-type junctions: no consistent assignment};
\coordinate (C) at (0,0);
\draw[red!80!black, line width=1.2pt] (C) -- (0,1.3) node[right, font=\scriptsize, text=black]{$+\tfrac12$};
\draw[violet, line width=1.2pt] (C) -- (1.5,0.45) node[above, font=\scriptsize, text=black]{$-\tfrac12$};
\draw[black!50, line width=1pt] (C) -- (-1.5,-1.25) node[midway, left=2pt, font=\scriptsize, text=black]{$-\tfrac32$};
\draw[black!50, line width=1pt] (C) -- (1.5,-1.25) node[midway, right=2pt, font=\scriptsize, text=black]{$-\tfrac32$};
\fill[red!80!black] (C) circle (2.6pt);
\node[font=\scriptsize, text=red!80!black] at (0.05,-0.5) {ice rule violated};
\foreach \x in {-1.5, 1.5}{
  \draw[red!80!black, line width=1.1pt] (\x,-1.25) -- ({\x-0.85},-1.75);
  \draw[red!80!black, line width=1.1pt] (\x,-1.25) -- ({\x-0.1},-2.15);
  \draw[red!80!black, line width=1.1pt] (\x,-1.25) -- ({\x+0.8},-1.8);
  \fill[blue!70!black] (\x,-1.25) circle (2.2pt);
}
\node[font=\scriptsize] at (-1.6,-2.45) {$+\tfrac12 \times 3$};
\node[font=\scriptsize] at (1.4,-2.45) {$+\tfrac12 \times 3$};
\end{scope}
\begin{scope}[xshift=7.6cm]
\node[font=\small] at (0,2.1) {(b) opposite-type junctions: consistent};
\coordinate (C) at (0,0);
\draw[red!80!black, line width=1.2pt] (C) -- (0,1.3) node[right, font=\scriptsize, text=black]{$+\tfrac12$};
\draw[violet, line width=1.2pt] (C) -- (1.5,0.45) node[above, font=\scriptsize, text=black]{$-\tfrac12$};
\draw[black!50, line width=1pt] (C) -- (-1.5,-1.25) node[midway, left=2pt, font=\scriptsize, text=black]{$-\tfrac32$};
\draw[black!50, line width=1pt] (C) -- (1.5,-1.25) node[midway, right=2pt, font=\scriptsize, text=black]{$+\tfrac32$};
\fill[teal!70!black] (C) circle (2.6pt);
\node[font=\scriptsize, text=teal!70!black] at (0.05,-0.5) {ice rule satisfied};
\draw[red!80!black, line width=1.1pt] (-1.5,-1.25) -- (-2.35,-1.75);
\draw[red!80!black, line width=1.1pt] (-1.5,-1.25) -- (-1.6,-2.15);
\draw[red!80!black, line width=1.1pt] (-1.5,-1.25) -- (-0.7,-1.8);
\fill[blue!70!black] (-1.5,-1.25) circle (2.2pt);
\draw[violet, line width=1.1pt] (1.5,-1.25) -- (2.35,-1.75);
\draw[violet, line width=1.1pt] (1.5,-1.25) -- (1.6,-2.15);
\draw[violet, line width=1.1pt] (1.5,-1.25) -- (0.7,-1.8);
\fill[blue!70!black] (1.5,-1.25) circle (2.2pt);
\node[font=\scriptsize] at (-1.6,-2.45) {$+\tfrac12 \times 3$};
\node[font=\scriptsize] at (1.4,-2.45) {$-\tfrac12 \times 3$};
\end{scope}
\end{tikzpicture}
\caption{The minimal excluded configuration for $S = 3/2$ (numbers
denote $S^z$ values). The central vertex is a pass-through vertex
traversed by two fundamental strands ($+\tfrac12$, red, and
$-\tfrac12$, violet); the two lower vertices are cubic junctions,
and the gray links are the vacuum links whose values the junctions
force. (a)~When both junctions are of the same type (each absorbing
three $+\tfrac12$ defects and forcing its vacuum link to
$-\tfrac32$), the central vertex would require
$\tfrac12 - \tfrac12 - \tfrac32 - \tfrac32 = -3 \neq 0$: no vacuum
assignment satisfies the ice rule, although the modular
$\mathbb{Z}_3$ condition holds at every vertex. (b)~Opposite-type
junctions supply one $+\tfrac32$ and one $-\tfrac32$, and the
assignment is consistent.}
\label{fig:SM_jam}
\end{figure*}

\subsection{Identification of fugacities}
\label{app:corr-S32-fugacity}

Equating the fundamental-string fugacities in the two models
provides the natural identification
\begin{equation}\label{eq:app-corr-S32-t1x1}
    t_1 \;\longleftrightarrow\; \tilde{x}_1 = \tfrac{2}{3}\,w^{-2}\,,
\end{equation}
so that the large-$w$ limit of the spin-ice model corresponds to
the high-temperature limit ($K \to 0$, $t_1 \to 0$) of the clock
model.
Under this identification, both expansions are power series in the
same small parameter.

\subsection{Key difference: vertex fugacities}
\label{app:corr-S32-lambda}

Comparing Eqs.~\eqref{eq:app-clock-loopgas-q3}
and~\eqref{eq:app-ice-loopgas-S32-recall}, the spin-ice sum
carries vertex-dependent factors
$\lambda_3^{V_3(G)}\,\lambda_4^{V_4(G)}$ that are absent in
the clock model.
The difference is summarized as follows:
\begin{align}\label{eq:app-corr-S32-comparison}
    \left.\frac{Z}{Z_{\mathrm{vac}}}\right|_{\text{clock}}
    &= \sum_{G} 2^{N_{\mathrm{comp}}}\, t_1^{|G|},\notag\\
    \left.\frac{Z}{Z_{\mathrm{vac}}}\right|_{\text{ice}}
    &= \sum_{G} 2^{N_{\mathrm{comp}}}\, \tilde{x}_1^{|G|}\,
      \lambda_3^{V_3}\,\lambda_4^{V_4}.
\end{align}
The vertex fugacities act in \emph{opposite directions}:
$\lambda_3 = \sqrt{3}/2 < 1$ \emph{penalizes} cubic vertices
(junctions), while $\lambda_4 = 3/2 > 1$ \emph{enhances}
crossing vertices relative to the clock model.
For graphs with $V_3 = V_4 = 0$ (non-overlapping simple loops),
the two models are
\emph{term-by-term identical} under the substitution $t_1 = \tilde{x}_1$.

\paragraph{Physical origin of $\lambda_3$ and~$\lambda_4$.}
The physical origin of these vertex fugacities---the
degree-dependent ice-rule state counts and their ratios
relative to the degree-2 baseline---is explained in
Sec.~V\,B of the main text.
These local factors aggregate, via the degree-sum identity
$V_2 = |G| - 3V_3/2 - 2V_4$,
into the vertex fugacities
$\lambda_3^{V_3} = (\sqrt{3}/2)^{V_3}$ and
$\lambda_4^{V_4} = (3/2)^{V_4}$
as derived in Sec.~\ref{app:ice-S32-loopgas}.

In summary, the ice rule $\sum_\ell S^z_\ell = 0$
is a strictly stronger constraint than
$\mathbb{Z}_3$ conservation.
The constraints agree at degree-2 vertices, but differ at
degree-3 vertices (penalty $\lambda_3 = \sqrt{3}/2 < 1$)
and at degree-4 crossings (enhancement $\lambda_4 = 3/2 > 1$).
The $\mathbb{Z}_3$ clock model (at $h = 0$) may thus be viewed as the
\emph{$\lambda_3 \to 1$, $\lambda_4 \to 1$ limit} of the
$S = 3/2$ spin-ice loop gas.

\subsection{Inclusion of monopoles at leading order}
\label{app:corr-S32-monopoles}

We now extend the correspondence to include open strings
(monopoles in the spin-ice language, $\mathbb{Z}_3$-charged
vertices in the clock model), working at leading order in the
respective endpoint fugacities.

\paragraph{Clock model ($q = 3$, $h > 0$).}
At finite field strength $h > 0$, the exact HTE
formula~\eqref{eq:app-HTE} admits open strings whose endpoints
carry $\mathbb{Z}_3$ charge $D_{\bm{r}} \not\equiv 0 \pmod{3}$.
At a degree-1 vertex (one active bond), the net current is
$D_{\bm{r}} = \pm 1$, and the vertex fugacity is
$u_1 = u_2 = \hat{B}^h_1/\hat{B}^h_0$ (by the conjugation
symmetry $\hat{B}^\kappa_m = \hat{B}^\kappa_{q-m}$).
Thus, each endpoint contributes a factor~$u_1$.
Including the direction factor $2^{N_{\mathrm{comp}}}$ and the
bond weights $t_1^{|G|}$, and absorbing the direction factor of
open-string components pairwise into the endpoint fugacity
$\tilde{u}_1 \coloneqq \sqrt{2}\,u_1$ (exactly as in the
$S \ge 2$ case, Sec.~\ref{app:corr-Sge2-monopoles}), the
partition function becomes
\begin{equation}\label{eq:app-clock-loopgas-q3-h}
    \frac{Z_{\mathrm{clock}}}{Z_{\mathrm{clock,vac}}}
    = \sum_{G} n^{N_{\mathrm{loop}}(G)}\, t_1^{|G|}\,
      \tilde{u}_1^{|\partial G|}\,,
\end{equation}
where $|\partial G|$ denotes the number of degree-1 (endpoint)
vertices of~$G$ and $N_{\mathrm{loop}}$ counts only closed-loop
components.
Notice that the clock-model formula contains no
vertex-degree-dependent weight beyond $\tilde{u}_1$ at the
endpoints: cubic vertices ($d_0 = 3$) still enter with unit
fugacity.

\paragraph{Spin-ice model ($S = 3/2$, $v > 0$).}
At finite monopole fugacity, the spin-ice
formula~\eqref{eq:app-ice-S32-loopgas-v} reads
\begin{equation}\label{eq:app-ice-loopgas-S32-v-recall}
    \frac{Z_{\mathrm{ice}}}{Z_{\mathrm{ice,vac}}}
    = \sum_{G} n^{N_{\mathrm{loop}}(G)}\,
      \tilde{x}_1^{|G|}\,\lambda_3^{V_3(G)}\,
      \lambda_4^{V_4(G)}\,
      \tilde{y}^{|\partial G|}\,,
\end{equation}
with $\tilde{y} = v\sqrt{3/2} = y$~\eqref{eq:app-ice-S32-y},
the same endpoint fugacity as in the $S \ge 2$
formula~\eqref{eq:app-ice-y}.
Here $N_{\mathrm{loop}}$ counts only closed-loop components;
the direction factor of open-string components is absorbed into
$\tilde{y}$ (Sec.~\ref{app:ice-S32-monopoles}).

\paragraph{Comparison.}
Both formulas~\eqref{eq:app-clock-loopgas-q3-h}
and~\eqref{eq:app-ice-loopgas-S32-v-recall} share the same
graph ensemble and the same absorbed direction factor
$n^{N_{\mathrm{loop}}}$, but differ in the vertex fugacities:
\begin{equation}\label{eq:app-corr-S32-comparison-v}
    \underbrace{t_1^{|G|}\,
      \tilde{u}_1^{|\partial G|}}_{\text{clock}}
    \quad\longleftrightarrow\quad
    \underbrace{\tilde{x}_1^{|G|}\,\lambda_3^{V_3}\,
      \lambda_4^{V_4}\,
      \tilde{y}^{|\partial G|}}_{\text{spin ice}}.
\end{equation}
The fugacity identification $t_1 \leftrightarrow \tilde{x}_1$
[Eq.~\eqref{eq:app-corr-S32-t1x1}] carries over unchanged.
The endpoint fugacities are identified as
\begin{equation}\label{eq:app-corr-S32-u1y}
    \tilde{u}_1 = \sqrt{2}\,u_1
    \;\longleftrightarrow\;
    \tilde{y} = v\sqrt{3/2}\,.
\end{equation}
The vertex fugacities $\lambda_3^{V_3}\,\lambda_4^{V_4}$ again
appear only in the spin-ice formula and are absent in the clock
model, exactly as in the monopole-free case.

Physically, the endpoint fugacities have different
origins: $\tilde{u}_1$ is controlled by the symmetry-breaking
field~$h$ and vanishes as $h \to 0$, while $\tilde{y}$ is
controlled by the monopole fugacity $v = e^{-J/(2T)}$ and
vanishes as $T \to 0$.
Despite this, their structural roles in the graphical
expansion are identical---each penalizes the creation of
open-string endpoints.
The correspondence therefore extends cleanly to the monopole
sector at leading order: the only additional differences
between the two models are the vertex
fugacities~$\lambda_3$ and~$\lambda_4$.

A subtlety specific to $S = 3/2$ is that the spin-ice model
admits connected components containing \emph{both} cubic
vertices and endpoints simultaneously
(Sec.~\ref{app:ice-S32-monopoles}).
Such components can have $|\partial G_\alpha| \neq 2$, for
which the pairwise absorption of the direction factor into
$\tilde{y}^{|\partial G|}$ acquires a correction factor of
$2^{1 - |\partial G_\alpha|/2}$ per component.
All such components require at least one cubic vertex
\emph{and} at least one monopole endpoint; they are therefore
doubly suppressed and do not affect the leading-order
correspondence~\eqref{eq:app-corr-S32-comparison-v}.

\section{Correspondence between the $S \ge 2$ Spin-Ice Model
         and the $\mathbb{Z}_{2S}$ Clock Model}
\label{app:correspondence-Sge2}

We now establish the analogous correspondence for the general case
$S \ge 2$, where the spin-ice model maps to a $\mathbb{Z}_q$ clock
model with $q = 2S$.
As in the $S = 3/2$ case (Sec.~\ref{app:correspondence-S32}),
we begin with the monopole-free sector.

\subsection{Non-overlapping loop gas: exact agreement at leading order}
\label{app:corr-Sge2-loopgas}

\paragraph{Clock model ($q = 2S$, $h = 0$).}
The zero-field HTE~\eqref{eq:app-HTE-h0} sums over all
$\mathbb{Z}_{2S}$-conserving current configurations.
When every active bond carries the fundamental current
$n_\ell = 1$ (or the reversed $n_\ell = q - 1 = 2S - 1$),
the conservation law~\eqref{eq:app-conservation} at each
active vertex requires the number of incoming and outgoing
unit currents to balance modulo~$q$.
For $q = 2S \ge 4$, this admits two types of active
vertices using only fundamental currents:
degree-2 (one incoming, one outgoing; $1 - 1 = 0$) and
degree-4 (two incoming, two outgoing;
$1 + 1 - 1 - 1 = 0$).
Degree-3 vertices, by contrast, require at least one
higher-current bond $n_\ell = 2$ (since $1 + 1 + 1 = 3
\not\equiv 0 \pmod{q}$ for $q \ge 4$), with fugacity $t_2$
satisfying $t_2/t_1 \to 0$ as
$K \to 0$~[Eq.~\eqref{eq:app-clock-ratio-gen}]; such
configurations are \emph{energetically subleading}.
Degree-4 vertices (loop crossings) carry weight
$t_1^4$ and are not energetically suppressed, but in the
dilute-gas limit ($t_1 \to 0$) they arise at higher total
graph order $|G| \ge |C_1| + |C_2|$ and are therefore
excluded in the leading-order expansion.
Since $t_1 = t_{q-1}$ (conjugation symmetry), each
non-overlapping loop of length~$|C|$ has weight $t_1^{|C|}$ and
admits two orientations.
The leading-order HTE is therefore
\begin{equation}\label{eq:app-clock-loopgas-q2S}
    \frac{Z_{\mathrm{clock}}\big|_{h=0}}{Z_{\mathrm{clock,vac}}}
    = \sum_{\{C\}} 2^{N_{\mathrm{loop}}(C)}\, t_1^{|C|}
    + O(t_1\,t_2)\,,
\end{equation}
where the sum runs over non-overlapping closed loop
configurations on the diamond lattice.

\paragraph{Spin-ice model ($S \ge 2$, $v = 0$).}
Eq.~\eqref{eq:app-ice-loopgas} gives
\begin{equation}\label{eq:app-ice-loopgas-Sge2-recall}
    \frac{Z_{\mathrm{ice}}}
         {Z_{\mathrm{ice,vac}}}
    = \sum_{\{C\}} 2^{N_{\mathrm{loop}}(C)}\, \tilde{x}_1^{|C|}
    + O(\tilde{x}_1\,w^{-(2S-3)})\,,
\end{equation}
with $\tilde{x}_1 = (2/3)\,w^{-(2S-1)}$~\eqref{eq:app-ice-x1}.
The sum runs over the same set of non-overlapping loop
configurations.

\paragraph{Comparison.}
Under the fugacity identification
\begin{equation}\label{eq:app-corr-Sge2-t1x1}
    t_1 \;\longleftrightarrow\; \tilde{x}_1 = \tfrac{2}{3}\,w^{-(2S-1)},
\end{equation}
the two formulas~\eqref{eq:app-clock-loopgas-q2S}
and~\eqref{eq:app-ice-loopgas-Sge2-recall} are
\emph{term-by-term identical} at leading order.
The direction factor $n = 2$ is the same in both models
(since $q = 2S \ge 4$ implies $1 \neq q - 1$, giving two
orientations per loop),
and the underlying lattice (diamond, $z = 4$) is common.
There is no analogue of the cubic vertex fugacity~$\lambda_3$
that distinguished the two models in the $S = 3/2$ case:
since only $d_0 \le 2$ vertices appear at leading order,
the Pauling weight $z_{\mathrm{dir}}(2)/z(0) = 1/3$ plays
the same role as the clock-model fugacity $t_1$, and both
simply contribute $\tilde{x}_1^{|C|}$ or $t_1^{|C|}$ per loop.

The correspondence between the $S \ge 2$ spin-ice model and
the $\mathbb{Z}_{2S}$ clock model at leading order is thus
\emph{exact} and \emph{trivial}---the two models generate the
same non-overlapping loop gas on the same lattice with the
same weights.

\subsection{Subleading corrections: exponential vs.\ polynomial}
\label{app:corr-Sge2-subleading}

The leading-order correspondence is exact, but
the $\mathbb{Z}_{2S}$ clock model is \emph{not} the correct
dual model beyond leading order.
Two independent obstructions arise for $q = 2S \ge 4$, both
absent for $q \le 3$~\cite{Watanabe2026}:
\begin{enumerate}
\item \emph{Ice-rule incompatibility.}
  The $\mathbb{Z}_q$ Gauss law
  $\sum_{\ell \in \bm{r}} \phi_\ell \equiv 0 \pmod{q}$ does not guarantee the ice
  rule $Q = 0$ for $q \ge 4$.
  For example, at $q = 4$, the degree-4 configuration
  $(\phi_1, \phi_2, \phi_3, \phi_4) = (1,1,1,1)$ satisfies
  $4 \equiv 0 \pmod{4}$ but carries monopole charge $Q = -4$;
  the clock model counts it with the same weight as the
  ice-rule-satisfying crossing $(1,3,1,3)$, whereas the
  spin-ice model suppresses it by $v^{Q^2}$.
\item \emph{Bond-fugacity mismatch.}
  For $q \ge 4$ the clock model has a single coupling~$K$ that
  constrains the $\lfloor q/2 \rfloor$ independent bond
  fugacities (e.g.\ $t_2 = t_1^2$ for $q = 4$).
  The spin-ice model, by contrast, has an independent energy
  scale $\mu$ controlling each defect level, so the fugacity
  ratios $x_d/x_1 = w^{-(d-1)(2S-d-1)}$ cannot in general be
  reproduced by any single~$K$.
  Physically, the clock model's single coupling~$K$ rigidly
  locks all higher-current fugacities to powers of~$t_1$
  (via the Bessel-function hierarchy $t_d \sim t_1^d/d!$),
  whereas the spin-ice model's independent stiffness~$\mu$
  allows higher-defect bonds to be exponentially more costly
  ($x_d/x_1 \sim e^{-(d-1)(2S-d-1)\mu/T}$) than any power of
  $x_1$ alone.
  This mismatch means that the clock model systematically
  \emph{overestimates} the weight of subleading (overlapping)
  graphs, explaining why the $XY$ model---not the clock
  model---is the correct effective theory at subleading order.
\end{enumerate}
These obstructions do not affect the leading non-overlapping loop
gas (which involves only degree-2 vertices with a single
fugacity~$\tilde{x}_1$), but they preclude an exact matching at
subleading order.
The physically correct dual model for $S \ge 2$ is the $XY$ model
(continuous $U(1)$ symmetry), whose HTE employs exact conservation
$\sum_{\ell \in \bm{r}} n_\ell = 0$ and thereby avoids the ice-rule incompatibility
automatically~\cite{Watanabe2026}.

Beyond these obstructions, the two models also differ
in the geometry and cost of the annihilation cascade for
higher-current (overlapping) bonds.

\paragraph{Cascade geometry.}
In the spin-ice model, the exact ice rule $\sum_{\ell \in \bm{r}} \phi_\ell = 4S$ forbids any vertex from absorbing more than two fundamental strings, so the $2S$ fundamentals fuse pairwise into composite strands, forming a fusion tree with at least $2S - 3$ intermediate bonds of charges $2 \le \phi \le 2S-2$; the linear chain, absorbing one fundamental at a time and carrying $\phi = 2, 3, \dots, 2S - 2$, is the simplest channel and the one evaluated in Sec.~\ref{app:ice-hierarchical}. [Other fusion trees carry slightly different exponents, but every channel is bounded by $w^{-4(S-1)(2S-3)}$: starting from $2S$ fundamental strands and ending at a terminal vertex that absorbs $k \in \{2, 3\}$ strands, the number of pairwise fusions is $2S - k$, each producing one intermediate bond, and an output can carry the un-suppressed charge $2S-1$ at most once [only in the $k = 2$ terminal $(2S-1, 1)$]; hence every channel contains at least $2S - k - \delta_{k,2} = 2S - 3$ bonds with $2 \le \phi \le 2S-2$, each costing at least $w^{-2(2S-2)}$.]
In the clock model, the weaker modular conservation
$\sum_{\ell \in \bm{r}} n_\ell \equiv 0\pmod{q}$ permits a qualitatively shorter cascade.
For $q \le z = 4$ ($S \le 2$), all $q$ fundamentals annihilate
at a single vertex with no intermediate bonds
(Sec.~\ref{app:graph-class}).
For $q > 4$ ($S > 2$), the modular conservation allows up to three
fundamentals to fuse at the first vertex and two more at each
subsequent vertex, so the cascade requires only
$\lceil S - 2 \rceil$ intermediate bonds carrying odd currents
$n = 3, 5, 7, \ldots$---roughly half the spin-ice cascade length
(Table~IV of the main text).

\paragraph{Per-bond costs.}
The two models also differ in the cost of each intermediate bond.
In the clock model, the bond carrying current $n$ has fugacity
$t_n$, with
$t_n / t_1 \sim t_1^{n-1}/n!$~[Eq.~\eqref{eq:app-clock-ratio-gen}],
a \emph{polynomial} function of the expansion parameter $t_1
\propto K$.
In the spin-ice model, the bond carrying effective charge~$\phi$ has
fugacity $x_\phi = w^{-\phi(2S-\phi)}$, giving the excess cost
\begin{equation}\label{eq:app-ice-ratio}
    \frac{x_\phi}{x_1} = w^{-(\phi-1)(2S-\phi-1)},
\end{equation}
which is an \emph{exponential} function of the Boltzmann
parameter $\ln w = \mu/T$.
The qualitative distinction ``polynomial vs.\ exponential''
refers to the dependence on the respective natural control
parameters ($K$ for the clock model, $\mu/T$ for the spin-ice
model).

\paragraph{Total cascade penalties.}
In the spin-ice model, the $2S - 3$ intermediate bonds
$\phi = 2, \dots, 2S-2$ yield the total bridge penalty
\begin{equation}\label{eq:app-total-bridge}
    \mathcal{P}_{\mathrm{bridge}}^{\mathrm{ice}}
    = \prod_{\phi=2}^{2S-2} x_\phi
    = w^{-(2S-3)(2S-1)(2S+4)/6},
\end{equation}
whose exponent grows as $\sim S^3$.
In the clock model, the cascade involves only
$\lceil S - 2 \rceil$ intermediate bonds with odd currents
$n = 3, 5, \ldots, 2\lceil S{-}2\rceil + 1$, giving
\begin{equation}\label{eq:app-total-bridge-clock}
    \mathcal{P}_{\mathrm{bridge}}^{\mathrm{clock}}
    = \prod_{k=1}^{\lceil S-2 \rceil} t_{2k+1}\,.
\end{equation}
Using $t_m \sim t_1^m/m!$~[Eq.~\eqref{eq:app-clock-ratio-gen}],
the leading asymptotic behavior is
$\mathcal{P}_{\mathrm{bridge}}^{\mathrm{clock}}
\sim t_1^{\,\sum_{k=1}^{\lceil S-2\rceil}(2k+1)}$,
where the exponent in $t_1$ grows as $\sim S^2$ for large~$S$.
Thus the clock model benefits from both a shorter cascade
and individually cheaper (polynomial) intermediate bonds.

The annihilation cost structures are summarized in
Table~IV of the main text.

Despite the milder cascade cost, the spin-ice model's subleading
corrections carry an important structural distinction.
In the clock model, all corrections are controlled by a single
expansion parameter $t_1 \propto K$: the cost of any
intermediate bond is a power of~$t_1$ itself, so subleading
graphs are merely higher-order terms in the same Taylor series.
In the spin-ice model, the Boltzmann weight $w^{-(d-1)(2S-d-1)}$
introduces a \emph{separate} energy scale---the spin stiffness
$\mu$---that suppresses intermediate bonds independently of the
loop fugacity~$\tilde{x}_1$.
As a consequence, the non-overlapping loop
gas~\eqref{eq:app-ice-loopgas-Sge2-recall} becomes an
exponentially accurate approximation at large $w$, controlled by
the Boltzmann gap $\Delta = (2S - 3)\mu$ rather than by the loop
fugacity alone.
In the clock model, the analogous corrections vanish only as
powers of $t_1 = \tilde{x}_1$ and therefore enter at every polynomial
order in the HTE.
This structural difference is the physical origin of the
stronger suppression of the bare coupling of the dangerously irrelevant operator in the
spin-ice model compared to the standard clock model.

\subsection{Inclusion of monopoles at leading order}
\label{app:corr-Sge2-monopoles}

At finite monopole fugacity, the two models again admit open
strings (monopole endpoints).

\paragraph{Clock model ($q = 2S$, $h > 0$).}
The exact HTE~\eqref{eq:app-HTE} with finite $h$ admits
open strings whose endpoints carry $\mathbb{Z}_{2S}$ charge
$D_{\bm{r}} \not\equiv 0 \pmod{2S}$.
At leading order, each bond carries the fundamental current
$n_\ell = 1$ (or $q-1$), and endpoint vertices (degree~$1$) have
net charge $D_{\bm{r}} = \pm 1$ with vertex fugacity~$u_1$.
Each open-string component admits two orientations, contributing
a direction factor of~$2$.
This factor is absorbed pairwise into the endpoint fugacity:
$2 = (\sqrt{2})^2$ per pair of endpoints, defining an effective
endpoint fugacity $\tilde{u}_1 \coloneqq \sqrt{2}\,u_1$.
The partition function at leading order is then
\begin{equation}\label{eq:app-clock-loopgas-q2S-h}
    \frac{Z_{\mathrm{clock}}}{Z_{\mathrm{clock,vac}}}
    = \sum_{G} 2^{N_{\mathrm{loop}}(G)}\, t_1^{|G|}\,
      \tilde{u}_1^{|\partial G|}
    + O(t_1 t_2),
\end{equation}
where the sum runs over non-overlapping graphs~$G$ with
vertex degrees $d_0 \le 2$ (closed loops and open strings).

\paragraph{Spin-ice model ($S \ge 2$, $v > 0$).}
The generalized loop-gas formula~\eqref{eq:app-ice-loopgas-v}
gives
\begin{equation}\label{eq:app-ice-loopgas-Sge2-v-recall}
    \frac{Z_{\mathrm{ice}}}{Z_{\mathrm{ice,vac}}}
    = \sum_{G} n^{N_{\mathrm{loop}}(G)}\,
      \tilde{x}_1^{|G|}\, y^{|\partial G|}
    + O(\tilde{x}_1\,w^{-(2S-3)}),
\end{equation}
with $y = v\sqrt{3/2}$~\eqref{eq:app-ice-y}.
The sum runs over non-overlapping graphs with vertex degrees
$d_0 \le 2$.

\paragraph{Comparison.}
At leading order, the two formulas are again identical under
the identifications
\begin{equation}\label{eq:app-corr-Sge2-idents}
    t_1 \;\longleftrightarrow\; \tilde{x}_1
    = \tfrac{2}{3}\,w^{-(2S-1)},
    \qquad
    \tilde{u}_1 \;\longleftrightarrow\; y = v\sqrt{3/2}\,.
\end{equation}
In both models the direction factor for open-string components
is absorbed into the endpoint fugacity in the same way
($2 = (\sqrt{2})^{|\partial G_\alpha|}$ for each two-endpoint
component), so the structural agreement is exact.
Since the graph ensemble at leading order consists of
non-overlapping loops and open strings (all vertex degrees
$d_0 \le 2$), no cubic vertex fugacity appears, and the
correspondence is exact.
The only difference, as in the monopole-free case, arises
at subleading order through the annihilation mechanism
and through the obstructions identified in
Sec.~\ref{app:corr-Sge2-subleading}.

Comparing with the $S = 3/2$ correspondence
(Sec.~\ref{app:correspondence-S32}), the $S \ge 2$ case
is structurally simpler: the leading-order correspondence is
exact with no additional vertex fugacity, because the
non-overlapping loop gas with $d_0 \le 2$ is sufficient for
both models.
The vertex fugacities $\lambda_3$ and~$\lambda_4$ that
distinguish the $S = 3/2$ spin-ice model from the
$\mathbb{Z}_3$ clock model have no counterpart for $S \ge 2$,
where the analogous higher-degree vertices are exponentially
suppressed rather than entering at the same order as simple loops.

\section{Decomposition framework for the
spin-ice universality class}
\label{app:sandwich}

In this section, working throughout in the monopole-free limit $v = 0$, we construct an exact decomposition of the
vacuum-normalized partition function of the spin-$S$ pyrochlore spin-ice model
into a continuous $U(1)$ sector (captured by a ``decorated $XY$ model'')
and an exponentially suppressed discrete $\mathbb{Z}_{2S}$ correction
(the fusion-cascade contribution).
This decomposition provides a \emph{quantitative} bound
on the bare coupling of the discrete $\mathbb{Z}_{2S}$
anisotropy, establishing that it is exponentially suppressed.
Combined with the renormalization-group (RG) irrelevance of this
perturbation at the 3D~$XY$ fixed point, we argue that the
$S \ge 2$ spin-ice model belongs to the 3D~$XY$ universality
class.

\subsection{Vacuum-normalized partition functions}
\label{app:sandwich-def}

We define the vacuum-normalized partition function of the
spin-ice model in terms of its high-temperature expansion (HTE) on the
diamond lattice.
The ``vacuum'' reference state is the
configuration with no active bonds, and we normalize so that
$\tilde{Z} = 1$ for that state.

\paragraph{Spin-ice model ($S \ge 2$).}
From Eq.~\eqref{eq:app-ice-loopgas-Sge2-recall},
\begin{equation}\label{eq:app-ice-loopgas-recall}
    \tilde{Z}_{\mathrm{ice}}
    = \sum_{\{C\}} 2^{N_{\mathrm{loop}}(C)}\,\tilde{x}_1^{|C|}
    + O(\tilde{x}_1\,w^{-(2S-3)})\,.
\end{equation}
In the full (non-leading-order) expansion, the effective charge
$\phi_\ell \in \{0,1,\dots,2S\}$ on each bond can reach
any value consistent with the ice rule $Q = 0$ at every vertex.
The bond fugacity for charge~$\phi$ is
$x_\phi = w^{-\phi(2S - \phi)}$.
Note that the ice rule is an \emph{exact} $U(1)$ conservation
law---the same Gauss law as in the $XY$ model---rather
than the modular $\mathbb{Z}_{2S}$ conservation of the clock
model.

\subsection{Decorated $XY$ model and the decomposition}
\label{app:sandwich-decorated}

To analytically isolate the continuous $U(1)$ sector of the
ice-model partition function, we construct a
``decorated $XY$ model'' that shares the $XY$ graph ensemble
(exact $U(1)$ conservation) but allows modified weights
at crossing vertices.

\subsubsection{Definition}
\label{app:sandwich-decorated-def}

For a parameter $\beta \ge 1$, define the \emph{decorated
$XY$ model} by
\begin{equation}\label{eq:app-decorated-XY}
    \tilde{Z}_{XY}^{(\beta)}
    \coloneqq
    \sum_{\{n_\ell\}}^{U(1)}
    \Bigl[\prod_\ell \tilde{x}_{|n_\ell|}\Bigr]\,
    \beta^{V_4(\{n_\ell\})}\,,
\end{equation}
where $V_4(\{n_\ell\})$ denotes the number of
degree-4 (crossing) vertices in the current configuration
$\{n_\ell\}$, i.e.\ vertices at which four bonds carry
nonzero current.
This model sums over the same $U(1)$-conserving graph ensemble
as the standard $XY$ model, but employs the ice-model bond
fugacities $\tilde{x}_{|n|}$ and enhances each loop crossing by
an additional fugacity factor~$\beta$.

The key feature of this model is that it sums over the same
$U(1)$-conserving configurations as the standard $XY$ model---and
hence excludes the $\mathbb{Z}_{2S}$ cascade configurations that
violate continuous current conservation---while using the
ice-model bond fugacities $\tilde{x}_{|n|}$ that faithfully
reproduce the microscopic weights of non-cascade graphs.

\subsubsection{Decomposition of the ice model}
\label{app:sandwich-upper-ice}

We now show that the ice-model partition function decomposes
exactly into the decorated model (which captures all
non-cascade configurations) plus exponentially
small cascade corrections.

Since $\tilde{Z}_{XY}^{(\beta)}$ now
employs the same ice-model bond fugacities $\tilde{x}_{|n|}$,
every non-cascade configuration in the ice model
is represented with identical weight in
$\tilde{Z}_{XY}^{(3/2)}$: the bond fugacities
match by construction, and the crossing enhancement
$\lambda_4 = 3/2$ (the universal Pauling weight ratio at degree-4 vertices, derived in
Secs.~\ref{app:ice-S32-zdir} and~\ref{app:ice-S32-loopgas}) is universal for all
$S \ge 3/2$.

The converse inclusion requires one further bookkeeping
step. The unrestricted $U(1)$ sum in
Eq.~\eqref{eq:app-decorated-XY} admits local current
configurations that no spin-ice configuration can realize: the
exact ice rule forbids a vertex from carrying more than two links
whose $S^z$ deviates from the vacuum in the same direction
(Sec.~\ref{app:corr-Sge2-subleading}), so a vertex with currents
$(+1, +1, +1, -3)$ conserves the $U(1)$ charge yet has no spin-ice
counterpart. We therefore understand the sum in
Eq.~\eqref{eq:app-decorated-XY} to be restricted to the
ice-realizable vertex configurations. This restriction acts on a
single vertex and treats all currents on the same footing, so
imposing it is a $U(1)$-symmetric, short-range modification of the
vertex weights---of exactly the same type as the crossing
decoration $\beta^{V_4}$, and irrelevant at the 3D $XY$ fixed point
for the same reason (Sec.~\ref{app:sandwich-irrelevance}). With
this restriction, the configurations summed in
$\tilde{Z}_{XY}^{(3/2)}$ are precisely the non-cascade
configurations of the ice model, with identical weights, and
$\delta Z_{\mathrm{cascade}}$ below is the only remaining
difference.

For $S \ge 2$, the ice model additionally includes cascade
configurations (vertices of degree $d_0 = 3$
connected by intermediate bonds of higher defect numbers).
These cascade contributions are exponentially suppressed:
as shown in Eq.~\eqref{eq:app-total-bridge}, the total
cascade penalty scales as
$w^{-(2S-3)(2S-1)(2S+4)/6}$, which is exponentially small
in $\mu/T$ for $S \ge 2$.

We denote the total cascade contribution by
$\delta Z_{\mathrm{cascade}}$.
Separating the crossing and cascade contributions, we
obtain the exact decomposition for all $S \ge 3/2$:
\begin{equation}\label{eq:app-sandwich-ice-upper}
    \tilde{Z}_{\mathrm{ice}}
    \;=\;
    \tilde{Z}_{XY}^{(3/2)}
    + \delta Z_{\mathrm{cascade}}\,,
\end{equation}
where the crossing enhancement $\lambda_4 = 3/2$ is
universal (independent of~$S$).
For $S = 3/2$, the fusion events require no intermediate bonds ($2S - 3 = 0$): the $U(1)$-breaking sector (the cubic junctions) carries no exponential suppression, consistent with the first-order 3-state Potts transition derived in Sec.~V\,B of the main text. Only for $S \ge 2$ is $\delta Z_{\mathrm{cascade}}$ exponentially suppressed, which is the regime in which the decomposition is useful.

\subsubsection{Irrelevance of the crossing decoration}
\label{app:sandwich-irrelevance}

The decorated $XY$ model~\eqref{eq:app-decorated-XY}
differs from the standard $XY$ model by the local factor
$\beta^{V_4}$ at crossing vertices.
We now argue that this decoration is \emph{irrelevant}
at the 3D $XY$ fixed point in the RG sense.

The crossing factor $\beta^{V_4}$ is a local interaction
that acts at single vertices.
In the loop-gas language, it modifies the weight of
configurations in which two loops cross at the same site.
Such a crossing can be written as a product of two
loop operators at the same vertex, which has the form
of a composite operator with scaling dimension
$\Delta_{\mathrm{cross}} = 2\Delta_1 + \ldots$, where $\Delta_1$ is
the scaling dimension of the fundamental loop operator.

More concretely, consider the operator
$\mathcal{O}_\beta \coloneqq (\beta - 1)\sum_{\bm{r}}
\delta_{V_4(\bm{r}),1}$
that generates the crossing decoration perturbatively.
At the 3D $XY$ (i.e.\ $O(2)$ Wilson--Fisher) fixed point, the
crossing operator corresponds to a local modification of the
quartic coupling $u(\phi^*\phi)^2$ away from its fixed-point
value.
Being a local, analytic perturbation, $\mathcal{O}_\beta$ can be
decomposed into contributions from singlet scalar operators
$s, s', s'', \ldots$ of the conformal field theory.
The leading singlet $s$ (the ``energy'' operator, with scaling
dimension $\Delta_s \approx 1.51$~\cite{Chester2020}) is relevant,
but it merely shifts the critical temperature $T_c$ without
changing the universality class.
The first operator capable of altering the nature of the
transition is the subleading singlet $s'$, whose scaling
dimension is $\Delta_{s'} = 3 + \omega \approx 3.79 > d = 3$,
where $\omega \approx 0.79$ is the leading correction-to-scaling
exponent determined by Monte Carlo
studies~\cite{Hasenbusch2019},
making $s'$ an \emph{irrelevant} perturbation.
All higher singlet scalars ($s'', \ldots$) have even larger
scaling dimensions and are therefore more strongly irrelevant.

The physical reasoning is straightforward: at the 3D $XY$
critical point, the correlation length diverges and the
long-distance physics is governed by the fluctuating phase
field~$\theta$.
A local modification of the crossing weight affects only the
short-distance structure of the loop gas, producing at most a
shift in $T_c$ (from the relevant operator~$s$) and subleading
corrections to scaling (from the irrelevant operator~$s'$).
The universality class itself is determined by the RG fixed
point, which is insensitive to the value of~$\beta$.

Therefore, the decorated $XY$ model
$\tilde{Z}_{XY}^{(\beta)}$ belongs to the
3D $XY$ universality class for all finite~$\beta$.
In particular,
$\tilde{Z}_{XY}^{(3/2)}$ is in the 3D $XY$
universality class.

\subsection{Complete decomposition and universality}
\label{app:sandwich-conclusion}

Combining the results of the previous subsections, we
obtain the exact decomposition, valid for
all $S \ge 3/2$:
\begin{equation}\label{eq:app-sandwich-complete}
    \boxed{
    \tilde{Z}_{\mathrm{ice}}
    \;=\;
    \tilde{Z}_{XY}^{(3/2)}
    + \delta Z_{\mathrm{cascade}}\,.
    }
\end{equation}
The crossing enhancement $\lambda_4 = 3/2$ is universal
(independent of~$S$), so the exact decomposition
involves the \emph{same} decorated $XY$ model
$\tilde{Z}_{XY}^{(3/2)}$ for all~$S$.
The only $S$-dependent quantity is the cascade correction $\delta Z_{\mathrm{cascade}}$: for $S = 3/2$ it is unsuppressed (the cubic junctions require no intermediate bonds and drive the first-order Potts transition), while for $S \ge 2$ it is exponentially suppressed.

For $S \ge 2$, the free energy per site therefore satisfies
\begin{equation}\label{eq:app-sandwich-free-energy}
    f_{\mathrm{ice}}
    \;=\;
    f_{XY}^{(3/2)} + O(e^{-c/T})
\end{equation}
for a positive constant~$c$ that depends on~$S$,
where $f_{XY}^{(3/2)}$ is the free energy of the
decorated $XY$ model with $\beta = 3/2$.
Since the crossing decoration is irrelevant at the 3D $XY$
fixed point (Sec.~\ref{app:sandwich-irrelevance}),
$f_{XY}^{(3/2)}$ has the same critical
singularity as the standard 3D $XY$ model.

\paragraph{From exact decomposition to universality.}
The exact decomposition~\eqref{eq:app-sandwich-complete}
establishes that the bare coupling of the
$\mathbb{Z}_{2S}$ discrete anisotropy---carried
by the cascade contribution
$\delta Z_{\mathrm{cascade}}$---is exponentially
suppressed: $O(e^{-c/T})$.
This is a \emph{non-perturbative} bound on the initial
condition of the RG flow.
We stress that the exact decomposition alone does not
constitute a mathematical proof that the critical
singularity is identical to that of the $XY$ model;
in principle, the exponentially small correction could
harbor a distinct (e.g.\ weakly first-order)
singularity.
However, the RG analysis of
Sec.~\ref{app:sandwich-irrelevance} shows that the
crossing decoration~$\beta^{V_4}$ and the cascade-induced
$\mathbb{Z}_{2S}$ anisotropy are both \emph{irrelevant}
operators ($\Delta > d = 3$) at the 3D $XY$ fixed point.
Combined with the exponential suppression of their bare
couplings, the RG flow drives the system to the $XY$
fixed point with no competing instabilities.

The synergy between the exact decomposition
and the RG irrelevance argument
is the key strength of this approach:
the exact decomposition provides a \emph{quantitative} bound on
the bare perturbation strength, while the RG provides
the \emph{qualitative} statement about the flow
topology.
Together, they establish that the $S \ge 2$ spin-ice
model belongs to the 3D $XY$ universality class
($\nu \approx 0.672$, $\eta \approx 0.038$)
with a degree of confidence that neither argument
alone could achieve.

\section{Bethe-Peierls Estimate of the $S=3/2$ Transition Point}
\label{app:bethe}

The mapping of the $S=3/2$ spin-ice partition function to the 3-state Potts model [Eqs.~(25) and~(26) of the main text]---established in Sec.~\ref{app:correspondence-S32} at the level of the graph ensemble, up to the short-range junction--junction exclusion of Sec.~\ref{app:corr-S32-jam}, and within Pauling's independent-vertex approximation for the vertex weights---enables an analytical estimate of the deconfinement fugacity $w_c$ via the Bethe--Peierls (cavity) approximation~\cite{Baxter1982}. The diamond lattice has coordination number $z=4$, and the Bethe lattice (Cayley tree) with the same coordination provides a controlled mean-field estimate. The Bethe approximation incorporates nearest-neighbor correlations exactly---unlike simpler Weiss-type mean-field theory---and becomes exact in the limit $z \to \infty$. For a first-order transition, which is driven by local energetics rather than by divergent critical fluctuations, the correlation length remains finite at the transition point; the only structural deficiency of the tree (the absence of loops) therefore produces only a minor quantitative correction, making the Bethe approximation particularly well-suited to this problem.

\subsection{Cavity recursion and fixed points}

Consider a $q$-state Potts model on a Bethe lattice with branching ratio $b = z - 1 = 3$. Each site has $z = 4$ neighbors, but on the tree one of these is the ``parent'' and the remaining $b = 3$ are ``children.'' The cavity method proceeds by computing the marginal probability that a given site takes spin state $\sigma$, conditional on the state of its parent, by iterating inward from the boundary of the tree. On the Bethe lattice, translational invariance reduces these belief-propagation equations to a single self-consistency condition. Specifically, let $r \coloneqq P(\sigma = \sigma_0)/P(\sigma = \sigma_1)$ ($\sigma_1 \neq \sigma_0$) denote the ratio of the probability that a site adopts a reference ordered state $\sigma_0$ to that of any other state; by the $S_q$ symmetry of the Potts model, the $q - 1$ non-reference states are equivalent. A site's cavity message $r$ is determined by the $b$ messages arriving from its children, each of which contributes a factor $(e^K r + q - 1)/(r + e^K + q - 2)$ via the Boltzmann weight $e^{K\delta_{\sigma,\sigma'}}$; the self-consistency condition is therefore
\begin{equation}
    r = \left( \frac{e^{K} r + q - 1}{r + e^{K} + q - 2} \right)^{b}.
    \label{eq:bethe_cavity}
\end{equation}
The \emph{disordered fixed point} $r = 1$ is always a solution and describes the paramagnetic phase in which all $q$ states are equally populated.

Linearizing Eq.~\eqref{eq:bethe_cavity} about $r = 1$ yields the Jacobian
\begin{equation}
    \left. \frac{dg}{dr}\right|_{r=1} = b \, t, \qquad t \coloneqq \frac{e^{K}-1}{e^{K}+q-1},
    \label{eq:bethe_jacobian}
\end{equation}
where $t$ is the standard high-temperature expansion parameter appearing in Eq.~(26) of the main text. The disordered fixed point becomes locally unstable (the \emph{disordered-phase spinodal}) at
\begin{equation}
    t_{\text{sp}} = \frac{1}{b} = \frac{1}{z - 1} = \frac{1}{3}.
    \label{eq:bethe_spinodal}
\end{equation}

\subsection{First-order coexistence}

For $q = 2$ (Ising), the ordered fixed point $r^* > 1$ bifurcates continuously from $r = 1$ at $t = t_{\text{sp}}$, giving a standard second-order transition. For $q > 2$, however, the cubic invariant $\sim \Phi^3$ in the Landau free energy tilts the free-energy landscape asymmetrically, and the nontrivial (ordered) fixed point $r^{*} > 1$ appears \emph{discontinuously} at the \emph{ordered-phase spinodal} $t_{\text{ord}} < t_{\text{sp}}$---i.e., the ordered solution does not grow smoothly from $r = 1$ but jumps to a finite value, the hallmark of a first-order transition. To locate the thermodynamic coexistence point (as opposed to the spinodal), we must compare the free energies of the two solutions. We evaluate the Bethe free energy per site via the Kikuchi cluster-variation formula~\cite{Kikuchi1951, Yedidia2001},
\begin{equation}
    -\beta f = \frac{z}{2}\ln Z_{\text{bond}} - (z-1)\ln Z_{\text{site}},
    \label{eq:bethe_free_energy}
\end{equation}
where the single-message ratio $R \coloneqq r^{1/(z-1)}$ and the site and bond partition functions are determined by the cavity messages:
\begin{align}
    Z_{\text{site}} &= R^{z} + (q-1), \label{eq:Z_site}\\[4pt]
    Z_{\text{bond}} &= e^{K}(r^{2}+q-1) + 2r(q-1) + (q-1)(q-2).
    \label{eq:Z_bond}
\end{align}

Numerical solution for $q = 3$, $z = 4$ shows that the ordered solution appears discontinuously at the \emph{ordered-phase spinodal} $t_{\text{ord}} \approx 0.3188$ ($e^{K} \approx 2.404$), where the cavity message jumps from $r = 1$ to $r^{*} \approx 2.18$. Immediately above this point, the ordered solution is still \emph{metastable} ($f_{\text{ord}} > f_{\text{dis}}$); the thermodynamic coexistence point lies slightly above it, where the two free energies cross. Remarkably, this crossing can be located in closed form. At
\begin{equation}
    e^{K_c} = 1 + \sqrt{2}, \qquad r^{*} = 2\sqrt{2}, \qquad R^{*} = \sqrt{2},
\end{equation}
one verifies from Eqs.~\eqref{eq:bethe_cavity}--\eqref{eq:Z_bond} that (i) $r^{*} = 2\sqrt{2}$ is a fixed point of the cavity recursion [the message ratio is $(e^{K_c} r^{*} + 2)/(r^{*} + e^{K_c} + 1) = \sqrt{2}$, and $(\sqrt{2})^{3} = 2\sqrt{2}$], and (ii) the free energies of the ordered and disordered solutions are \emph{exactly} equal: with $Z_{\text{bond}}^{*} = 12 + 18\sqrt{2}$, $Z_{\text{bond}}^{(1)} = 9 + 3\sqrt{2}$, $Z_{\text{site}}^{*} = 6$, and $Z_{\text{site}}^{(1)} = 3$, the free-energy difference is
\begin{equation}
    \frac{z}{2}\ln\frac{Z_{\text{bond}}^{*}}{Z_{\text{bond}}^{(1)}} - (z-1)\ln\frac{Z_{\text{site}}^{*}}{Z_{\text{site}}^{(1)}}
    = 2\ln\!\big(2\sqrt{2}\big) - 3\ln 2 = 0.
\end{equation}
The equilibrium first-order transition is therefore located at
\begin{equation}
    t_{c} = \frac{\sqrt{2}}{3+\sqrt{2}} = \frac{3\sqrt{2}-2}{7} \approx 0.3204 \;\, (\text{Bethe, } q=3,\, z=4).
    \label{eq:bethe_tc}
\end{equation}

\subsection{Mapping to the spin-ice fugacity}

The identification of the spin-ice loop-gas fugacity $\tilde{x}_1 = \frac{2}{3}w^{-2}$ with the Potts high-temperature parameter $t$ yields the Bethe-lattice estimate of the deconfinement fugacity, in closed form:
\begin{equation}
    w_c^{2} = \frac{2}{3\,t_c} = \sqrt{2} + \frac{2}{3} \approx 2.081, \qquad w_c \approx 1.443.
    \label{eq:bethe_wc}
\end{equation}
This overestimates the Monte Carlo result $w^2 \approx 2.02$ \cite{Pandey2026_S32} (where Ref.~\cite{Pandey2026_S32} uses $w$ to denote what corresponds to $w^2$ in our convention) by approximately 3.0\%, consistent with the known tendency of the Bethe approximation to overestimate the stability of the ordered phase.

A simpler---but less precise---bound follows from the disordered-phase spinodal alone:
\begin{equation}
    w_c \ge \sqrt{\frac{2}{3\,t_{\text{sp}}}} = \sqrt{2} \approx 1.414.
    \label{eq:bethe_lb}
\end{equation}
Within the same Bethe--Peierls/Pauling framework, this lower bound excludes the value $w_c \approx 1.36$ that the continuous-transition scaling formula $w_c \approx (1.859)^{1/(2S-1)}$ of the main text would give if (incorrectly) continued to $S = 3/2$, corroborating the first-order character.

\subsection{Critical endpoint in a monopole field}

At finite temperature, thermal monopoles generate string endpoints with per-endpoint fugacity $\sqrt{3}\,v/2$ [Sec.~VI\,B of the main text]. In the Potts language, this corresponds to a symmetry-breaking field $h$ that explicitly breaks the $\mathbb{Z}_3$ symmetry by favoring one of the three Potts states. The relationship between $h$ and the monopole fugacity $v$ is determined by the $\mathbb{Z}_3$ Fourier decomposition of the Potts field weight, giving the high-temperature parameter
\begin{equation}
    s \coloneqq \frac{e^{h}-1}{e^{h}+2} = \frac{\sqrt{3}\,v}{2}.
    \label{eq:s_v_map}
\end{equation}

At $h = 0$, the Bethe free energy~\eqref{eq:bethe_free_energy} has two degenerate minima as a function of the cavity message $r$: the disordered solution $r = 1$ and the ordered solution $r = r^* > 1$, separated by a free-energy barrier. As $h$ increases from zero, the field tilts the free-energy landscape, making one minimum deeper than the other. However, as long as the barrier persists, the first-order coexistence line survives [with a shifted coexistence coupling $K_c(h)$]. The critical endpoint is the value of $h$ at which the barrier just vanishes and the two minima merge into a single minimum---beyond this point, the transition is replaced by a smooth crossover.

On the Bethe lattice, the field modifies the cavity equation~\eqref{eq:bethe_cavity} to
\begin{equation}
    r = e^{h}\left(\frac{e^{K}r + q - 1}{r + e^{K} + q - 2}\right)^{b}.
    \label{eq:bethe_cavity_h}
\end{equation}
The prefactor $e^h$ biases the cavity message: when $h > 0$, the iterative map favors $r > 1$ (the field-aligned state), breaking the symmetry between the $q$ Potts states.

To locate the critical endpoint, we seek the point in the $(K, h)$ plane where the free-energy barrier between the ordered and disordered solutions disappears. In the language of the iterative map $g(r) \coloneqq e^{h} f(r)^{b}$ with $f(r) = (e^{K}r + q - 1)/(r + e^{K} + q - 2)$, this requires three simultaneous conditions:
\begin{enumerate}
    \item[(i)] \emph{Fixed point:} $g(r) = r$. The cavity message must be self-consistent.
    \item[(ii)] \emph{Marginal stability:} $g'(r) = 1$. For $h < h_c$, the map $g(r) - r$ has two stable zeros (the ordered and disordered fixed points) and one unstable zero between them. At the critical endpoint, two of these zeros---one stable and one unstable---coalesce, so the slope of $g$ at the merged fixed point equals unity.
    \item[(iii)] \emph{Vanishing inflection:} $g''(r) = 0$. This ensures that the coalescence is complete: the function $g(r) - r$ has a cubic tangency (rather than a simple tangency with a residual barrier), confirming that the free-energy barrier has fully disappeared.
\end{enumerate}
These three equations determine the three unknowns $(K_c, h_c, r_c)$ at the critical endpoint.

From condition (iii), writing $g''(r) = e^h b f^{b-1}[(b-1)(f')^2 + f f''] = 0$ and noting $e^h f^{b-1} \neq 0$, the inflection condition $(b-1)(f')^{2}+f\,f'' = 0$ (with $b = 3$, i.e., $2(f')^2 + ff'' = 0$) reduces to
\begin{equation}
    r = \frac{e^{2K}+e^{K}-4}{e^{K}}.
    \label{eq:rc_cond}
\end{equation}
Substituting into the remaining two conditions and eliminating $r$ and $h$ yields, remarkably, a single quadratic equation for $x \coloneqq e^{K}$:
\begin{equation}
    x^{2}+x-8 = 0 \;\implies\; e^{K_c} = \frac{\sqrt{33}-1}{2}.
    \label{eq:bethe_Kc_ep}
\end{equation}
The remaining critical-endpoint parameters are determined analytically:
\begin{align}
    r_c &= \frac{\sqrt{33}+1}{4}, \quad f_c = \frac{\sqrt{33}-1}{4}, \label{eq:bethe_rc_fc}\\[4pt]
    e^{h_c} &= \frac{r_c}{f_c^{\,3}} = \frac{16(\sqrt{33}+1)}{36\sqrt{33}-100} \approx 1.010.
    \label{eq:bethe_hc}
\end{align}
The critical field $h_c \approx 0.010$ is small, reflecting the weakness of the first-order transition at $q = 3$ (marginally above the tricritical value $q_c = 2$).

Translating back to the spin-ice monopole fugacity via Eq.~\eqref{eq:s_v_map}:
\begin{equation}
    v_c = \frac{2}{\sqrt{3}}\,\frac{e^{h_c}-1}{e^{h_c}+2} \approx 0.004, \quad \frac{T_c}{J} = \frac{1}{2|\ln v_c|} \approx 0.09.
    \label{eq:bethe_vc}
\end{equation}
This estimate indicates that the $S = 3/2$ first-order transition survives only in a narrow low-temperature window $T \lesssim J/11$, consistent with the expectation that the $q = 3$ Potts first-order transition is weak. Beyond $v_c$, the transition is rounded into a crossover, restoring the generic behavior shared by all other spin values.

\end{document}